\documentclass[twocolumn,rmp,superscriptaddress, showpacs]{revtex4}

\usepackage{amsmath}
\usepackage{amsthm}
\usepackage{framed}
\usepackage{soul}
\usepackage[caption=false]{subfig} 

\def\bra#1{\mathinner{\langle{#1}|}}
\def\ket#1{\mathinner{|{#1}\rangle}}

\newcommand{\be}{\begin{equation}}
\newcommand{\ee}{\end{equation}}
\newcommand{\bq}{\begin{eqnarray}}
\newcommand{\eq}{\end{eqnarray}}

\newtheorem*{lemma}{Cleaning lemma}

\usepackage{amsfonts}
\usepackage{amsthm}
\usepackage{graphicx}
\usepackage{enumerate}
\usepackage{xcolor}

\usepackage{hyperref}
\definecolor{darkblue}{RGB}{0,0,127}
\definecolor{darkgreen}{RGB}{0,150,0}
\hypersetup{breaklinks, colorlinks, linkcolor=darkblue, citecolor=darkgreen, filecolor=red, urlcolor=blue}

\begin{document}
\title{Quantum memories at finite temperature}

\author{Benjamin J. Brown}
\affiliation{Quantum Optics and Laser Science, Blackett Laboratory, Imperial College London, Prince Consort Road, London, SW7 2AZ, United Kingdom}
\affiliation{Niels Bohr International Academy, Niels Bohr Institute, Blegdamsvej 17, 2100 Copenhagen, Denmark}
\author{Daniel Loss}
\affiliation{Department of Physics, University of Basel, Klingelbergstrasse 82, CH-4056 Basel, Switzerland}
\author{Jiannis K. Pachos}
\affiliation{School of Physics and Astronomy, University of Leeds, Leeds LS2 9JT, United Kingdom}
\author{Chris N. Self}
\affiliation{Quantum Optics and Laser Science, Blackett Laboratory, Imperial College London, Prince Consort Road, London, SW7 2AZ, United Kingdom}
\affiliation{School of Physics and Astronomy, University of Leeds, Leeds LS2 9JT, United Kingdom}
\author{James R. Wootton}  
\affiliation{Department of Physics, University of Basel, Klingelbergstrasse 82, CH-4056 Basel, Switzerland}


\begin{abstract}
To use quantum systems for technological applications we first need to preserve their coherence for macroscopic timescales, even at finite temperature. Quantum error correction has made it possible to actively correct errors that affect a quantum memory. An attractive scenario is the construction of passive storage of quantum information with minimal active support. Indeed, passive protection is the basis of robust and scalable classical technology, physically realized in the form of the transistor and the ferromagnetic hard disk. The discovery of an analogous quantum system is a challenging open problem, plagued with a variety of no-go theorems. Several approaches have been devised to overcome these theorems by taking advantage of their loopholes. Here we review the state-of-the-art developments in this field in an informative and pedagogical way. We give the main principles of self-correcting quantum memories and we analyze several milestone examples from the literature of two-, three- and higher-dimensional quantum memories. 
\end{abstract}

\maketitle
\tableofcontents

\section{Introduction}

Quantum mechanics holds the potential for performing computational simulations~\cite{Feynman82} and information processing tasks~\cite{Deutsch85, Shor94} much faster than classical technologies. To outperform modern classical machines, a quantum computer must manipulate hundreds of computational qubits to follow a program of millions of quantum logical operations~\cite{Wecker14, Hastings14, Poulin14}. To realize such a feat using a real physical system, we must preserve a vast entangled quantum state over a long duration while computations are executed. Recognizing that any device exists in an ambient environment at non-zero temperature, we see that probabilistic errors will continually disrupt experimental efforts to coherently control quantum states. It is widely understood that the problem of decoherence is among the largest obstacles impeding the realization of quantum technologies.

The breakthrough that validated the practical possibility of quantum computation was the discovery of quantum error correction~\cite{Shor96, Steane1996, LidarBrun}. The principle behind quantum error correction is to use a redundancy of physical quantum systems to encode a small number of logical computational qubits. We are then able to realize error-correcting protocols by measuring auxiliary physical systems to identify errors affecting the encoded information, and subsequently correct for them. Provided the incident noise is suitably low, we can achieve robust quantum states for an arbitrarily long time using a suitably large redundancy of ancillary systems~\cite{Preskill98}. While we have recently seen impressive experimental progress in the direction of realizing quantum error-correcting codes~\cite{Reed, Nigg, Barends, Kelly15, Corcoles15}, such active quantum error correction schemes present an uphill challenge in preparing the complex entangled states that encode quantum information, and in reducing incident laboratory noise to the encoded states below a threshold value.



A novel alternative to active error correction would be the discovery of a {\em self-correcting quantum memory}~\cite{Kitaev03, Bacon06}; a physical system that is able to reverse the effects of errors by itself. This could be achieved using a many-body Hamiltonian whose energy landscape suppresses large errors that directly affect encoded quantum information. This suppression can be increased indefinitely by increasing the size of the system, allowing arbitrarily large storage times without the need to manually repair the memory. 

Ideally, a self-correcting memory must be robust to all forms of physical noise including both finite-temperature effects and small imperfections on the ideal system. Our main focus here will be the study of many-body quantum systems coupled to a thermal bath. To the best of our knowledge, there are no known quantum systems that can preserve coherent quantum information for arbitrarily long timescales at finite temperature, and as such, this will be the main consideration of this Review~\cite{Bacon06}. It is also important to recognise that no system will ever be free from weak perturbations such as, for instance, an external magnetic field. Such imperfections may also affect the ability of a system to preserve quantum information so to this end we discuss in parallel known results on the effects of local perturbations on the considered many-body models.

Self correction is the principle that lies behind the storage of classical information in magnetic media. Here, classical bits of information are encoded in the magnetic orientation of some ferromagnetic material. In such a system, thermal noise can cause individual spins to flip, but they will be reoriented quickly by the macroscopic effect of their neighbouring spins. As such, the ferromagnet is robust to a spontaneous change of orientation due to the collective behavior of some Avogadro's number of physical spins.

It is an exciting and fundamental question of nature, and indeed the topic of this Review, as to whether we can find macroscopic quantum systems to maintain coherent quantum information while simultaneously equilibrating with its surrounding environment. The discovery of such a system will provide a beautiful solution for one of the largest puzzle pieces required to achieve scalable quantum computation. In addition to the remarkable practical applications, the realization of a self-correcting universal quantum computer is also of significant fundamental interest. A macroscopic system that is capable of simulating arbitrarily complex quantum phenomena would provide a powerful demonstration that quantum mechanics is not restricted to only the microscopically accessible parts of the Universe~\cite{Farrow}.

Many physical systems have been considered for the storage of qubits, for instance spin qubits in quantum dots~\cite{Loss98,Kloeffel13}, the ground space of ions~\cite{Harty14}, superconducting systems~\cite{Devoret13} or other solid state devices~\cite{Fuchs11, Saeedi13}. For a concise review and comparison of different schemes see Ref.~\cite{Schoelkopf08}. A constant challenge for these schemes is increasing coherence times using mechanisms such as an energy gap to separate excited states from their ground space. Nevertheless, this timescale will always be finite in nature, placing a limit on the computations that can be performed without some error-correction protocol. Ideally, we would like to construct a quantum memory that can store quantum states for times that can be tuned arbitrarily using a variable parameter, such as system size.

Furthermore, mediating interactions that entangle qubits encoded in atomic systems requires a coupling bus, for example the vibrational modes when considering trapped-ion quantum computation or an optical cavity in the case of neutral atoms. These additional structures are subject to thermal errors and are therefore prone to decoherence while performing computational tasks. Ultimately we seek a system that is able to preserve coherent quantum states for timescales much longer than the time it takes to perform logical operations on the encoded states. This would allow the execution of arbitrarily long quantum algorithms given sufficient quantum resources.

Topologically ordered many-body systems~\cite{Wenmanybody} play an important role in the study of self correction. They posses degenerate ground states that cannot be distinguished by local observables. In this feature lies the appeal of topological models as candidate systems for quantum memories; if quantum information is locally indistinguishable, local noise cannot have irreversible effects. Moreover, properties of topologically ordered models have been shown to be stable against weak local perturbations acting on the ideal model Hamiltonian at zero temperature~\cite{Kitaev03, BravyiHastingsMichalakis10}. This is an important feature, as we hope that the robust features of topological phases will still be present under realistic conditions where the system will certainly be subject to small imperfections.

In addition to their locally inaccessible degrees of freedom, topological quantum systems are of further interest due to their amenable features for realizing fault-tolerant quantum computation. This has been the subject of intense study~\cite{Nayak08, Pachos} for two-dimensional anyonic systems, where quantum information can be stored in collective states of anyons and processed through their braiding. The topological nature of the models again ensures a degree of protection against local noise as long as the anyons are kept well separated. Models that achieve universal fault-tolerant quantum computation by anyon braiding are well known~\cite{Freedman02, Kitaev03, BrennenPachos2007}.

The study of topological quantum computation has extended far beyond the study of anyon braiding. Fault-tolerant computational operations are also realized by the manipulation of holes~\cite{Raussendorf06, Bombin09, Wootton_rev}, twist defects~\cite{Bombin10a, BarkeshliJianQi13, Barkeshli14} or by other means~\cite{WoottonLahtinenDoucotPachos}. It is also noteworthy that research in the direction of computation using experimentally amenable anyon models~\cite{Bravyi06, Zilberberg08} that do not support a universal set of topological computational operations has led to schemes to supplement such systems with non-topological operations to complete their computational gate set~\cite{BravyiKitaev05, WoottonLahtinenPachos2010}. Consideration of topologically ordered systems as a basis for quantum memories therefore allows us to draw from this wealth of established knowledge to realize a fault-tolerant computational model.

In spite of many known interesting and attractive models, we are yet to rigorously prove the existence of a low-dimensional passively protected quantum memory. It is the purpose of this Review to highlight the challenges involved in finding systems that maintain their quantum character at finite temperatures, and to discuss new models that come towards a solution to this problem. The present Review is separated into two distinct parts. In the first part we introduce the field, and paint a picture that demonstrates the difficulty in discovering a stable memory. We show this by means of explicit introductory examples, as well as discussions of rigorously proved no-go theorems for the finite temperature stability of large classes of systems. In the second part we discuss new models that come some way towards finite-temperature quantum stability over macroscopic timescales. We offer insights into how such models are discovered and we assess their favorable features, and their drawbacks. In doing so we identify underlying open problems and discover established tools that can be used to approach this actively studied and exciting field.

The present Review takes the following structure. In Secs.~\ref{Sctn:HamiltoniansBasics} and~\ref{Sctn:FiniteTemperatures} we introduce a common notation, concepts in quantum error correction and the analytical and numerical methods for examining finite temperature. We conclude Sec.~\ref{Sctn:FiniteTemperatures} with a rigorous set of conditions that we demand of a quantum memory, together with a list of attractive features that would make a model suitable for quantum computation and plausible for experimental realization. In Sec.~\ref{Sctn:NoGoTheorems}, we review the plethora of no-go theorems established so far with respect to passive error correction. We use this study to chart the landscape of the proposed models. The latter Sections, Secs.~\ref{Sctn:HighD}, \ref{Sctn:InteractingAnyons}, \ref{Sctn:ThreeDimensions}, \ref{Sctn:EntropicProtection} and~\ref{Sctn:Subsystems} offer a comprehensive review of current actively studied models that demonstrate favorable properties for self correction. In Sec.~\ref{Sctn:Outlook} we conclude with an overview of the current state of the field where we discuss open problems that remain unsolved.

\section{Local Hamiltonians and Quantum Error Correction}
\label{Sctn:HamiltoniansBasics}

The study of quantum memories at finite temperature lies at the intersection of the fields of quantum error correction, condensed-matter physics and statistical mechanics. We therefore require a unifying language that captures the breadth of physics covered by all of these fields. We find such a language in the {\em stabilizer formalism}. This formalism, initially introduced as an efficient description of quantum error correcting codes~\cite{GottesmanThesis}, provides a natural way of understanding the Hamiltonian models considered here. 

The stabilizer formalism efficiently describes quantum error correcting codes using a list of commuting Pauli operators, known as {\em stabilizers}. We can use this operator description from quantum error correction to write down a large class of degenerate Hamiltonians. The Hamiltonians we obtain this way have ground state subspaces that correspond to the code space of some quantum code, and its excited states reflect the errors the code suffers.

We remark that the stabilizer formalism does by no means describe general many-body Hamiltonians that are capable of robust information storage. Indeed, quantum-double models~\cite{Kitaev03}, string-net models~\cite{Levin05, WalkerWang}, subsystem codes~\cite{Poulin05}, Turaev-Viro codes~\cite{Koenig10} and non-Abelian stabilizer codes~\cite{Xi14} are only a few of the classes of interesting Hamiltonian models that are not represented by the stabilizer formalism. In this Review however we largely restrict our attention to stabilizer models as they provide an analytically tractable class of Hamiltonians upon which many of the developments in this field have been based.

In Subsec.~\ref{Sctn:CPHamiltonians} we begin by introducing the class of models that we will mainly be concerned with here, namely {\em commuting Pauli Hamiltonians}. In Subsec.~\ref{Sctn:StabilCodes} we give a comprehensive overview of the stabilizer formalism that enables us to identify error-correcting procedures for the considered Hamiltonians. We review how one might perform error correction on either a quantum code or a commuting Pauli Hamiltonian in Subsec.~\ref{Sctn:ErrorCorrection}. We then study an explicit and extensively studied example of a commuting Pauli Hamiltonian in Subsec~\ref{Sctn:2dTC}, namely, {\em Kitaev's toric code model}. In addition to providing a straight-forward example of a commuting-Pauli Hamiltonian, the toric code model also exhibits topological order and gives rise to anyonic quasi-particle excitations. We discuss at length the topological nature of the toric code in Subsec.~\ref{Sctn:AnyonicPicture}. We conclude this Section by discussing the stability of the gap at zero temperature in Subsec.~\ref{subsec:stability}; a feature presented naturally by topologically ordered systems, and an important feature to consider while searching for stable quantum memories.

\subsection{Commuting Pauli Hamiltonians}
\label{Sctn:CPHamiltonians}

We first define the Pauli group $\mathcal{P}_n = \mathcal{P}_1^{\otimes n}$ acting on $n$ distinct two-level quantum systems that we refer to as qubits. The Pauli group, $\mathcal{P}_1$, includes, up to phases, the Pauli matrices
\begin{equation}
X = 
\begin{pmatrix}
0 & 1 \\
1 & 0
\end{pmatrix}, \,\, Y = 
\begin{pmatrix}
0 & -\text{i} \\
\text{i} & 0
\end{pmatrix}, \,\, Z = 
\begin{pmatrix}
1 & 0 \\
0 & -1
\end{pmatrix},
\end{equation}
and identity, $\openone$. We will often use indices with elements of $  \mathcal{P}_1$ to describe the elements of $\mathcal{P}_n$ that act on single qubits. For instance, we can write the operator $U \in \mathcal{P}_1$ that acts on the $j$th physical qubit where $1 \le j \le n$ using the notation $U_j \in \mathcal{P}_n$. Written explicitly, we have 
\begin{equation}
U_j \equiv \underbrace{\openone \otimes \openone \otimes \dots \otimes \openone}_{j-1} \otimes U \otimes \underbrace{\openone \otimes \openone \otimes \dots \otimes \openone}_{n-j}.
\end{equation}
This notation is particularly convenient as we can generate the group $\mathcal{P}_n$ up to phases using only the single qubit operators $X_j$ and $Z_j$. We finally remark that all elements of $ \mathcal{P}_n$ necessarily have eigenvalues $\pm 1$, which is seen from the fact that $U^2 = \openone$ for all $U \in  \mathcal{P}_n$.

\begin{figure}
\includegraphics{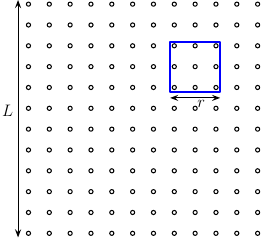}
\caption{\label{LocalInteraction}(Color online) A regular two-dimensional lattice of qubits with linear size $L$. All Hamiltonian interactions are contained within a box of linear size $r$, shown in blue.}
\end{figure}

Having introduced the Pauli group acting on $n$ qubits, we are able to write down Pauli Hamiltonians that describe interactions between the qubits of a regular lattice. Consider a $D$-dimensional lattice of qubits of linear size $L$, as shown in Fig.~\ref{LocalInteraction}. The $ n \sim L^D$ qubits of the lattice are arranged in a structure that depends on the model we introduce. We write down Hamiltonians of the type
\begin{equation}
H = - \frac{\Delta}{2} \sum_{j} S_j, 
\label{Eqn:StabilHam}
\end{equation}
where we sum over a set $\mathcal{I}$ of Hermitian interaction terms $S_j \in \mathcal{I}$ such that $\mathcal{I}$ is a subset of $\mathcal{P}_n$.

We must impose physical constraints on Hamiltonian~(\ref{Eqn:StabilHam}). We demand that the Hamiltonian interactions are local. We therefore constrain all elements of $\mathcal{I}$ to have non-trivial i.e., non-identity, support only on qubits that can be contained within a box on the lattice of linear size no greater than $r$, where $r$ is independent of the lattice size. We show a box of linear size $r=3$ in Fig.~\ref{LocalInteraction}. Additionally we must bound the interaction strength of the Hamiltonian. To this end we impose that $\Delta$ is a positive constant independent of system size. Similarly, we enforce that each qubit supports only a constant number of interaction terms independent of the size of the system.

In general, Hamiltonians that are the sum of local elements of $\mathcal{P}_n$ are intractable for study. We are able to impose additional restrictions that enable us to find solvable classes of Hamiltonians. We first demand that elements $S_j \in \mathcal{I}$ commute, i.e.
\begin{equation}
[S_j,S_l] \equiv S_j S_l - S_lS_j  =0, \quad \forall j,\, l. \label{AbelianStabilizers}
\end{equation} 
In addition to this, we consider only frustration-free Hamiltonians. Specifically, for all Hamiltonian ground states, $\ket{\psi}$, all elements $S_j \in \mathcal{I}$ satisfy the condition
\begin{equation}
S_j\ket{\psi}=(+1)\ket{\psi} , \label{EQN:stabilcond}
\end{equation}
where ground states are described by an orthonormal basis $\ket{\psi_\mu}$ whose states are indexed by integers $\mu$ such that
\begin{equation}
\ket{\psi} = \sum_\mu c_\mu \ket{\psi_\mu}, \label{Eqn:GroundState}
\end{equation} 
with $\sum_\mu|c_\mu|^2 = 1$. Conditions~(\ref{AbelianStabilizers}) and~(\ref{EQN:stabilcond}) enable us to employ the stabilizer formalism that is described in the following Subsection.  

We finally remark on the excited eigenstates of Hamiltonian~(\ref{Eqn:StabilHam}). The Hamiltonian terms $S_j \in \mathcal{I}$ are elements of $\mathcal{P}_n$, and therefore all satisfy the property that $S_j^2 = \openone$. It follows from this that the eigenvalues, $M_j$, of operators $S_j$ take only values $ \pm 1$. We are therefore able to specify excited states of Hamiltonian~(\ref{Eqn:StabilHam}) using the list of eigenvalues $M_j$. The Hilbert space can then be described completely by the ground eigenspace of $H$, together with the list of eigenvalues $\{ M_j \}$. The excited eigenstates of $H$ are achieved by applying operators $E\in \mathcal{P}_n$ to states $\ket{\psi}$ in the ground eigenspace of $H$. It is therefore convenient to write Hamiltonian eigenstates $E \ket{\psi}$ and omit any $M_j$ notation. The values $M_j$ are determined by commutation relations $S_j E = M_j E S_j$. The energy eigenvalue $\varepsilon_E$ for eigenstate $ E \ket{\psi}$ follows immediately from the values $M_j$ such that
\begin{equation}
\varepsilon_E = -\frac{\Delta}{2} \sum_j M_j. 
\end{equation}

\subsection{The Stabilizer Formalism}
\label{Sctn:StabilCodes}

Many quantum error-correcting codes can be described using the stabilizer formalism~\cite{GottesmanThesis}. This formalism shares many parallels with the commuting Pauli Hamiltonians introduced in the previous Subsection. Quantum error-correcting codes describe a subspace of states called the {\em code space}. We denote an orthonormal basis of states in the code space with vectors $\ket{\psi_\mu}$ for $1 \le \mu \le 2^k $. The code space of a stabilizer code is specified by the {\em stabilizer group}. The stabilizer group $\mathcal{S}$ is an Abelian subgroup of $\mathcal{P}_n$ where we have defined $\mathcal{P}_n$, the Pauli group for $n$ qubits, in the previous Subsection. A stabilizer group thus defines a quantum error-correcting code such that the code subspace is the common $+1$ eigenspace of all the elements of the stabilizer group. Formally, we write this property such that all elements $S_j \in \mathcal{S}$ satisfy the condition
\begin{equation}
S_j \ket{\psi_\mu} = (+1) \ket{\psi_\mu}, 
\end{equation}
for all encoded states $\ket{\psi_\mu}$, thus specifying the code space of the code. 

The stabilizer group can be described using a generating set of $m \leq n$ stabilizers, listed in angle braces
\begin{equation}
\mathcal{S} = \langle S_1, \, S_2,\,\dots,S_m \rangle,
\end{equation}
where all $S_j$ of the generating set are {\em independent} elements of the stabilizer group, i.e. the stabilizer generators satisfy the condition that $\prod_j S_j^{n_j} = \openone$ with $n_j \in \{ 0, 1 \}$ only for $n_j = 0$ for all $j$. A code of $n$ qubits that is generated by $m$ independent stabilizer generators will encode $k = n - m$ logical qubits. 

Encoded logical qubits are manipulated by the group of logical operators $\mathcal{L}$. The group $\mathcal{L}$ is denoted concisely by a generating set of operators $\overline{X}_j,\, \overline{Z}_j \in \mathcal{P}_n $ for $j = 1,2, \dots ,k$. Operators $\overline{X}_j$ and $\overline{Z}_j$ commute with all elements of $\mathcal{S}$, and with logical operators $\overline{X}_l$ and $\overline{Z}_l$ for $ l \not= j $. Operators $\overline{X}_j$ and $\overline{Z}_j$ mutually anticommute. The logical operators therefore generate the Pauli group over the $k$ encoded logical qubits.

We remark that logical operators are not unique with respect to their action upon the code space, but are only unique {\em up to multiplication by stabilizer operators}. We consider logical operators $\overline{L}, \overline{L}' \in \mathcal{L}$ that differ only by multiplication by an arbitrary element $S_j \in\mathcal{S}$, i.e. $\overline{L}' = S_j\overline{L}$. Then, using the commutation relation $[S_j, \overline{L}] = 0$ we observe that
\begin{equation}
\overline{L}' \ket{\psi_\mu} = S_j \overline{L} \ket{\psi_\mu} =  \overline{L} S_j  \ket{\psi_\mu} = \overline{L} \ket{\psi_\mu}, \quad \forall \mu,
\end{equation}
thus demonstrating that the action of $\overline{L}'$ and $\overline{L}$ on the code space are equivalent.

We finally introduce the definition of the weight of an operator, and the distance of a quantum error-correcting code. These are useful terms when comparing different error-correcting codes. The {\em weight} of operator $U$, denoted $\text{wt}(U)$, is the number of qubits that $U$ has non-trivial support over. For instance, the operator $U = X_2 X_{3}$ has $\text{wt}(U) = 2$, as it acts non trivially on qubits $2$ and $3$. We use the weight to find the {\em distance}, $d$, of a quantum error-correcting code. To define the distance we consider {\em least-weight} non-trivial i.e., non-identity, logical operators of a code $\overline{L}^* \in \mathcal{L}$ that satisfy the inequality $\text{wt}(S_j \overline{L}^*) \ge \text{wt} (\overline{L}^*) $ for all elements $S_j \in \mathcal{S}$. The distance of a code is then defined as the weight of the least-weight non-trivial logical operator $ \overline{L}^* \in \mathcal{L}$ with the lowest weight. We write this definition concisely such that
\begin{equation}
d = \min_{S_j \in \mathcal{S}}\min_{\overline{L}\in \mathcal{L}}\text{wt}(S_j \overline{L}). 
\end{equation}
A quantum error-correcting code is able to tolerate and correct for as many as $d/2-1$ errors on distinct physical qubits with certainty. In general however a code can probabilistically tolerate errors with weight greater than $d/2$, provided the errors incident to the system do not find adversarial configurations with respect to the {\em error-correction protocol}, as discussed below. The quantum error-correcting codes we review are typically designed to correct low-weight errors with high probability. Here, where {\em correctable errors} are discussed, we will typically consider errors $E$ such that $\text{wt}(E) / n \ll 1$.

Having introduced the stabilizer formalism, we are now able to explicitly see the correspondence between stabilizer quantum error-correcting codes and frustration free commuting Pauli Hamiltonians. The ground states of Hamiltonian~(\ref{Eqn:StabilHam}) are the common $+1$ eigenspace of the set of commuting local interaction terms, $\mathcal{I}$. We are therefore able to identify the ground space of Hamiltonian~(\ref{Eqn:StabilHam}) with the code space of a stabilizer group $\mathcal{S}$, whose generators are included in $\mathcal{I}$. In general, $\mathcal{I}$ can be an over-complete generating set and include some extra elements that are not independent of the others. In Subsec.~\ref{Sctn:CPHamiltonians}, we specified only that elements of $\mathcal{I}$ are local with respect to the geometry of its underlying lattice.

\subsection{Quantum Error-Correction Protocols}
\label{Sctn:ErrorCorrection}

The stabilizers of a quantum error-correcting code are designed to detect the typical errors suffered by encoded quantum states. Provided noise incident to a code occurs at a suitably low rate, we can correctly identify errors with a probability that increases with the distance of the code. This is due to the celebrated {\em accuracy threshold theorem}~\cite{Shor96, AharonovBen-Or, Kitaev97, Preskill98, Knill98, Aliferis06, Aliferis07,Terhal13}. Once we have identified an error, we can subsequently find an operator that reverses the error and thus corrects for the incident noise. Here we elaborate on the quantum error-correction procedure.

We consider encoded states $\ket{\psi}$ decohering due to a local quantum channel. Given the vast space of realistic noise channels a physical quantum system can suffer, we might suspect that one cannot possibly expect to reverse incident noise. However, if the noise acting on the system is local, and occurs at a sufficiently low rate, then the act of measuring stabilizer operators projects the encoded state onto a state arbitrarily close to $E \ket{\psi}$, where $E\in \mathcal{P}_n$ is some correctable low-weight Pauli error acting on the state. Having measured the stabilizer operators, attempting to determine and correct for the discrete set of Pauli errors $E$ becomes a much more palatable challenge.

In addition to projecting local noise onto an error from the Pauli group $ \mathcal{P}_n$, stabilizer measurements $S_j$ also furnish us with information that we can use to estimate the Pauli error $E$. The set of measurement outcomes, $M_j = \pm 1$, for stabilizers $S_j$ are referred to as the {\em error syndrome}. Values $M_j$ are determined by the commutation relation
\begin{equation}
S_j E = M_j E S_j, \label{Eqn:Commutator}
\end{equation} 
which is seen by consideration of the eigenvalue equation $S_j E \ket{\psi} = M_j E S_j \ket{\psi} = M_j E \ket{\psi} $ that corresponds to the measurement of operator $S_j$. Obtaining the syndrome data greatly restricts the possible errors that could have occurred, as the incident errors must be consistent with the syndrome information.

There are many errors that can give rise to a given syndrome. To reverse an error $E$, we consider correction operators, $C\in \mathcal{P}_n$, that are consistent with the measured syndrome, i.e., such that $S_jC = M_j C S_j$. If the correction operator satisfies the condition $CE \in \mathcal{S}$, then application of $C$ will restore the quantum error-correcting code to its initial state since $CE \ket{\psi} = \ket{\psi} $ if and only if $CE \in \mathcal{S}$. Alternatively, we may find a correction operator such that $CE$ is a non-trivial logical operator. In this case, we introduce errors that effect the encoded information. We use a {\em decoder} to attempt to find a correction operator that returns the code to its initial state.

In addition to the error syndrome, the decoder uses information about the error model to find a correction operator that will most-likely return the code to its initial state. Specifically, a decoder evaluates the probability $P(\overline{L})$ that the error that caused the syndrome was a member of an {\em equivalence class} of errors, where each member is equivalent in the sense that they all have the same effect on the encoded information. Explicitly, the probability that an error is a member of a given equivalence class is determined by the equation
\begin{equation}
P(\overline{L}) = \sum_j \text{prob}( S_j C \overline{L}) \label{Eqn:DecoderCalculation}
\end{equation}
where $\text{prob}(E)$ is the probability that Pauli error $E$ is introduced by the known noise model, $C$ is an arbitrary choice of correction operator consistent with the error syndrome, $\overline{L} \in \mathcal{L}$ are the logical operators of the code, and where we sum over all elements $ S_j \in \mathcal{S}$. The decoder will then choose the correction operator $C\overline{L}$ as a representative member of the most likely equivalence class to attempt to recover encoded information. 

In general it is not always an efficient task to find the most likely equivalence class for which the true error is a member. Instead, we can devise efficient decoding algorithms that approximately determine the most likely class of errors of which the error incident to the code is a member. In App.~\ref{Sctn:Decoders} we describe in detail a specific implementation of an efficient decoder, namely the clustering decoder, which is commonly used throughout this Review. The clustering algorithm is very versatile for decoding the quantum error-correcting codes defined by local commuting Pauli Hamiltonians.

The correspondence between the syndrome of a quantum error-correcting code and the excited states of commuting Pauli Hamiltonians means that all the error-correction procedures explained here can be adapted to correct errors suffered by states encoded in the ground space of commuting Pauli Hamiltonian models.

\subsection{The Toric Code}
\label{Sctn:2dTC}

\begin{figure}
\includegraphics[scale=2]{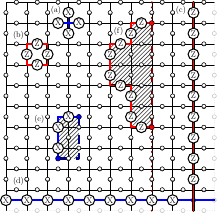}
\caption{\label{ToricCodeStabilizers}(Color online) The toric code lattice. Qubits, shown by white points, are arranged on the edges of a square lattice. The left and dotted right boundary are unified and similarly the top and dotted bottom edges are unified. (a) and (b) show a star and plaquette operator, respectively. (c) and (d) show logical operators $\overline{Z}_1$ and $\overline{X}_1$, respectively. The Pauli operator in the bottom right corner of each operator is omitted to show the crossing point. (e) A small error that is easily corrected. The error syndromes are marked by points at the end of the error string. (f) A string error with syndromes separated by half the code distance. We cannot reliably correct this error as there are two available correction operators with equal weight, one of which will lead to a logical error on the code space. The support of the two distinct lowest-weight correction operators, $C'$ and $C''$, are shown in red dashed and dotted lines respectively.}
\end{figure}

We illustrate the concepts introduced above using {\em Kitaev's toric code model}~\cite{Kitaev03}. A comprehensive study of the toric code model from the point of quantum error correction can be found in Ref.~\cite{DennisKitaevLandahlPreskill}. Qubits are arranged on the edges of a two-dimensional square lattice of linear size $L$ with periodic boundary conditions, i.e. a torus, as shown in Fig.~\ref{ToricCodeStabilizers}. The stabilizer group is generated by star, $A_v$, and plaquette, $B_p$, operators for each vertex, $v$, and face, $p$, of the lattice. The star and plaquette operators are defined such that
\begin{equation}
A_v = \prod_{\partial j \ni v} X_j, \quad B_p = \prod_{ j \in \partial p  } Z_j.
\end{equation} 
Written colloquially, a star operator $A_v$ is the tensor product of Pauli-X operators supported on the edges $j$ that include vertex $v$ in its boundary, $\partial j$, and the plaquette operator is the tensor product of Pauli-Z operators acting on the edges that bound a face $p$, where the boundary of face $p$ is denoted $\partial p$. We show examples of such operators in Fig.~\ref{ToricCodeStabilizers}(a) and~(b). Star and plaquette operators share either zero or two common qubits and therefore commute. The set of all the star and plaquette operators generate the stabilizer group.

When defined on a torus, the toric code model encodes two logical qubits. The logical operators of the model correspond to extensive strings of Pauli-X and Pauli-Z operators that wrap around non-trivial cycles of the torus. We show two such operators in Fig.~\ref{ToricCodeStabilizers}(c) and~(d), respectively. One can see from the digram that these logical operators have distance $d = L$, the linear size of the system. It is easily checked that these operators commute with all the stabilizers of the code, but mutually anticommute. The displayed logical operators overlap at a single edge of the lattice where, in the diagram, the Pauli operators are omitted.

Error correction for the toric code is particularly intuitive as its syndrome follows a simple geometrical structure. Errors can be regarded as `strings' on the lattice. We show two such errors composed of Pauli-X and Pauli-Z operators in Fig.~\ref{ToricCodeStabilizers}(e) and~(f) respectively. We cannot detect the positions of the string-like errors. Instead, the syndrome measurements identify the end points of the string-like errors. The decoding procedure then consists of using the known end points of the strings and trying to determine the least-weight operator $E$ that may have caused the syndrome. The decoder subsequently returns a string-like correction operator $C$, that corresponds to a string that reconnects all the stabilizers that returned a $-1$ measurement outcome. If the errors are very few and error strings are very short with respect to the size of the lattice, it is straight forward to identify likely correction operators. This path will connect the end points of string errors such that $CE$ will correspond to a stabilizer operation with high probability, i.e. $CE \in \mathcal{S}$. In Fig.~\ref{ToricCodeStabilizers}(e) we show a dotted line that supports a suitable correction operator. 

In general, the product of an error and its corresponding correction operator will form closed loops on the toric code lattice. The action of these operators will trivially affect the code space {\em only if} the loops formed by $CE$ are the boundaries of regions on the lattice. We shade a bounded region in Fig.~\ref{ToricCodeStabilizers}(e). 

In the case that either the error strings are very long, such as those shown in Fig.~\ref{ToricCodeStabilizers}(f), or there are many error strings scattered over the lattice, it becomes very difficult to unambiguously find the correction operator $C$ such that no logical error is introduced to the system. For the example given in Fig.~\ref{ToricCodeStabilizers}(f), there are two possible least-weight correction operators of weight $d/2$, which we call $C'$ and $C''$, whose trajectories are marked by red dashed and dotted lines, respectively. Operator $C'$ is such that $C'E \in \mathcal{S}$. In the diagram we shade the region enclosed by the error and the dashed red line that marks the correction operator. The action of $C''E \in \mathcal{L}$ on the other hand does not enclose a region of the lattice. Instead, as we see, the correction has non-trivial support over an odd number of qubits that support the logical operator shown in blue at Fig.~\ref{ToricCodeStabilizers}(d). Such a correction will therefore cause a logical error on the code space. It is with this example that we see that determining the correct correction operator becomes difficult once the weight of the error becomes large.

Error correction on the toric code and the structure of its stabilizers can be understood at the fundamental level of homology. This topic goes beyond the scope of the present Review, but the interested reader is advised to read Ref.~\cite{Nakahara} or Appendix A of Ref.~\cite{Anwar14} to find a discussion of homology in the context of quantum error correction.

 \begin{figure}
\includegraphics[width=\columnwidth]{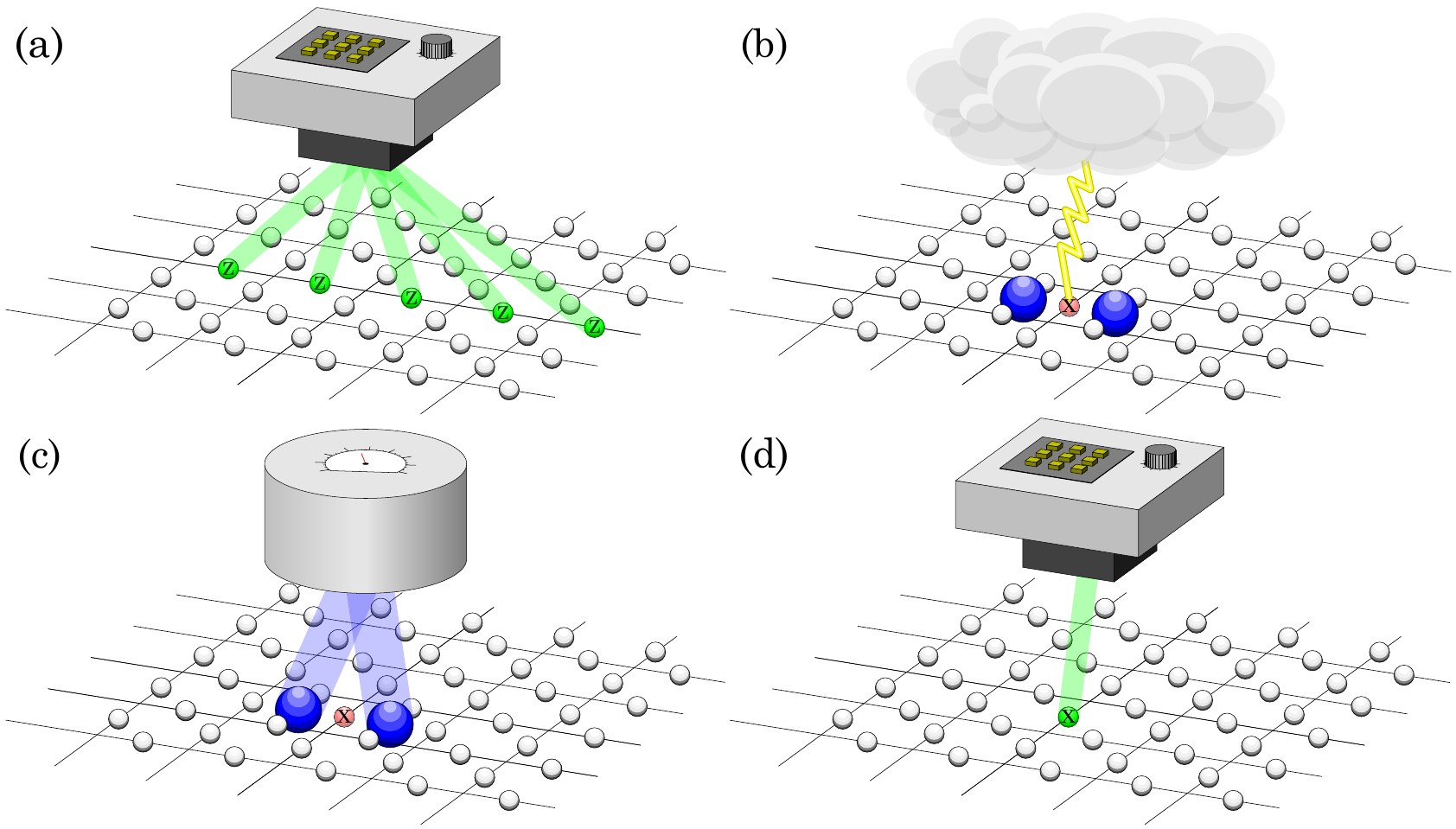}
\caption{(Color online) The error-correcting protocol for the toric code model. (a) The system is initialized. (b) An error occurs due to unavoidable coupling to the environment. (c) The syndrome is measured and fed to the decoding software. (d) The decoding algorithm determines a correction to recover the encoded state and in turn corrects the error. \label{chapDefectfigFail}}
\end{figure}

We briefly summarize the quantum error-correcting protocol for the toric code. The system is initialized in a code state by applying appropriate operations, Fig.~\ref{chapDefectfigFail}(a). While the quantum information is stored, errors might occur on the system, as shown in Fig.~\ref{chapDefectfigFail}(b). To identify these errors, stabilizer measurements are performed to obtain syndrome information. The locations of the stabilizers that returned a $-1$ outcome are recorded in Fig.~\ref{chapDefectfigFail}(c). A decoder subsequently uses the syndrome information to attempt to find a correction operator. A suitable correction operator that successfully corrects the incident error is applied in Fig.~\ref{chapDefectfigFail}(d), thus enabling the reliable readout of encoded quantum information.

\subsection{Topological Order and Anyons in the Toric Code}
\label{Sctn:AnyonicPicture}
The Hamiltonian of the toric code model~\cite{Kitaev03} gives rise to a $\mathbb{Z}_2$ lattice gauge theory~\cite{Wegner71, KogutSusskind75, Kogut79, Kitaev03, Nussinov, Nussinov09, Wen03}. Here, we consider the model as a prototypical example of a {\em topologically ordered lattice model} with {\em anyonic quasiparticle excitations}~\cite{Leinaas77, Wilczek82}. Its Hamiltonian
\begin{equation}
H_{\text{toric}} =-{1 \over 2}\sum_v A_v - {1\over 2}\sum_p B_p , \label{Ham:Toric}
\end{equation}
has degenerate ground states $\ket{\psi _\mu}$ as defined previously. We take interaction strength $1/2$ such that quasiparticles have unit mass. Its anyonic excitations are a special class of particles that exist in two-dimensional systems. Anyons are of particular interest due to their exchange statistics that are neither fermionic nor bosonic. Interestingly, the concepts of topological quantum error correction and of anyons are intrinsically connected, as will become apparent from the toric code example. For a comprehensive explanation of the general theory of anyons we advise the reader to see Appendix E of Ref.~\cite{Kitaev06}, or alternatively~\cite{Pachos, BrennenPachos2007, Nayak08, PreskillsLecture} for an introductory overview. In this Subsection we review the anyonic picture of the excitations of the toric code as it will often provide an efficient description of the dynamics of certain models presented in this Review.

The toric code has four types of quasiparticle excitations. The first, known as the {\em vacuum particle}, is denoted $1$. The vacuum particle describes no anyons. All models, topologically ordered or otherwise, support the vacuum particle. Excited eigenstates of Hamiltonian~(\ref{Ham:Toric}), $ \ket{ \phi} = E\ket{\psi}$, have electric charges, labelled $e$, on vertices $v$ that satisfy $ A_v\ket{\phi}  =  - \ket{\phi}$. Similarly, the toric code supports magnetic charges, $m$, on faces $p$ whenever $B_p \ket{\phi}  =  - \ket{\phi} $. The fourth particle of the toric code is known as the {\em dion}, labelled $ \epsilon$, that is the combination of an $e$ and an $m$ particle.

Anyonic systems have {\em fusion rules} to describe the combination of pairs of particles. We write the fusion product of particles $a$ and $b$ as $ a \times b $. The fusion product is commutative and associative. For the toric code we have
\begin{eqnarray} 
a \times 1 &= & a, \nonumber \\
e \times m &=&\epsilon, \nonumber \\ 
a \times a & = & 1, \label{Eqn:PairCreation} 
\end{eqnarray}
for all $a = 1,\, e,\, m,\, \epsilon $. This anyon model, and all others for which the fusion product always leads to a definite result, are called Abelian.

In full generality, we can define {\em non-Abelian} anyon models, where pairs of anyons can have multiple fusion outcomes. Like Abelian models, these also require error correction. However, the corresponding error correction problem is quite distinct to that of Abelian anyons, as discussed in~\cite{Wootton14}. This would have important consequences for the related problem of self-correction. Unlike Abelian anyons, no current proposals for self-correction have been based on non-Abelian models. As such they are beyond the scope of this review. However, the recent work on quantum error correction with non-Abelian models can be found in Refs.~\cite{Wootton14, Brell14, Hutter14, Burton15}.

Interestingly, quasiparticle excitations of the toric code are created in pairs.  We witness this at the microscopic level of the lattice as anyons are created at the two endpoints of string-like operators. This feature is reflected by the fusion rule~(\ref{Eqn:PairCreation}), which shows that we require two anyons to recover the vacuum state. Conversely, we can only create anyons from the vacuum in particle-antiparticle pairs.

\begin{figure}
\includegraphics[scale=2]{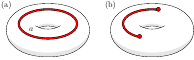}
\caption{ \label{Fig:JiannisBookDecoder}(Color online) The ground space of the toric code and its low-energy excitations. (a) The ground space of the toric code is naturally described with a basis labeled by anyonic charges, $a$, wrapping around a non-trivial cycle of the torus and then annihilating with its antiparticle, such as that shown in red. (b) Two anyonic excitations created that can propagate at no energy cost to affect the ground space of the system.}
\end{figure}

By using the anyonic description of error operators we find an alternative understanding of the logical operators of the toric code. As described in the previous Subsection, the logical operators are string-like operators that wrap around non-trivial cycles of the torus. In the anyonic picture string-like operators correspond to the trajectories of anyons. A logical operator corresponds to the creation of a pair of anyonic particles. One such particle then follows some non-trivial trajectory around the torus and subsequently annihilates with the other pair-created anyon that remained at its initial point. With this point we can define a natural basis for the ground space of the toric code, where orthogonal ground states correspond to different particle fluxes that pass around some arbitrarily selected non-trivial cycle of the torus. We show such a cycle in Fig.~\ref{Fig:JiannisBookDecoder}(a), where the flux of anyon $a$ wraps around the torus along the red line. In the case that many anyonic excitations move around the torus, the ground state is well defined according to the fusion rules of the different particle types. If we change the number of handles, or {\em genus}, $g$, of the surface where Hamiltonian~(\ref{Ham:Toric}) is embedded, then we change its degeneracy to $2^{2g}$ and we are able to encode $2g$ qubits there. This is attributed to the extra non-trivial cycles that can be traversed by anyons on the topologically deformed surface.

As an aside remark, the non-trivial braiding statistics between anyons can be obtained from the commutation relations of logical operators \cite{Einarsson90} of the toric code, as the commutation relations between crossing logical operators that follow different non-trivial cycles of the torus correspond to the braiding of anyonic quasiparticles.

Errors can also be interpreted in the anyonic picture. Errors occur when energy is introduced to the system which then creates anyons. Two such anyons are shown in Fig.~\ref{Fig:JiannisBookDecoder}(b). Anyons that propagate around non-trivial cycles on the torus introduce logical errors to the ground space of the system. Unfortunately, once anyons are created on the toric code, it is possible for them to propagate across the system via some suitable mechanism with no additional energy cost. We find this by observing that string-like operators can be introduced to a ground state of Hamiltonian~(\ref{Ham:Toric}) at constant energy, independent of the length of the string. This insight is the underlying problem that makes it very difficult to design two-dimensional topologically ordered passive quantum memories.

The low-energy excitations of the toric code is an example of a {\em topological quantum field theory}~\cite{Witten98}. Models that support topological field theories can be identified by their anyonic statistics, non-trivial ground state degeneracy, or by means of order parameters such as topological entanglement entropy~\cite{KitaevPreskill06, LevinWen06} that developed from earlier studies of entanglement entropy in topologically ordered lattice models~\cite{Hamma05a, Hamma05}. Topological quantum field theories, and extensions thereupon~\cite{HammaZanardiWen, WalkerWang, Haah, Yoshida13} give rise to classes of models that are of interest in the field of quantum memories. In the following Subsection we discuss the stability of the gap that is exhibited by topologically ordered systems at zero temperature in the presence of stray perturbations.

\subsection{Zero-Temperature Stability}
\label{subsec:stability}

In addition to considering the stability of memories against thermal noise, we must also be mindful of the effects of perturbations when designing quantum memories. Any deviation of a Hamiltonian from our idealized expectations will cause differences in energies as well as dynamics. This can have deleterious effects for any quantum information stored and processed in the system. Fortunately, topologically ordered systems are naturally adept at suppressing the effects of such perturbations at zero temperature. Even so, for arbitrary suppression we require systems of arbitrarily large size, so we will be interested in the thermodynamic limit for true stability.

Probably the most well known, but also most misinterpreted result regarding perturbations in topological ordered systems is that of \cite{BravyiHastingsMichalakis10}. In that work gapped Hamiltonians made up of local, frustration-free and commuting terms are considered at zero temperature where the degenerate ground state space is topologically ordered. Local perturbations of general form with finite but sufficiently low strength with respect to the unperturbed Hamiltonian gap are then introduced. It is shown that the splitting of the topologically originated ground state degeneracy is at most exponentially small with the system size. It is further shown that the gap between the ground state space and its excited states is also stable against small local perturbations. The topologically ordered phase is then preserved in the thermodynamic limit, and any given degree of suppression can be efficiently realized. The explicit example of the toric code Hamiltonian perturbed by magnetic fields has been well studied in Refs.~\cite{Dusuel, Trebst07, Vidal09, Tupitsyn}.

This result suggests that the timescale at which decoherence by dephasing is induced will diverge exponentially with system size. However, this conclusion is too readily adopted. It is very likely that the system will typically not be in the ground state of the perturbed Hamiltonian. One reason is that, for an arbitrarily large system, it will become a certainty that localized excitations will exist somewhere. Another reason is that the ground state may need to be prepared rather than achieved by cooling. Since the perturbations are not known in general, and since the resultant perturbed ground states may be too complex to prepare, we would expect to use the ground state of the unperturbed Hamiltonian. Finally, perturbations will be time-dependent in general.

Since the state of the system will not typically be an eigenstate of the Hamiltonian, the effects of dynamics must be considered. For the case of the toric code, it has been shown that the coherence time will be at most $\mathcal{O}(\log{L})$ in the presence of certain local perturbations, including a simple magnetic field \cite{Kay11}. Much of this is due to the perturbations enabling anyons to hop across the lattice. It has been shown that this effect can be suppressed by randomizing the couplings of the toric code Hamiltonian, thus introducing Anderson localization \cite{WoottonPachos,Stark}. The lifetime then improves to $\mathcal{O}(\rm{poly}(L))$ \cite{Kay11}. These dynamical effects are also studied in~\cite{Kay09,Pastawski,Tsomokos,BravyiKonig,Rothlisberger}.

Properties such as these are not necessarily limited to topologically ordered systems. In principle, other types of order may possess equal or perhaps better ground state stability against unknown perturbations. However, topologically ordered phases are currently the only known means for such suppression, and thus forms the backbone of current proposals for self-correcting memories.

\section{Memories at Finite Temperature}
\label{Sctn:FiniteTemperatures}

In this Section we consider the physics of a quantum memory coupled to a thermal environment. We introduce the necessary mathematical and numerical tools needed to analyze the effects of finite temperature on specific quantum memory models. As a concrete example we analyze the toric code coupled to a finite-temperature environment. We study both qualitatively and quantitatively the time evolution of this system and we identify when the stored information decoheres as a function of bath temperature. We conclude this Section with a list of criteria we demand from an experimentally amenable quantum memory.

The present exposition is motivated as a search for systems with quantum properties that are robust at finite temperatures. Nevertheless, the generic thermal dynamics we consider here are widely applicable to other instances of many-body physics as well as to quantum error correction. For example, it has been understood that quantum error-correcting codes based on self-correcting quantum memories can be decoded {\em locally} by using an algorithm based on thermal evolution~\cite{DennisKitaevLandahlPreskill, Pastawski11}. In a similar spirit, the study of self-correcting memories has also led to the discovery of {\em single-shot error correction}~\cite{Bombin14}. This is a remarkable discovery that could allow us to construct improved quantum error-correcting codes. We discuss local decoders and single-shot error correction in Subsecs.~\ref{Sctn:4Dtoric} and~\ref{Subsctn:GCC}, respectively.

The study of finite-temperature quantum systems is further motivated by the work of Pastawski~{\it et al.}~\cite{Pastawski09}. They showed that an error-correcting code can be passively protected by coupling the code to an auxiliary clock system whose qubits are maintained at infinite temperature. We also point out the work of Kapit~{\it et~al.}~\cite{Kapit14}, mentioned in Subsec.~\ref{Sctn:AVinteractions}. They showed that photon loss at zero temperature in superconducting systems can be modelled as an infinite-temperature noise model in the weak-coupling limit. The authors draw on this analogy to discover a medium that passively protects quantum information from photon loss in systems where temperature is neglected. From these examples it becomes apparent that the tools and models we develop in this Review are broadly applicable to many areas of quantum information and many-body physics. 

\subsection{Modeling a Finite-Temperature Environment}
\label{Sctn:RateEqn}

Formally, to model a quantum memory at finite temperature, we introduce an auxiliary system, which we call the {\em thermal bath}. We couple the bath to the memory system using some appropriate interaction terms that have non-trivial support on both systems. During evolution the interaction terms entangle the memory and the bath. In this way, information stored in the initial state of the memory is shared with the bath and as such it becomes difficult to recover by only accessing the memory.

In general the evolution of a many-body quantum system interacting with a thermal bath is very complicated. In fact it is unknown if the model describing the full thermal evolution is even analytically solvable~\cite{Terhal2005}, so to study a memory evolving in a thermal environment we need to make some simplifying assumptions. We assume that the memory interacts weakly with the environment and that the thermal bath is {\em Markovian}. A Markovian heat bath is such that the state of the bath is unmodified by interactions with a memory. A consequence of this assumption is that information transferred from a memory to the bath becomes unrecoverable. Additionally, we assume that the bath acts locally on the physical degrees of freedom of the memory. We model the thermal environment such that each qubit is independently coupled to a bath of harmonic oscillators. With this assumption, each event that occurs during the thermal evolution will affect only one physical qubit of the memory system at a time.

In principle, a thermal evolution is accurately described by the bipartite system of the quantum memory and the auxiliary bath. In practice, we need only model the dynamics of the memory subsystem. For this simplification to be valid the time evolution needs to satisfy certain criteria. In particular, the dynamics must evolve the memory towards its {\em Gibbs state}
\begin{equation}
\rho_\beta = \sum_j \frac{e^{-\beta \varepsilon_j}}{\mathcal{Z}}  \ket{e_j} \bra{e_j} , \label{Eqn:GibbsState}
\end{equation}
where $\mathcal{Z} = \sum_j \bra{ e_j } e^{-\beta H} \ket{ e_j }$ is the canonical partition function. The vectors $\ket{ e_j }$ comprise an orthonormal basis of eigenstates of the memory Hamiltonian, whose corresponding energy eigenvalues are $\varepsilon_j$. We denoted by $\beta = 1/T$ the inverse temperature of the heat bath and we took Boltzmann's constant equal to one. 

An extensive programme of research has shown that we can model thermal dynamics of a many-body quantum memory with a simple rate equation~\cite{Davies, Alicki2007a, AlickiFannes09,  AlickiFannesHorodecki, AlickiHorodeckiHorodeckiHorodecki, Alicki09, ChesiRothlisbergerLoss, ViyuelaRivasMartin-Delagado,  Weiss2012}. These methods are built from the discovery of exact master equations to study dissipation; a study initially pioneered by Caldeira and Leggett~\cite{Caldeira81, Leggett1987, DiVincenzoLoss}. We summarise the derivation of the dynamical model below.

The rate equation evaluates the rate at which an event, described by operator $V$, occurs during a thermal evolution, such that $\langle e_\text{f} | V | e_\text{i} \rangle = 1$, where $|e_\text{i} \rangle$ and $|e_\text{f} \rangle$ are the initial and final eigenstates of the memory with respect to event $V$. The rate at which event $V$ occurs depends on the difference in energy of the initial and final eigenstate, which we denote as $\omega_V =  - ( \varepsilon_\text{f} -  \varepsilon_\text{i} ) $. We thus have the rate equation that describes the frequency at which event $ V $ occurs under thermal evolution
\begin{equation}
\gamma( \omega_V ) = \frac{\omega_V}{1-e^{-\beta \omega_V}}  .
\label{secIIpartCeqnRate}
\end{equation}
Intuitively Eqn.~(\ref{secIIpartCeqnRate}) dictates that processes that increase the energy of the system are exponentially suppressed compared with processes that do not increase the energy of the memory. It is guaranteed that the memory system will evolve towards the Gibbs state if rate Eqn.~(\ref{secIIpartCeqnRate}) satisfies {\em detailed balance}~\cite{Kossakowski77}. Namely it must satisfy
\begin{equation}
\gamma( \omega_V ) = e^{\beta \omega_V} \gamma( -\omega_V ) \label{Eqn:DetailedBalance},
\end{equation} 
for all events $V$. It is easily verified that Eqn.~\eqref{secIIpartCeqnRate} satisfies the detailed balance condition. 

The open quantum dynamics we have described are derived from a {\em Lindbladian master equation}~\cite{Kossakowski72, Lindblad76}. The master equation is obtained by considering the closed dynamics of the system evolving under the Hamiltonian acting on both the memory and the bath subsystems
\begin{equation}
H = H_{\text{M}} \otimes \openone_{\text{B}} + \openone_{\text{M}} \otimes H_{\text{B}} + \sum_\alpha W_\alpha \otimes f_\alpha.\label{Ham:ClosedDynamics}
\end{equation}
The last term of Hamiltonian~(\ref{Ham:ClosedDynamics}) describes the interactions between the memory and the bath. Local Hermitian operators $W_\alpha$ and $f_\alpha$ act only on the memory subsystem and the bath subsystem, respectively. Given certain assumptions that we specify shortly, the evolution of the memory is well described by the master equation
\begin{equation}
\dot{\rho} = i [ H_{\text{M}}, \rho ] + \mathcal{L}(\rho), \label{Eqn:LindbladMasterEqn} 
\end{equation}
where $\rho$ is the density matrix describing the state of the memory subsystem and $\mathcal{L}$ is the {\em Liouvillian}. The Liouvillian describes the dynamics due to the interactions between the memory and the bath. It takes the form
\begin{equation}
\mathcal{L}( \rho ) = \sum_{\alpha,\omega \ge 0} \mathcal{L}_{\alpha \omega} (\rho), \label{Eqn:Liouvillian}
\end{equation} 
where the individual terms of the Liouvillian are written
\begin{eqnarray} \nonumber 
&\mathcal{L}_{\alpha \omega} (\rho) = \hat{g}_\alpha(\omega) \left\lbrace V_\alpha(\omega)^\dagger [\rho, V_\alpha(\omega)] + [V_\alpha(\omega)^\dagger, \rho] V_\alpha(\omega) \right. & \\
&+ \left. e^{-\beta \omega} \left( V_\alpha(\omega) [ \rho, V_\alpha(\omega)^\dagger ] + [ V_\alpha(\omega), \rho ] V_\alpha(\omega)^\dagger \right) \right\rbrace .& 
\end{eqnarray}
In this expression $\hat{g}_\alpha(\omega)$ is the power spectrum of the bath and $V_\alpha(\omega)$ are the Fourier components of $W_\alpha$, i.e.
\begin{equation}
U(t) W_\alpha  U^\dagger(t) = \sum_\omega V_\alpha(\omega) e^{-i \omega t}.
\end{equation} 
where $U(t) = e^{- \text{i} H_{\text{M} } t  } $. 

It can be checked that if we take the interaction terms acting on the memory $W_\alpha $ as single qubit Pauli matrices, $X_j$ and $Z_j$, then the density matrix is diagonal in the energy eigenbasis at any given point in the evolution  $\rho(t) = \sum_j p_j(t) \ket{e_j} \bra{e_j}$. It can then be shown that Eqn.~\eqref{Eqn:LindbladMasterEqn} reduces to a much simpler form 
\begin{equation}
\dot{p_k} = \sum_{j \neq k} \left( \, \Gamma_{j \rightarrow k} \, p_j - \Gamma_{k \rightarrow j} \, p_k \, \right),
\end{equation}
where $\Gamma_{j \rightarrow k}$ are the rates of transition from state $j$ to $k$. Let $V$ be an error process that takes the system from eigenstate $j$ to $k$ with energy cost $\omega_V$. Then the the rates are expressed as
\begin{equation}
\Gamma_{j \rightarrow k} = \gamma( \omega_V ) \propto \frac{ \hat{g}( \lvert \omega_V \rvert )}{\lvert 1-e^{-\beta \omega_V} \rvert} \, .
\label{secIIpartCeqnRateSECOND}
\end{equation}
Making the additional assumption that the bath has Ohmic spectral density and large cut-off energy i.e. $\hat{g}_\alpha(\omega_V) \propto \omega_V$~\cite{Leggett1987, Weiss2012} then, up to a normalization, Eqn.~\eqref{secIIpartCeqnRateSECOND} yields the rate equation Eqn.~\eqref{secIIpartCeqnRate}.

In order to achieve thermalization it is important to require that the interaction operators $W_\alpha$ are {\em ergodic}. This means that the thermal bath is able to address all eigenstates of the memory system. It is known that ergodicity is assured if the only operators that commute with both the memory Hamiltonian and the set of interaction terms $W_\alpha$ are proportional to the identity operator~\cite{Spohn77, Frigerio07}. It is easily checked that if the $W_\alpha$ terms are single-qubit Pauli operators then ergodicity is assured. 

\subsection{Coherence Time of Memories} \label{Sctn:CoherenceTimes}

To determine how well a candidate memory performs at finite temperature we need to introduce a suitable figure of merit. To this end we define the {\em coherence time}, $\tau$, as the maximum amount of time information encoded in a system can undergo thermal evolution and remain recoverable with high probability. To evaluate the coherence time of a system, we encode information in a system of interest and evolve the system under the thermal dynamics introduced in the previous Subsection. To recover the information, we allow the use of active error correction techniques {\em at the time of readout}. 

To understand the capacity of a system to support quantum information at finite temperature we will primarily be interested in the dependence of the coherence time on parameters such as the system size and the inverse temperature of the bath. Naturally, the coherence time will also depend on microscopic details of the system such as the natural units that describe the strength of the local Hamiltonian interactions. These details are overlooked as they are fixed by Eqn.~(\ref{secIIpartCeqnRate}), but will always take constant values independent of system size.

When evaluating coherence times, we will often assume that we can initialize a specific ground state of a system to encode information. This choice is in the interest of providing a fair comparison between different memory systems, and also to conceptually simplify our exposition. In general, we expect it to be very hard to prepare a many-body Hamiltonian in its ground state as this will require cooling the system to zero temperature. Alternatively, we might consider manually preparing ground states by means of controlled laboratory operations. However, manual preparation of ground states will also introduce small errors as in general laboratory equipment is fallible~\cite{Lodyga15}. To this end, the ground state preparation we assume here is unreasonable. However, we do not expect the results we discuss under this assumption to be fundamentally different from the realistic case. Indeed, it is shown in Ref.~\cite{Bombin13} that random local errors will only adjust the phase transition point of a self-correcting memory.

We also assume that we can realise Hamiltonians that are free from small imperfections such as weak local perturbations as has been discussed in Subsec.~\ref{subsec:stability}. Once again, this is not a realistic assumption, as we would typically expect stray fields and other imperfections to alter system Hamiltonians. We make this assumption because the present Review is primarily concerned with the finite-temperature behavior of quantum memories. Moreover, this assumption greatly simplifies the computational methods we use to analyse different models. In general, the simultaneous consideration of both temperature and local perturbations makes calculations notoriously difficult, and as such, our overview of the field will typically discuss these two forms of noise independently. 

We consider now a simple example where we find explicilty the coherence time of a small four-qubit toric code. The four qubits of the model, indexed $j = 1,\,2,\,3,\,4$, are subject to the Hamiltonian
\begin{equation}
H_{\text{4Qu. toric}} = -\frac{\Delta}{2}(A + B), \label{Ham:TinyToric}
\end{equation}
with stabilizers
\begin{equation}
A = X_1 X_2 X_3 X_4, \,\,\,B = Z_1 Z_2 Z_3 Z_4.
\end{equation}
The code states are given by the four-qubit Greenberger-Horne-Zeilinger states~\cite{Greenberger89, Bouwmeester99}, which are commonly known as GHZ states, with an even number of qubits in the 1 state. Logical operators for this code act on only two qubits. For example, we can choose $\overline{X} = X_1 X_2$. The thermal error model defined above applies single qubit Pauli operators one at a time. It can therefore apply a logical $\overline{X}$ by first applying $X_1$ and then $X_2$. The first operator anticommutes with $B$, and so costs an energy $ \Delta$ as dictated by Hamiltonian~(\ref{Ham:TinyToric}). According to the rate equation, Eqn.~\eqref{secIIpartCeqnRate}, this process will take a time of around $ e^{\beta\Delta}$. The next flip, $X_2$, required to introduce a logical error is a relaxation process, and so occurs much more quickly. The coherence time of this small toric code then is $\tau \sim e^{\beta\Delta}$. This is exactly that obtained from Arrhenius' law
\begin{equation}
\tau \sim e^{\beta \varepsilon} .
\label{Eqn:Arrhenius}
\end{equation} 
This law asserts that the coherence time scales exponentially with the energy cost $\varepsilon$ of introducing a logical error into the system. In the case of the small toric code, we have that the energy cost of introducing a logical error is equal to the gap of the system $\varepsilon = \Delta$. 

Exponential coherence time scaling with inverse temperature, determined by the constant Hamiltonian interaction strength, is common to all memories of small size. As such, we must look to macroscopic models to find systems with extended coherence tim es. We must therefore ask, what happens to the coherence time of the memory against thermal noise as we increase the system size? It is useful to compare with the benchmark $\tau \sim e^{ \beta \Delta }$ for small systems obtained from Arrhenius' law, Eqn.~(\ref{Eqn:Arrhenius}). The worst possible case for a memory would be sub-Arrhenius scaling of coherence time with $\beta$. This would mean that large system sizes have entropic effects that cause the memory to fail faster than for small system sizes. A memory with Arrhenius scaling allows the same protection as one would get against thermal errors for a small system size. Although the resilience of the model to thermal errors may not increase, it may be beneficial to increase the size of the system to improve perturbative stability of a model. We might even expect systems of larger sizes to have greater coherence times than we expect of small system sizes. In which case we might expect super-Arrhenius scaling in coherence time as the temperature is reduced in the limit of large system sizes. Examples of such models are discussed in later Sections.

\subsection{The Energy Barrier} \label{Sctn:EnBar}

A useful concept in the study of quantum memories is the energy barrier. The four-qubit toric code discussed in the previous Subsection gave an example where Arrhenius' law can be directly applied. In that case we obtained that the lifetime of the memory is correlated in a simple way to its gap, $\Delta$. However, the corresponding process for larger many-body systems is much more complicated. Thermal errors act locally, and so it is not possible to transition between ground states via a single excited state. Instead, errors must navigate a highly degenerate landscape of excited states to modify the ground space of the system. For this reason there is typically no simple generalization to find the value $\varepsilon$ that can be used to estimate the lifetime via Arrhenius' law. Nevertheless, we can gain some intuition about the thermalization process by identifying the dominant energy scale of the evolution.

Consider the case that the commuting Pauli Hamiltonian system is initially in a logical ground state $\ket{\psi}$. We wish to determine how easy it is for thermal errors to rotate the encoded state to, for instance, ${\overline{X}}\ket{\psi}$, by introducing the logical error $\overline{X}$. As we consider physically motivated local noise models, the logical error operator $\overline{X}$ is decomposed into a sequence of the single qubit operators $U_t$ that the thermal bath can apply, such that
\be
\overline{X} = \prod_{t=1}^N U_t \, .
\ee
Here $U_t$ denotes the $t$-th operator to be sequentially applied, and $N$ is the total number of operators required to construct $\overline{X}$. Note that this decomposition of the logical operator into single qubit operators is not unique. For instance, a permutation in the ordering of the $U_t$ will result in the same action $\overline{X}$ upon state $\ket{\psi}$. Indeed, choosing a different sequence of $U_t$ with different $N$ can yield the same logical operator in the case of commuting Pauli models.

Unlike the initial and final states, the states where the first $0<t<N$ steps of the error sequence have been applied will be an excited state. Let us use $\varepsilon_t$ to denote its energy. For each decomposition of $\overline{X}$ into $U_t$ operators we can consider the energy $\max_t\varepsilon_t$, the maximum energy cost incurred during the sequence. This may be artificially high simply due to a badly chosen sequence. We therefore minimize the energy over all possible sequences to obtain the energy barrier of the model, $\varepsilon_\text{B}$. This is the minimum energy that the system must achieve in order for a logical error to occur.

We take the toric code Hamiltonian~(\ref{Ham:Toric}) as an example to calculate its energy barrier. A logical operator can be applied by first creating a pair of anyons and then transporting them around a non-contractible loop. The first operation incurs the energy cost for creating a single pair but no subsequent operation will increase the energy of the system, therefore $\max_t \varepsilon_t = 2$. Alternatively, one could generate a logical operator by first applying rotations on every other qubit around a non-contractible loop, and then annihilating all the generated anyons by rotating the remaining qubits around the loop. After the creation of all the anyons, the system reaches a state of energy $ L$. We thus have $\max_t \varepsilon_t =L$. Clearly the former logical error path is energetically favorable as it has the smallest energy. Hence, it will be the most common process that introduces logical errors at low temperature. We therefore find the energy barrier of the toric code to be $\varepsilon_{\text{B}} = 2$.  

Much of the study of finite-temperature quantum memories has sought complex systems that achieve $\varepsilon_{\text{B}}$ that scales with the size of the system. For such systems we should expect their coherence times to scale favorably with system size according to Arrhenius' law. Examples of such models are studied in Sec.~\ref{Sctn:ThreeDimensions}. 
We must bare in mind that in general it is not clear that models can realize logical operators via an ordered sequence of local unitary operators~\cite{Haah12, LandonCardinalPoulin}. For this reason we reemphasise that the discussion given here is restricted only to commuting Pauli Hamiltonian models. 

\subsection{Free Energy and the Curie-Weiss Model}
\label{Subsctn:CurieWeiss}

In general we cannot completely characterize the coherence time scaling of a memory by only considering its energy barrier. It is possible that entropic effects can modify the predictions obtained by Arrhenius' law~\cite{Temme14,Yoshida14}. A more accurate characterization of a system is obtained by consideration of its {\em free energy}. 

With few exceptions, careful analysis of the free energy of a system is intractable due to its computational complexity. Such an analysis involves a careful consideration of an exponentially large number of micro states of the given system. However, evaluating the free energy of a system sheds significant light on its behavior at non-negligible temperatures. In particular analysis of the free energy enables us to identify low-temperature ordered phases where we expect self correction to be possible. 

The free energy $F$ is the energy cost of an event, $E$, offset by an entropic contribution, $S$, such that
\begin{equation}
F = E - S / \beta. \label{Eqn:FreeEnergy}
\end{equation}
It provides a more accurate estimate of coherence time compared with the energy barrier as it includes the effect of entropy. We obtain the coherence time using the expression
\begin{equation}
\tau \sim e^{\beta F}. \label{Eqn:FreeEnergyCoherence}
\end{equation}
In this Subsection we consider a toy model, namely the {\em Curie-Weiss model}, for which the free energy can be evaluated. While unphysical due to its non-local Hamiltonian interactions, the Curie-Weiss model is a simple classical model that enables a detailed analysis of the contribution of both the energy barrier and the free energy that can be used to determine its coherence time~\cite{Alicki06}. For a detailed discussion on the Curie-Weiss model we recommend Ref.~\cite{Kochmanski13} and references therein.

The Curie-Weiss model is a two-fold degenerate model comprised of $n$ classical spins, $\sigma_j $, with $1 \le j \le n$ that take values $ \sigma_j =  \pm 1$. We denote a configuration of the spins of the system as $\boldsymbol{\sigma}$. The energy of a state of the Curie-Weiss model is 
\begin{equation}
E_{\text{CW}}(\boldsymbol{\sigma}) = - \frac{\Delta}{n}E_{\text{para}}({\boldsymbol{\sigma}})^2, \label{Ham:CurieWeiss}
\end{equation}
where $\Delta$ is a constant independent of system size and $E_{\text{para}}({\boldsymbol{\sigma}}) = -\sum_j \sigma_j $ is the classical Hamiltonian that describes a paramagnet. Specifically, $E_{\text{para}}$ assigns one unit of energy to each spin in the $-1$ state and negates one unit of energy for each spin otherwise. We point out that the non-local nature of the Hamiltonian is such that each spin is involved in a number of interaction terms that scales with $n$. To compensate for this, the $1/n$ factor in Hamiltonian~(\ref{Ham:CurieWeiss}) ensures that the energy cost of a single spin flip does not scale with the size of the system. 

The only relevant quantity when studying configurations $\boldsymbol{\sigma}$ is $x = n_\downarrow / n$, where $n_\downarrow$ is the number of spins of configuration $\boldsymbol{\sigma}$ in the $-1$ state. The two ground states take values $x=0$ and $x = 1$, and the energy of a typical configuration is
\begin{equation}
E_{\text{CW}}(x) = -\Delta n (1-2x)^2.
\end{equation}
It is also important to notice that we have $C = n! / (n-n_\downarrow)! n_\downarrow!$ unique configurations that give rise to a particular $x$. Rearranging, and making use of Sterling's approximation, we obtain
\begin{equation}
C(x) = e^{S(x)},
\end{equation} 
where 
\begin{equation}
S(x) = - n [ x \log x  + (1-x) \log (1-x)], 
\end{equation} 
is the entropy of the system. The probability that a system is in configuration $\boldsymbol{\sigma}$ is found using a Boltzmann weight,
\begin{equation}
\text{prob}(\boldsymbol{\sigma} ) = e^{-\beta E_{ \text{CW} } ( \boldsymbol{\sigma} )  } /  \mathcal{Z},
\end{equation} 
where $\mathcal{Z} = \sum_{\boldsymbol{ \sigma }} e^{ - \beta E_{ \text{CW} } ( \boldsymbol{ \sigma } ) } $ is the partition function of the system. We thus find the probability that a system is in a configuration that takes value $x$
\begin{equation}
\text{prob}(x) =   C(x) e^{-\beta E_{\text{CW}}(x)} = e^{- \beta F(x)} / \mathcal{Z}. \label{Eqn:CWFreeEnergy}
\end{equation}
where now $F(x) = E_{\text{CW}}(x) - S(x) / \beta$.

\begin{figure}
\includegraphics[width=\columnwidth]{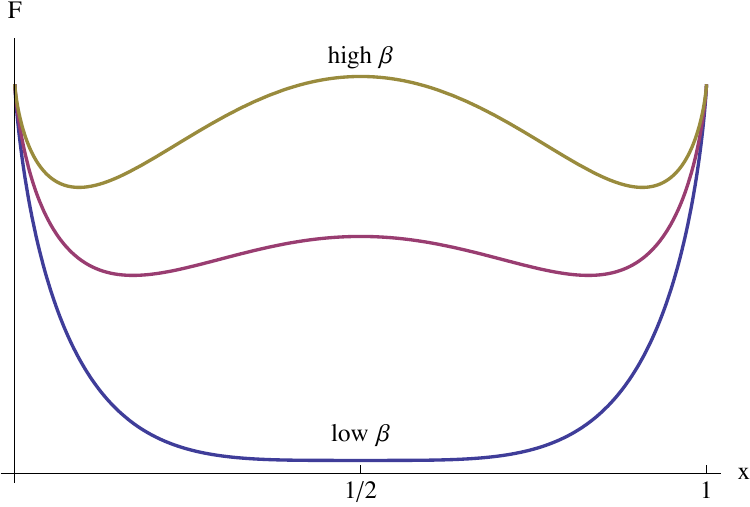}
\caption{(Color online) Free energy plotted as a function of $x$ for low, intermediate, and high $\beta$, shown by the bottom blue line, the intermediate red line and the top yellow line, respectively. With decreasing temperature the local minima become more pronounced. At high temperature the entropic contribution is dominant.  \label{Fig:FreeEnergy}}
\end{figure}

We can use Eqn.~(\ref{Eqn:CWFreeEnergy}) to understand the behavior of the Curie-Weiss model as a classical memory. We do not require that the memory remains in the ground space to encode a state. We only require that $x$ remains close to its encoded value, where either $x \ll 1/2$ or $x \gg 1/2$. Provided the value of $x$ remains far away from $x\sim 1/2$, we can recover the state of the encoded bit by measuring the magnetization of the system. Finding the magnetization is physically equivalent to taking a majority vote over all the spins of the system. We plot the free energy as a function of $x$ for various $\beta$ in Fig.~\ref{Fig:FreeEnergy}. The probability that a state takes value $x$ is inversely proportional to the exponent of the free energy, as shown in Eqn.~(\ref{Eqn:CWFreeEnergy}). Therefore, we can regard the free energy plot of Fig.~\ref{Fig:FreeEnergy} as a potential landscape, where the system will preferentially find local minima, and is unlikely to achieve states with large free energy. 

At low temperatures, Fig.~\ref{Fig:FreeEnergy} shows that the system has two potential minima, one for $x \ll 1/2$, and one at $x \gg 1/2$. At suitably low temperatures, we can increase the depth of the two potential minima by increasing $n$. As such, it is highly unlikely for a state to achieve a configuration with $x \sim 1/2$ in the thermodynamic limit, as states with a large free energy are achieved very infrequently. Therefore, if we encode a state by preparing it in, for example, a configuration with $x \ll 1/2$, it is very unlikely that the thermal environment will evolve the state to one of $x \gg 1/2$ via a sequence of local spin flips, as the evolution must pass through highly improbable states where $x \sim 1/2$. To this end, in the thermodynamic limit, and at suitably low temperatures, the Curie-Weiss model is able to robustly encode a classical bit of information for arbitrarily long timescales. This is shown in Ref.~\cite{Alicki06} by taking $x$ to the continuum limit and applying Kramer's formula~\cite{Gardiner83}.

The Curie-Weiss model is one of the simplest examples of a model that is in an ordered phase at finite temperature. As we have observed, if the temperature is suitably low, we can robustly encode classical information for arbitrarily long timescales. Conversely, if we increase the temperature, the model undergoes a phase transition into the disordered phase where the storage of information is no longer possible. We observe this in Fig.~\ref{Fig:FreeEnergy}. Specifically, at high temperatures, the free-energy curve no longer has two well-resolved minima that are separated by a large potential. Ideally, we seek to find quantum systems that have an ordered phase at finite temperature. We expect that such a system will be able to robustly encode quantum information for long durations. We discuss systems with ordered phases at finite temperature in more depth in Sec.~\ref{Sctn:HighD}.

We finally remark that while the free energy offers a much more accurate description of the behavior of a system including evaluation of the phase diagram of a system, it is often difficult to evaluate. As such, we often resort to using simpler concepts such as energy barriers to evaluate the behavior of a system. Recent work has been conducted in this area in Refs.~\cite{Temme14, Temme15, Komar16} where it is shown that Arrhenius' law gives an {\em upper bound} on the coherence time of a memory for a large class of local commuting Hamiltonians. Indeed, the resulting coherence times achieved by the use of Arrhenius' law, Eqn.~(\ref{Eqn:Arrhenius}), is often considered as a widely applicable~\cite{Laidler72} rule of thumb.

\subsection{Simulating Finite-Temperature Effects}
\label{Sctn:NumericalMethods}

Monte Carlo methods are frequently used to numerically analyze the evolution of a system where analytical methods are intractable, or to find data that supports theoretical conjecture. Here we give an overview of a general method to conduct finite-temperature Monte Carlo simulations for commuting Pauli Hamiltonians~\cite{ChesiRothlisbergerLoss, Bortz}.

The noise model approximates the thermal evolution of an eigenstate $\ket{\psi(t)}$ with respect to commuting Pauli Hamiltonian $H$ when interacting with a thermal bath of inverse temperature $\beta$. As we have mentioned in Subsec.~\ref{Sctn:CPHamiltonians}, eigenstates of commuting Pauli Hamiltonians are easily described using a list of eigenvalues, $M_j$. We simulate the noise model as a sequence of discrete events that map between eigenstates of $H$. At each event we look to obtain some $V$ and $\delta t$ such that $\ket{\psi(t +\delta t )} = V \ket{\psi(t)}$. At $t=0$ we typically initialize the state to a ground state of $H$ and we simulate the thermal evolution up to some time $t_\text{max}$.

For commuting Pauli Hamiltonians the random incident errors $V$ are hopping operators that act like $X_j$, $Y_j$ or $Z_j$ on the state. Importantly, the action of operators $V$ ensure that $\ket{\psi(t+\delta t)}$ is an eigenstate of $H$. The relative probability of error event $V$ is determined using Eqn.~(\ref{secIIpartCeqnRate}). Rates $\gamma(\omega_V)$ are evaluated with respect to $H$ and $\ket{\psi(t)}$. Explicitly, we select error event $V$ by calling from the distribution 
\begin{equation}
p_V ={ \gamma(\omega_V) \over  R },
\end{equation} 
where we normalize using the total rate $ R = \sum_V \gamma( \omega_V) $ with the summation running over all errors $V$ realizable by the noise model.

The time $\delta t$ that passes between each step as $V$ is applied is determined using $R$. Since each $V$ occurs as a random process at rate $\gamma(\omega_V)$, the time step $\delta t$ is a random variable distributed as an {\em exponential} distribution with parameter $R$, the total rate. We numerically generate values of $\delta t$ such that
\be
\delta t = -{\ln(\mathtt{rand}) \over  R}  \, ,
\ee
where $\mathtt{rand}$ is a random variable chosen uniformly from the interval $(0,1]$. We thus obtain the new eigenstate after a time $\delta t$, which passes during the event, such that $\ket{\psi(t + \delta t )} = V \ket{\psi(t)}$. 

At the end of each event, we check the total time of the system. If $t + \delta t < t_\text{max}$ we perform another event using the new eigenstate $\ket{\psi(t+\delta t)}$. Otherwise we stop the simulation and use $\ket{\psi(t+\delta t)}$ and the total incident error to collect sample data.

By averaging over many trials of this process we can obtain estimates of many non-equilibrium thermal quantities, such as the coherence time $\tau$. We use two different methods to estimate $\tau$ in this Review. In the first we apply the decoder repeatedly as we evolve the system and define $\tau$ as the average time it takes for the decoder to fail once. Alternatively we define $\tau$ as the time that the decoder success rate falls below some threshold e.g. 99\%. The values obtained with these different methods may differ by a constant factor, but ultimately will both reveal the coherence-time dependence of the system on its size and temperature.

\subsection{Toric Code at Finite Temperature}
\label{SEC:TCfinitetemp}

The thermal dynamics of the toric code have received extensive study Ref.~\cite{AlickiFannesHorodecki, Jouzdani2014b, Hutter14a, Freeman14}. In particular, in Ref.~\cite{AlickiFannesHorodecki} it is shown analytically using the dynamical model reviewed in Subsec.~\ref{Sctn:RateEqn} that the coherence time of the model will not exceed $\tau \sim e^{\beta \Delta}$. In this Subsection we provide a self-contained study of the thermal dynamics of the toric-code model. We make use of the numerical tools we have discussed throughout this Section to complement known analytical results. 
Importantly, the thermalization dynamics of the toric code provide an explicit example of the behavior of the thermal dynamics of a Hamiltonian system that we aim to defend against.

It is a generic feature of quantum memories that their finite-size behavior differs from their behavior at the thermodynamic limit. Both of these regimes are important. The latter considers properties relevant to the scalability of the system, and is pertinent for many theoretical considerations. The former however is more relevant to current experimental efforts. Here we use the toric code as a specific example to identify and compare the two different regimes.  We identify a critical system size below which finite size effects are apparent. This size is a function of temperature and the energy cost required to create excitations.

\begin{figure}
\includegraphics[width=\columnwidth]{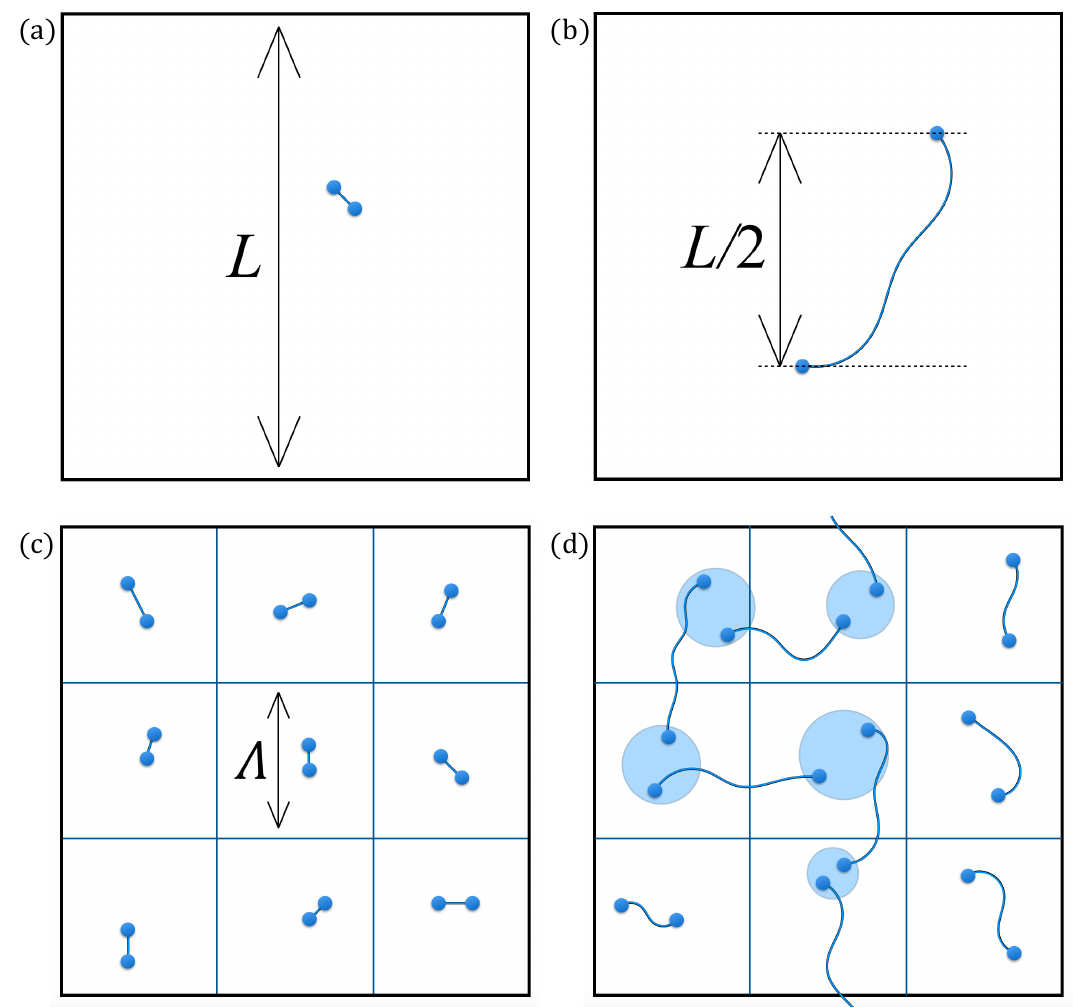}
\caption{(Color online) The thermal dynamics of the excitations of the toric code in both the small~(a) and~(b), and large~(c) and~(d), system size limits. Typical error configurations quickly achieve excitation density $\rho \approx e^{-\beta \Delta}$ for excitations with mass $\Delta$ at inverse temperature $\beta$. 
In the small system size limit where $ L^2 \rho /2 \lesssim 1$, it is common for only a single pair of excitations to be created,~(a), which then rapidly propagate to cause an uncorrectable error,~(b). 
In contrast, in the large system size limit a uniform distribution of anyon pairs is quickly created,~(c). These pairs diffuse and overlap eventually creating a chain that percolates over the lattice causing an uncorrectable error,~(d).
\label{Fig:TCregimes}}
\end{figure}

The energy cost to create a single excitation is often referred to as its mass. The mass equates to the interaction strength, $\Delta$, of the toric code Hamiltonian~(\ref{Ham:Toric}). In thermal equilibrium we expect the average density of anyons in the toric code to scale like $\rho \sim e^{-\beta \Delta}$. Therefore the number of anyon pairs present in a thermalized system of size $L$ is
\be
\langle N \rangle \sim  {L^2 \rho \over 2} = {L^2 e^{-\beta \Delta} \over 2}  .
\label{EQN:TCLowTempNumber}
\ee
Using Eqn.~(\ref{EQN:TCLowTempNumber}) we see that for systems smaller than $L \lesssim e^{\beta \Delta/2}$ we have $\langle N \rangle \lesssim 1$. It follows that the probability that there is more than one single pair of anyons on the lattice is negligible. Therefore, in this regime the thermal decoherence of encoded information will most likely occur due to the creation of a single pair of anyons that rapidly propagate across the lattice and introduce a logical error to the memory. We will demonstrate that in the small-size limit the coherence time can be approximated by Arrhenius' law applied to the minimum energy barrier up to system size dependent corrections, similar to the four-qubit toric code discussed in Subsec.~\ref{Sctn:CoherenceTimes}. This behavior differs from that of larger system sizes where $L^2 \rho / 2  \sim \langle N \rangle \gg 1$ such that many anyon pairs are uniformly distributed over the lattice. For this case we observe that the coherence time is exponentially shorter than Arrenhius' law predicts, and is no longer dependent on system size. The two different limits for the toric code are demonstrated in Fig.~\ref{Fig:TCregimes}.

\subsubsection{Small system size limit}

For small systems, and at a temperature low enough for us to expect a good memory, we can typically expect a single pair of excitations to cause the toric code memory to fail. We show such a configuration in Fig.~\ref{Fig:TCregimes}(b). We estimate the coherence time $\tau_{\textrm{small}} = \tau_c + \tau_m $, where $\tau_c$ is the time it takes for this pair to be created and $\tau_m$ the time it takes for the anyons to diffuse across the lattice up to a separation $L/2$. When the anyons have crossed a distance $L/2$ then the encoded information is irrecoverable by quantum error correction. Only the separation between the anyons is important, so we treat one as fixed and just consider the relative motion. This motion is an unbiased random walk allowing us to easily estimate the diffusion time $\tau_m$. The typical number of steps required for a two-dimensional random walker to reach a distance $L/2$ from its starting point is $(L/2)^2$. Under the noise model discussed in Subsec.~\ref{Sctn:NumericalMethods}, the typical time of random walk steps is $(8\gamma(0))^{-1}$, where the factor of 8 counts the number of processes on the lattice that move anyons. We thus obtain
\be
\tau_m \simeq \frac{1}{8 \gamma(0)} \left(\frac{L}{2}\right)^2 = \frac{\beta L^2}{32}  .
\label{EQN:taum}
\ee
To estimate $\tau_c$ we note that the energy cost of pair creation is $2 \Delta$. Again applying the noise model from Subsec.~\ref{Sctn:NumericalMethods}, pairs are created from the vacuum at rate $R_0 = 2 L^2 \gamma(-2 \Delta)$. This implies the time we wait to see a creation event is $1/R_0 \sim e^{2 \beta \Delta}/L^2$. However, not all pairs diffuse to the required distance. Some pairs will instead fuse back to the vacuum at some point later in time, we can quantify the effect this has on the coherence time by considering the random walks of the pairs. 

We denote by $\Pi(L,\beta)$ the probability that a pair does not self-annihilate before reaching separation $L/2$. The motion of the pair is described by an unbiased random walk. It is a standard result for a two-dimensional random walk on a square lattice that the probability of a walker not returning to the origin in the first $K$ steps scales as $1 / \ln(K)$. On average we need the walker to avoid self anihilation for $(L/2)^2$ steps. We therefore expect a factor of $[2 \ln(L/2)]^{-1}$ in $\Pi$. Additionally, in order to begin the random walk the pair must avoid fusing back to the vacuum immediately. This is a relaxation process so happens at a higher rate than beginning the walk. We therefore include a factor $\sim 1/(1 + A\beta)$ in $\Pi$, where $A$ is some constant and $A\beta$ is the relative chance the anyons annihilate when they are nearest neighbors. The probability $\Pi(L,\beta)$ then takes the form
\be
\Pi(L,\beta) \sim \frac{1}{1+A\beta} \frac{1}{\ln(L/2)} \label{EQN:TCPi}  .
\ee
Combining these elements, we expect a creation timescale
\be
\tau_c \simeq \frac{1}{R_0} \frac{1}{\Pi} \sim \frac{e^{2 \beta \Delta}}{L^2} (1+A\beta) \ln(L/2) \label{EQN:tauc} .
\ee
The total coherence time is $\tau_{\textrm{small}} = \tau_c + \tau_m$. In the small-system limit time clearly $\tau_c$ is the dominant contribution to $\tau_{\textrm{small}}$ due to its exponential dependence on $\beta$.

\begin{figure}
\includegraphics[width=0.92\columnwidth]{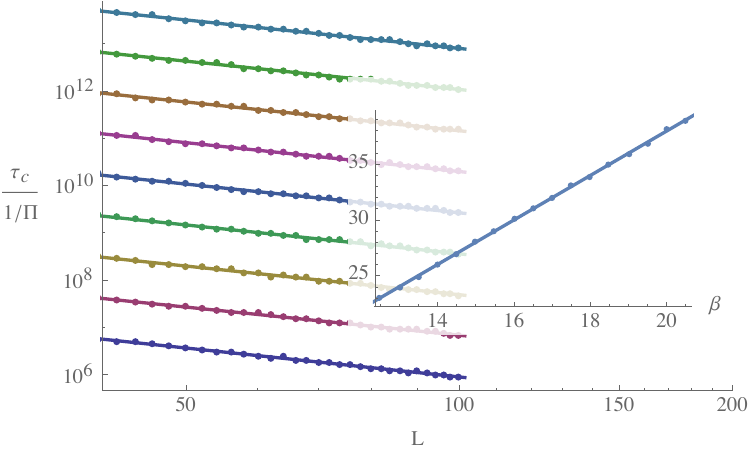}
\caption{ \label{Fig:TCLowTempTauC}(Color online)
Pair creation times for toric code excitations in the small system size limit. 
The average time of pair creation from initialization in the ground state $\tau_c$ is shown as a function of system size $L$, for a range of values from $\beta = 12 $ (bottom line) to $\beta = 22$ (top line). The inset shows the values of the fittings in the main plot at the y-axis intersection point, plotted as a function of $\beta$. 
Times $\tau_c$ are obtained by averaging over 1000 simulations. Here the values of $\tau_c$ are divided by a factor of $1/\Pi(L,\beta)$, which we determine numerically independent of $\tau_c$. 
The gradient of the linear fits, averaged over $\beta$, is -2.01, giving an overall scaling of $\tau_c = 0.150 ( e^{1.99\beta}/L^{2.01} ) /\Pi $. This is in agreement with the behavior predicted by Eqn.~(\ref{EQN:tauc}).}
\end{figure}

To test these predictions we rigorously study the system evolving in this limit using different numerical experiments with various initial conditions and some variations to the physical noise model. Key technical calculations involved in finding the coherence-time scaling are discussed at length in App.~\ref{Sctn:NumericsForTC}. Here we present the main results concerning the most significant contribution to the coherence time. We separately estimate $\Pi(L,\beta)$, $\tau_c$ and $\tau_m$ using numerical simulations. We find good agreement with the predictions of Eqns.~\eqref{EQN:taum},~\eqref{EQN:TCPi} and~\eqref{EQN:tauc} with the constant $A\approx 5$ for function $\Pi(L,\beta)$. The most significant contribution to $\tau_{\textrm{small}}$ comes from $\tau_c$, and the scaling of $\tau_c$ is dominated by the factor $1/R_0$. Fig.~\ref{Fig:TCLowTempTauC} shows just the $1/R_0$ scaling of $\tau_c$. This observation matches the predicted values of the key parameters very well, demonstrating a dependence on $1/L^2$ and an exponential growth with $2 \beta \Delta$.

The minimum energy barrier of the toric code is $ 2\Delta$, giving an Arrhenius' law estimate of the coherence time $\tau \sim e^{2 \beta \Delta}$. We have shown that in the small-size limit the leading contribution to the coherence time is $\tau_c$, given by Eqn.~\eqref{EQN:tauc}. If we ignore the sub exponential $\beta$ dependence inside $1/\Pi$, we can approximate the coherence time by
\be 
\tau_{\textrm{small}} \sim e^{2 \beta \Delta} \frac{\ln(L/2)}{L^2} .
\label{EQN:TCLowTempTauFinal}
\ee
We see that as $L$ becomes larger the lifetime of the toric code memory decreases polynomially up until $L \sim e^{\beta \Delta/2}$. This is a critical size above which the system starts to behave as it would in the thermodynamic limit. We now show that in the large-size limit the lifetime looses any dependence on system size and is determined only by $\beta$.

\subsubsection{Large system size limit}

In the large-size limit thermalization creates many anyons, with an equilibrium density of $\rho \sim e^{-\beta \Delta}$ for single anyons. On average, anyon pairs are created uniformly throughout the system and each occupy an area of $2/\rho$. We approximate the area as a square of linear size $\Lambda = (\rho/2)^{-1/2} \sim e^{\beta \Delta/2}$ as shown in Fig.~\ref{Fig:TCregimes}(c). The probability the decoder fails becomes appreciable once some fraction of pairs separate to distance $ \Lambda$ such that an error chain can percolate through the whole system, as shown in Fig.~\ref{Fig:TCregimes}(d).

Within a single $\Lambda \times \Lambda$ region the evolution proceeds like the small size case, i.e. there will be a creation event and the anyons subsequently diffuse apart. We say that a region fails once its anyons move close to anyons from neighboring regions. Assuming each region evolves independently, the time the system decoheres $\tau_{\textrm{large}}$ is estimated by the time the typical region fails. This is given by Eqns.~\eqref{EQN:taum} and~\eqref{EQN:tauc} where we set $L/2 = \Lambda$, giving diffusion and creation timescales $\tau_{m,\Lambda} \sim \beta \Lambda^2 \simeq \beta e^{\beta \Delta}$ and
\be
\tau_{c,\Lambda} \sim \frac{e^{2 \beta \Delta}}{\Lambda^2} (1+5\beta) \ln(\Lambda )  .
\label{EQN:TCHighTauCPart1}
\ee
To get an expression in terms of $\beta$ we write $\Lambda = C\,e^{\beta \Delta/2}$, where $C$ accounts for the constants absorbed into $\rho$. Then substituting $\Lambda$ into Eqn.~\eqref{EQN:TCHighTauCPart1} and rearranging
\be
\tau_{c,\Lambda} \sim e^{\beta \Delta} ( 1 + r\beta + s\beta^2 ) ,
\label{EQN:tauhc} 
\ee
where $r$ and $s$ are new constants related to those already introduced. In contrast to the small-size case, the creation and diffusion timescales have the same exponential dependence on $\beta$, and so we expect them both to contribute appreciably to the coherence time. Combining terms we obtain high-temperature coherence time
\be
\tau_{\textrm{large}} \simeq \tau_{c,\Lambda} + \tau_{m,\Lambda} \simeq e^{\beta \Delta} (1 + r'\beta + s\beta^2) \, ,
\label{EQN:tauh} \, 
\ee
where $r^\prime$ also includes the motional contribution. This simple modeling ignores thermal effects that can now take place such as pair fusion, where two anyons from different pairs fuse to vacuum combining their strings to form a longer string. Processes such as these will alter both the diffusion speed and the pair self-annihilation probability. As a result we do not expect this model to give good predictions of the values of $r^\prime$ and $s$. However, we do expect it to correctly predict the general features of the dynamics. In particular, the altered exponential dependance on $\beta$, and the system size independence of the lifetime. If we ignore its sub-exponential $\beta$ dependence the large-size coherence time is approximated by
\be
\tau_{\textrm{large}} \sim e^{\beta \Delta} \, .
\label{EQN:TCHighTempTauFinal}
\ee
This is also exponentially growing with $\beta$ but at a much slower rate than Arrhenius' law applied to the minimum energy barrier predicts.

\begin{figure}
\includegraphics[width=0.92\columnwidth]{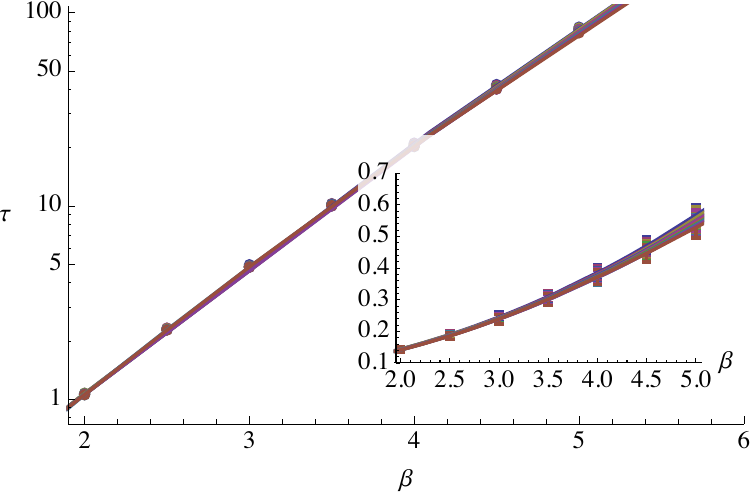}
\caption{ \label{Fig:TCHighTempTau}(Color online) 
Coherence time of the toric code in the large system-size limit. 
Times $\tau$ are averaged over 1000 simulations. They are shown here as a function of $\beta$ for a range of system sizes $L=100,\, 120,\dots,\,200$. The data is fitted to Eqn.~\eqref{EQN:tauh}. We observe only small variations in the fit parameters between system sizes and their averages give the expression $\tau_{\textrm{high}} = 0.56 \, e^{1.01\beta} (1 + 0.28\beta + 0.31\beta^2)$. Inset is the non-exponential part of the scaling obtained by dividing the values $\tau$ shown in the main plot by $e^{\beta \Delta}$.}
\end{figure}
We simulate the system evolving in this regime in order to test our assumptions and verify predictions about the dynamics. Here we present the key results. A thorough discussion of our methods and results is given in App.~\ref{Sctn:NumericsForTC}. We verify that anyon densities at the time the decoder fails obey $\rho \sim e^{-\beta \Delta}$ and that for the parameters we consider the typical number of anyons is always large, $\langle N \rangle \gg 1$. The average separation between anyon pairs that were either created together or joined by a fusion is seen to scale as $\Lambda \sim e^{\beta \Delta/2} $ as expected. In addition we see that the maximum separation between any pair is always much less than $L/2$ confirming that the decoherence results from the average motion of anyons in local regions. We give numerical data showing the scaling of $\tau_{\textrm{large}}$ predicted in Eqn.~\eqref{EQN:tauh} in Fig.~\ref{Fig:TCHighTempTau}. Our results are also seen to clearly demonstrate coherence time scaling that is independent of the size of the toric code.

\subsection{Characteristics of Self-Correcting Memories} \label{Sctn:Condit}

To properly compare and classify models that are proposed as self-correcting quantum memories, we must have a clear idea of what a self-correcting quantum memory is. We conclude this Section by presenting the general characteristics we require from a quantum memory at finite temperature. We use this criteria as a comparative tool to guide us through the presentation of a wide variety of models. We emphasise that the list we give should be regarded as a set of guidelines to be challenged. Other variations of the desiderata asked of a quantum memory are given in Refs.~\cite{Brell, Landon-Cardinal15}.

In the study of a quantum memory, we first require physically realistic systems. We are interested in Hamiltonian models with interaction terms that are defined locally in three or fewer dimensions. Additionally, we will only consider Hamiltonians whose local interaction terms have eigenvalues that are bounded by a constant independent of the size of the system. In a similar vein, we also require that each physical degree of freedom in the system only supports a constant number of Hamiltonian interactions. These conditions have been specified for commuting Pauli Hamiltonians precisely in Subsec.~\ref{Sctn:CPHamiltonians}.

We next ask what properties we expect of a self-correcting quantum memory. Importantly, we must be able to write information to a quantum memory. We can achieve this using external control during the preparation of the system. Then, encoded information should remain coherent without the application of any control for an arbitrary amount of time while the system is exposed to thermal errors. Ideally we hope that the coherence time of encoded quantum information will diverge to infinity as the size of the system is increased. We require this behavior to be present at some arbitrarily small but non-zero temperature. Such behavior is typically associated with a phase of matter that is ordered at finite temperature below some critical temperature, similar to the ordered phase we have observed in Subsec.~\ref{Subsctn:CurieWeiss} with the Curie-Weiss model. 

Further, to ensure that encoded information evolves coherently for an arbitrarily long time, we require that the orthogonal encoded states of a quantum memory are degenerate with respect to the system Hamiltonian. Otherwise, encoded quantum information decoheres rapidly. To this end, we require that the energy splitting between the orthogonal states of the encoded space of the memory vanishes as the size of the system diverges.

In general we do not expect to be able to realize an exact quantum Hamiltonian. Typically a physical system will be subject to minor perturbations due to stray fields or perhaps imperfections in their preparation. We therefore require that the properties that we ask of a quantum memory to be robust against arbitrary local Hamiltonian perturbations, provided the perturbations remain sufficiently weak.  

Finally, we require the ability to read out encoded information after the memory has suffered some errors. Even with a memory with self-correcting properties, we still expect to sustain some small errors that may affect the measurement of logical states. We thus require a decoding algorithm such as those discussed in Subsec.~\ref{Sctn:ErrorCorrection} and in App.~\ref{Sctn:Decoders} to identify and correct for small physical errors at the point of readout. Moreover, in order for the memory to scale in a practical manner, we require the execution time of the decoding algorithm to scale efficiently with the size of the system.

With these considerations the following list summarizes the criteria we ask of a self-correcting quantum memory

\begin{enumerate}
\item Locally embeddable in three or fewer dimensions with bounded Hamiltonian interactions and where qubits support a bounded number of interactions.
\item Encodes a quantum state whose coherence time diverges with system size at a sufficiently low non-zero temperature.
\item Splitting between the energy levels of the encoded subspace vanishes in the thermodynamic limit.
\item Memory properties are robust under local perturbations.
\item Efficiently decodable at readout.
\end{enumerate}

To the best of the knowledge of the authors there currently exists no model that has been proven to satisfy all of these criteria. Indeed, we will see throughout this exposition that there are many example models that achieve some of these properties, or perhaps weaker notions of these properties, and compromise others. Certainly, it remains very interesting to discuss the capability of a quantum memory that does not satisfy all of the listed criteria. We will see that some of the models discussed in this review have coherence times that increase with system size up to some cutoff. While not truly self-correcting by our criteria, such models may be very useful if the coherence time cutoff is large.

While it is difficult to discover a system that satisfies the proposed criteria for a self-correcting quantum memory, it is the goal of this research programme to realize a quantum memory in the laboratory that can ultimately be manufactured for the purposes of quantum technologies. Moreover, we require that a memory serves as a component of a larger information processing machine that must work and communicate with other components of a larger processor to complete computational tasks. We therefore append to list of criteria some additional desiderata that we might reasonably ask of a quantum memory.

We first consider further the feasibility of realising different Hamiltonians. Although it already presents a significant challenge to discover self correction among Hamiltonians with constant interaction terms, as we have phrased the problem, the constant weight of a Hamiltonian interaction can in general be a large constant. In reality, naturally occurring Hamiltonians typically have two-body interactions. It is therefore interesting to discover self-correcting Hamiltonians that have strictly two-body interactions. Similarly, imposing translational invariance is particularly exciting with respect to scalability as we could potentially engineer such a system by designing simple repeating units of the many-body system. We may even expect to find such a system in a regular strongly interacting crystal.

Further, in the interests of experimental amenability, although we can realize three-dimensional systems, such models may be difficult to manipulate. Specifically, we might expect that the quantum degrees of freedom in the center of a three-dimensional crystal will be difficult to access. Such accessibility is likely to be invaluable for encoding and reading out encoded quantum states, and for measuring syndrome data to identify errors suffered by the system. To this end, it is favorable to find a quantum memory in dimensions smaller than three.

We finally consider fault-tolerant computational abilities in our wish list. Indeed, although finding systems capable of preserving coherent quantum states at finite temperature already presents a considerable challenge, we may also wish to directly perform interesting computational tasks on information encoded within the memory. Such a property may help reduce computational overhead when we consider manipulating encoded information in a quantum circuit.

We summarize the discussed desiderata below

\begin{enumerate}
 \item Low-weight, ideally two-body, Hamiltonian interactions.
 \item Translational invariance.
 \item Embeddable in a low number of dimensions.
\item Compatible with a fault-tolerant universal quantum gate-set.
\end{enumerate}

\section{No-Go Theorems}
\label{Sctn:NoGoTheorems}

Before beginning the search for a quantum memory over the vast space of many-body lattice Hamiltonians, it is wise to rule out systems which we cannot expect to maintain quantum information at finite temperature. For this purpose we now consider no-go theorems that identify broad classes of models with physical characteristics that we cannot expect to lead to passively protected memories. 

The study of finite-temperature quantum memories requires a breadth of technical aspects, from the abstract mathematical theory of coding, to the more physically motivated field of study of finite-temperature effects on lattice Hamiltonians. To this end, no-go results can be broadly separated into two types. We label these {\em general no-go theorems} and {\em physically motivated no-go results}. The distinction is the following; general no-go theorems seek to exclude large classes of systems from possessing important properties that we expect to be necessary for self correction. Physically motivated no-go results take into consideration dynamics and microscopic thermal effects to show specific models that will fail to behave well as a quantum memory. Both approaches have complementary advantages, and are ultimately of equal importance.

The general no-go theorems typically eliminate the possibility of {\em energy barriers} in certain classes of systems. Macroscopic energy barriers between degenerate ground states are the basis of our current understanding of finite-temperature stability in classical models. The prototypical case of a classical stable model is the two-dimensional Ising model, which is presented in detail in Subsec.~\ref{Sctn:TheIsingModel}. Moreover, it has been shown that an energy barrier is required for a large class of commuting Hamiltonian models~\cite{Temme14, Temme15, Komar16}. It is therefore unlikely that we can expect to find a passive quantum memory with a model that does not support a macroscopic energy barrier.

Physically motivated results lose the generality of their counterpart class of no-go theorems. Instead, they model thermal effects acting on specific models. This approach offers new intuition to show that under physical considerations certain models fail to perform well as a quantum memory. Such results are typically obtained by studying the relevant order parameters that correspond to logical operations acting on the code space of quantum memories. Order parameters are then studied with respect to the dynamics of quantum system when interacting with an auxiliary environmental system, or in its Gibbs thermal equilibrium state. These results support known general no-go theorems for models where it is believed that finite-temperature stability cannot exist.

Known no-go theorems are most limiting in two dimensions. It has been shown  by Bravyi and Terhal~\cite{BravyiTerhal09} that two-dimensional commuting Pauli Hamiltonian models, including the toric code model, cannot support an energy barrier that scales with the size of the system. This result is generalized to two-dimensional topologically ordered commuting Hamiltonians by Landon-Cardinal and Poulin~\cite{LandonCardinalPoulin}. These results are supported by a wealth of physical no-go results, typically obtained using Kitaev's toric code model, where various order parameters are shown to rapidly decay at finite temperature.

Three-dimensional no-go theorems are significantly less restrictive when compared with their two-dimensional counterparts~\cite{Yoshida, Haah13, PastawskiYoshida}. These results leave more promise for the discovery of new models with a macroscopic energy barrier. Indeed, the assumptions necessary to prove the discovered three-dimensional no-go theorems describe a limited set of models when compared with the theorems known in two dimensions. Supporting these results, we also have various physical results showing that topological entanglement entropy~\cite{CastelnovoChamon08}, and the correlation functions of string-like logical operators~\cite{AlickiHorodeckiHorodeckiHorodecki} decay rapidly for the three-dimensional toric code model. We see in Sec.~\ref{Sctn:ThreeDimensions}, that there are many known physically feasible three-dimensional models that avoid the no-go assumptions we describe here, and present favorable properties for finite temperature stability. 

In this Section we begin by reviewing no-go theorems in two dimensions. We reproduce the proof of Bravyi and Terhal to show that two-dimensional stabilizer models cannot support a macroscopic energy barrier, and we discuss the supporting physically motivated no-go results. We follow the discussion by considering the no-go theorems in three dimensions. We conclude with possible avenues for avoiding the known no-go theorems. The final Subsection serves as motivation for the positive results that we will discuss later on, which include the various models that have been proposed as stable quantum memories. 

\subsection{No-Go Results in Two Dimensions}

In this Subsection we review no-go results in two dimensions. We consider in detail the general no-go theorem due to Bravyi and Terhal~\cite{BravyiTerhal09}, and we discuss the physically motivated no-go results. We motivate this no-go theorem by first considering the toric code, the prototypical model for quantum error correction. We have seen in Sec.~\ref{Sctn:2dTC} that the logical operators of the toric code are one-dimensional string-like operators. Models with logical operators of this type have a constant energy barrier.  To understand this from the anyonic picture of two-dimensional topologically ordered memories given in Subsec.~\ref{Sctn:AnyonicPicture}, these logical operations correspond to the creation of a pair of anyonic excitations at a constant energy cost which are then free to walk across the lattice at no additional energy penalty. This is discussed in detail in Refs.~\cite{NussinovOrtiz07, AlickiFannesHorodecki}, but ultimately follows from the fact that one can find a sequence of single qubit error operations that will realize a logical operator without increasing the energy of the system beyond a constant value that is independent of the system size. With the toric code in mind it becomes interesting to see if we can find a two-dimensional system with logical operators that are not supported along a one-dimensional line. We follow the proof of Bravyi and Terhal~\cite{BravyiTerhal09} to show that local two-dimensional stabilizer models necessarily have one-dimensional logical operators which are expected to be incompatible with finite temperature stability. In the exposition we show how a local noise model can construct a logical operation over its code space at no more than a constant energy cost with respect to the size of the lattice, thus completing the proof.

\begin{figure}
\includegraphics{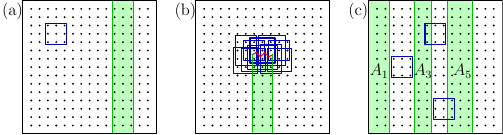}
\caption{(Color online) A sketch of the proof of the no-go theorem due to Bravyi and Terhal. Figures depict square lattices where qubits are marked by black points. (a) Local stabilizer generators are confined to small squares of size $r$ such as that shown in blue in the top left corner of the Figure where $r = 3$. For such a code Bravyi and Terhal show that there must exist a logical operator supported on a quasi one-dimensional strip of width $r$, such as the vertical strip which is shaded green in the Figure. (b) An error that forms part of a logical operator, supported on the shaded green vertical strip, do not violate more than a constant number of local stabilizers, marked by blue squares of width $r$, that extend no further than a distance of $r-1$ away from the broken end point of the shaded strip, shown as a red zig-zag line. (c) A high weight logical operator can be cleaned onto region $A = \bigcup_k A_k$ for odd values of $k$, where each stabilizer generator, examples of which are supported inside small squares such as those displayed on the Figure in blue, have common support with no more than one strip. The proof finally uses this fact to conclude that a logical operator must be supported on a single vertical strip. \label{NoGoDiagrams}}
\end{figure}

We consider a two-dimensional square lattice of size $L\times L$. The qubits interact via the local Hamiltonian 
\begin{equation}
H = -  \sum_j S_j,
\label{eqn:no-go}
\end{equation} 
where $S_j$ are in general an over-complete set of local generators for stabilizer group $\mathcal{S} \subset \mathcal{P}_n$ as defined in Sec.~\ref{Sctn:HamiltoniansBasics}. The ground space of $H$ is the code space of $\mathcal{S}$. Without loss of generality, each local stabilizer generator that acts on a small subset of qubits on the lattice can be contained in a square box no larger than constant linear size $r$. We show such a box in blue in Fig.~\ref{NoGoDiagrams}(a). Moreover, each box can contain no more than a bounded constant number of interaction terms. Hamiltonians with these properties are physically well-motivated as described in Subsec.~\ref{Sctn:CPHamiltonians}. We will observe that for the described stabilizer group with local generators $S_j$, there must exist a logical operator that is supported on a one-dimensional strip of width $r$, as shown in green in Fig.~\ref{NoGoDiagrams}(a). 

We now elaborate on the noise model that introduces a one-dimensional logical error without increasing the energy of the system above a constant independent of its size. The noise model of interest can introduce Pauli errors to single qubits of the lattice. We consider `segments' of a Pauli logical operator supported on the green region in Fig.~\ref{NoGoDiagrams}(b) which is of variable length $ 1 \le l \le L$, cutoff along the horizontal red zigzag line. Importantly, the minimum energy cost associated to this segment is upper bounded by $ \varepsilon \sim \mathcal{O} (r^2)$, independent of the system size or $l$. The part of the logical operator supported on the green region will only violate, i.e. anti-commute with, stabilizer generators of the physical Hamiltonian that are within a radius $\sim r$ from the cutoff point, shown by the blue area on the lattice. Violated stabilizers correspond to the energy cost of the error on the segment with respect to the Hamiltonian.

Given that we have shown that the energy cost of a logical operator segment is upper bounded by energy cost $\varepsilon$ independent of $l$, it suffices to demonstrate that the energy cost of moving the cutoff by a single unit via single qubit flips for a generic two-dimensional stabilizer model is constant. Indeed, we can change segment length from $l$ to $l+1$ by overcoming an energy barrier that does not exceed a constant energy cost, $\omega \sim \mathcal{O}(r^3)$, independent of system size, before returning to its energy minima $\varepsilon$ once the logical segment achieves length $l+1$. We bound $\omega$ by considering the introduction of a single qubit Pauli error on the lattice. Due to the locality of the terms of the physically constrained Hamiltonian, introducing a new Pauli error can only increase the energy of the system by a constant at most $\sim r^2$. To increase the logical operator segment from length $l$ to length $l+1$ the noise model introduces a specific set of $r$ single-qubit Pauli errors close to the red zig-zag line in Fig.~\ref{NoGoDiagrams}(b). This requires the addition of no more than $r$ single qubit Pauli operators, whose energy cost can be no more than $r^2$. We are therefore able to bound $\omega \sim r \times r^2$. The described argument holds in the generality of creating a logical operator segment from the ground space of the lattice by considering the increase of the size of a segment from $l  = 0$ to $l = 1$.

We have shown that a logical operator segment of length $l $ has energy at most $\varepsilon$, and that we can increase the length of the logical operator segment with energy cost no greater than $\omega$. It follows from this that the single-qubit Pauli error noise model can introduce a logical error to the ground space of the model using a logical operator segment with length $l = L$ with energy never greater than $\varepsilon+\omega$. This demonstrates that commuting Pauli Hamiltonians with a one-dimensional logical operator have a constant energy barrier.

It only remains to show that a stabilizer group generated locally in two dimensions necessarily has one-dimensional logical operators. The technical proof makes use of {\em the cleaning lemma}, proved in Ref.~\cite{BravyiTerhal09}. 

\begin{lemma}
Given a stabilizer group $\mathcal{S}$ that acts on a set of qubits $Q$, one of the following statements holds for any subset of qubits $A \subseteq Q$:
\begin{enumerate}
\item There exists a non-trivial logical operator $\overline{L} \in \mathcal{P}_n $ supported entirely on $A$.
\item All logical operators $\overline{L} \in \mathcal{P}_n$ can be deformed by a stabilizer $S \in \mathcal{S}$, such that $\overline{L}S $ is not supported on $A$. 
\end{enumerate}
\end{lemma}

We complete the proof using the cleaning lemma. We separate the lattice into an even number of strips of width either $r $ or $r-1$. We can check that a lattice of size $L = a (r-1) + b r$ can be decomposed into some even number of $a + b $ strips for $L \ge 2(r-1)^2$, for proof, see footnote
\footnote{Proof by induction. We obtain $L = 2(r-1)^2$ with solution $a = 2(r-1)$ and $b = 0$. Assume true for $L = a(r-1) + br $ for all $L \ge 2(r-1)^2$. In the case that $a > 0$ we obtain $L+1 = a'(r-1)+b'r$ with values $a' = a-1$ and $b' = b+1$. If $a = 0$ we choose $a' = 2r - 1$ and $b' = b-2(r-1) +1$, which satisfies $a'+b' = \text{even}$, since $b$ is even.}. 
We index the strips in order, and we consider the region of odd strips $A = \cup_{k \in \text{odd}}A_k$, as shown in Fig.~\ref{NoGoDiagrams}(c).

We now obtain this proof by contradiction. We assume that there exists a logical operator $\overline{L}' \in \mathcal{P}_n$ whose minimum support cannot be contained on a vertical strip of width $r$. Due to the width of the strips, such a logical operator cannot be deformed by stabilizers away from region $A$. Therefore, by the cleaning lemma, it must be possible to find a logical operator $\overline{L} = \overline{L}'S  $ for some $S \in \mathcal{S}$ such that $\overline{L}$ is supported entirely on region $A$. As the logical operator support is wider than a single strip, it must be supported on multiple odd strips $A_k$. Accordingly, we decompose the logical operator $\overline{L} = \prod_{k \in \text{odd}}\overline{L}_k$, where Pauli operators $\overline{L}_k$ are the support of $\overline{L}$ on strip $A_k$ for odd $k$.

To complete the argument, we consider operators $\overline{L}_k$. A logical operator will commute with all elements of $\mathcal{S}$. Given the choice of strip width, we observe that the support of any stabilizer overlaps with no more than one odd strip. We show examples of the supports of stabilizer generators within blue squares in Fig.~\ref{NoGoDiagrams}(c). It follows from this fact that, in addition to the logical operator $\overline{L}$, all operators $\overline{L}_k$ must also commute with the stabilizer group. The Pauli operators that commute with the stabilizer group are one of two types of operators. Either, they are elements of the stabilizer group, such that $L_k \in \mathcal{S}$, or, they themselves are logical operators. Given that $\overline{L}$ is a logical operator, there must be one $\overline{L}_k$ that is a logical operator with width less than or equal to $r$, providing the desired contradiction. With the observation that we necessarily have at least one logical operator with a one-dimensional support for a stabilizer group which is generated by local two-dimensional stabilizer generators, we conclude the proof that there exists a constant energy barrier between two orthogonal ground states of commuting Pauli Hamiltonian in two dimensions.

The discussed work of Bravyi and Terhal has been extended in a number of different directions. In Ref.~\cite{LandonCardinalPoulin}, it is shown that given a local {\em topologically ordered} commuting Hamiltonian, that there always exists a noise model that can locally create a logical operation on the ground space of the model at no more than a constant increase in system energy, thus extending the result of Bravyi and Terhal to a more general class of systems. We remark also on the work due to Haah and Preskill in Ref.~\cite{Haah12} where it is shown that two-dimensional stabilizer models do not support any logical errors that cannot be achieved with an energy cost that is independent of the size of the system under a local noise model.

Supporting the discussed general results in two dimensions, there is also a plethora of physically motivated results in the literature that typically consider the prototypical case; the toric code model. Among the approaches include a study of topological entanglement entropy~\cite{KitaevPreskill06, LevinWen06} for the toric code model in the thermal equilibrium state, see Refs.~\cite{CastelnovoChamon07, Iblisdir09, Iblisdir10}. In Refs.~\cite{Iblisdir09, Iblisdir10} it is identified that in realistic systems the topological entanglement entropy vanishes in the large system size limit at finite temperature for the general class of Kitaev quantum double models~\cite{Kitaev03}. Further physically motivated study includes the rigorous proof of instability in the toric code model using the Liouvillian open dynamics to show that the expectation values of the logical operators of the toric code model decay rapidly when weakly coupled to a Markovian environment~\cite{AlickiFannesHorodecki}. These results are generalized in Ref.~\cite{ChesiLossBravyiTerhal}, and are considered for the toric code with higher-dimensional spins in Ref.~\cite{ViyuelaRivasMartin-Delagado}. 

In addition to the physically motivated no-go results we also remark on the result due to Hastings~\cite{Hastings11} which shows that commuting two-dimensional models are unable to support topological order at finite temperature. Similar conclusions were derived by Nussinov and Ortiz by consideration of lattice models in Refs.~\cite{Nussinov, Nussinov09, NussinovOrtiz07} using methods that were later improved in Ref.~\cite{ChesiLossBravyiTerhal}. Results such as these are particularly important with respect to finite-temperature perturbative stability, which we regard as a required condition for a stable quantum memory. The results of Hastings are supported numerically as discussed by Wootton in Ref.~\cite{Wootton13}, where he compares the topological order of unstable memories at finite temperature with stable interacting models. These results are discussed later in Subsec.~\ref{Subsec:FiniteTempTO}.

\subsection{No-Go Results in Three Dimensions}

Thus far we have seen that no-go theorems are very restrictive against energy barriers in two-dimensional systems. It is shown that both commuting topologically ordered Hamiltonians, and commuting Pauli Hamiltonians in two dimensions necessarily support at most a constant energy barrier. In this Subsection we discuss known results in three dimensions. Here, the landscape of local Hamiltonians is much more rugged. Indeed, we will observe that no-go theorems for three-dimensional models are often superseded by models that avoid the assumptions of a given theorem in a physically sound way. 

The first general no-go result in three dimensions is given in Ref.~\cite{Yoshida}. Here, the methods of~\cite{KayColbeck08} and~\cite{YoshidaChuang10} are used to show that three-dimensional Pauli Hamiltonians that are translationally invariant, and have a constant ground state degeneracy, must have a one-dimensional logical operator, which can be produced with energy cost independent of the system size. 

The work of Yoshida has been supported by physically motivated results that study the three-dimensional toric code model~\cite{HammaZanardiWen} following various approaches. Castelnovo and Chamon study the topological contributions to the entanglement entropy of the three-dimensional model in its Gibbs equilibrium state in Ref.~\cite{CastelnovoChamon08}. They find that, while some topological order parameters remain robust up to a critical temperature, a qubit cannot be stored in the three-dimensional toric code at finite temperature because loop-like order parameters decay rapidly in the large system size limit. Similar results are anticipated using the methods of Alicki {\it et al.} in Ref.~\cite{AlickiHorodeckiHorodeckiHorodecki}, where they study the thermal dynamics of the three-dimensional toric code model by considering the model weakly coupled to a Markovian thermal reservoir.

Pleasingly, it has been explicitly shown that one can surpass the no-go theorem of Yoshida. In Ref.~\cite{Haah}, Haah constructs a translationally invariant three-dimensional Pauli Hamiltonians with a macroscopic energy barrier. The model, commonly known as the cubic code, does not have a constant ground state degeneracy, and thus avoids the no-go theorem of Yoshida. We discuss this model in detail in Sec.~\ref{Sctn:CubicCode}.

Following the discovery of the cubic code, Haah generalized Yoshida's no-go theorem in Ref.~\cite{Haah13}. He uses an elegant representation of the Pauli group to show that translationally invariant three-dimensional commuting Pauli Hamiltonians can support at best an energy barrier that scales logarithmically with the size of the system. Subsequently, Michnicki demonstrated an explicit example of a Pauli Hamiltonian model that supports a power-law energy barrier, which is not translationally invariant, known as the welded toric code model~\cite{Michnicki, Michnicki14}. In this model the energy barrier of the non-commuting logical operators are varied by changing the size of the system over different length scales. Notably, the welded code has a constant two-fold ground state degeneracy, and only violates the translational invariance assumption of the Yoshida proof. We discuss the welded toric code in Sec.~\ref{Sctn:WeldedToric}.

Finally, we remark on a recent result given in Ref.~\cite{PastawskiYoshida}. There the authors show that there is a trade off for three-dimensional commuting Pauli Hamiltonians between their capability to support an energy barrier and the fault-tolerant quantum gates that can be achieved by local operations within the ground space of the Hamiltonian. Specifically they considered commuting Pauli Hamiltonians that can perform a fault-tolerant non-Clifford logical operation  by local operations. They show Hamiltonians with such a property cannot support a macroscopic energy barrier. This result is obtained by extending the results of Ref.~\cite{BravyiKonig13}. An example of such a code which performs a non-Clifford gate, namely the $\pi/8$-gate, by applying the $\pi/8$-gate locally to each of the physical qubits, is the three-dimensional color code~\cite{Bombin07}. Indeed, this model is not expected to support finite temperature stability, as it falls into the class of models described by the no-go theorem due to Yoshida.

\subsection{On No-Go Results}
\begin{figure}
\begin{center}
\includegraphics{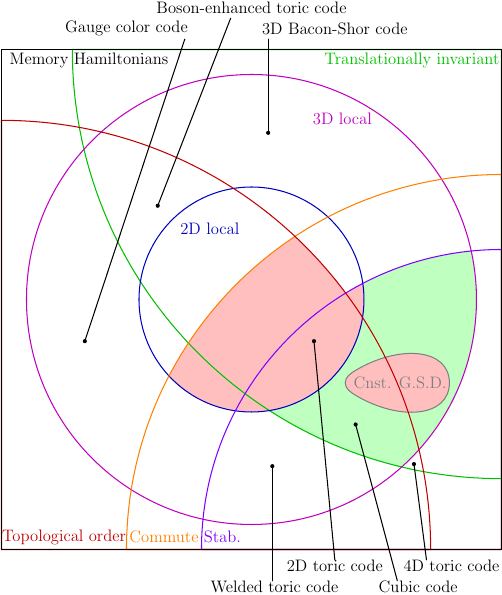}
\end{center}
\caption{(Color online) The landscape of no-go theorems in the space of candidate memory Hamiltonians. Two- and three-dimensional models are shown in magenta and blue circles which are centered in the middle of the diagram. Commuting and stabilizer models are shown inside the orange and purple circles that are centered in the bottom-right corner of the diagram. Translationally invariant models lie within the green circle which is centered at the top-right corner of the diagram. Models satisfying topological order conditions are shown inside the red circle centered at the bottom-left corner of the diagram. Three-dimensional models with a constant ground state degeneracy lie inside the grey `egg-shaped' region to the right of the Figure. Dark red shaded regions have been proven to support an energy barrier that does not scale with the size of the system. Light green shaded areas correspond to models that have energy barriers that scale at best logarithmically with system size. We mark some specific examples of models that we will discuss later in the Review.\label{NoGoVenn}}
\end{figure}

In this Section we have considered several no-go results. Known no-go theorems identify two large classes of two-dimensional models that cannot support a macroscopic energy barrier; commuting Pauli Hamiltonians, and topologically ordered commuting Hamiltonians. In addition we have discussed no-go theorems showing three-dimensional commuting Pauli Hamiltonians are constrained in their ability to support a power law energy barrier if they are translationally invariant. These results are summarized in the Venn diagram shown in Fig.~\ref{NoGoVenn}. Importantly, we provide specific models as examples of the general categories that demonstrate the corresponding behavior. 

The no-go results significantly restrict the models we might consider in two dimensions for finite-temperature quantum memories. In particular, it is shown that commuting two-dimensional models cannot support topological order at finite temperature~\cite{Hastings11}, and conversely that commuting topologically ordered models cannot support a macroscopic energy barrier~\cite{LandonCardinalPoulin}. These are very restrictive findings given that we demand perturbative stability for a quantum memory, which is assured with the condition of topological order. Nevertheless, it is not known that topological order is necessary for perturbative stability. As such, there may exist perturbatively stable models that are not topologically ordered that can support a macroscopic energy barrier~\cite{LandonCardinalPoulin}. With this in mind, there may still exist commuting two-dimensional models that are suitable as quantum memories.

Another approach to overcoming no-go results in two-dimensional topologically ordered systems is to simply violate their physical assumptions. One study that has attracted notable interest are {\em interacting anyon models}. In general, achieving such systems requires the violation of the locality assumption of the discussed no-go theorems. Considerable work has been conducted to find condensed-matter systems that give rise to an effective interacting anyon theory in a local setting. Interacting anyon models are the topic of Sec.~\ref{Sctn:InteractingAnyons}.

Further, as we have touched upon in this Section, we can obtain positive results for macroscopic energy barriers in three-dimensional commuting Pauli Hamiltonians. In Sec.~\ref{Sctn:ThreeDimensions} we review three-dimensional models including {\em the cubic code model}, a translationally invariant model with logarithmic energy barrier, and the {\em welded toric code model}, that obtains a power law energy barrier by breaking translational invariance.

Curiously, we see in Sec.~\ref{Sctn:ThreeDimensions} that the logarithmic energy barrier of the cubic code or the power-law energy barrier of the welded toric code do not satisfy the required conditions for a self-correcting quantum memory given in Subsec.~\ref{Sctn:Condit}. To this end, it is even an interesting point of study to try to better understand general necessary and sufficient conditions for models to satisfy the desiderata of a self-correcting memory. Work in this direction has been conducted in Refs.~\cite{Temme14, Temme15, Komar16}.

Beyond commuting Pauli Hamiltonians there may exist stable quantum memories based on two- or three-dimensional non-commuting Hamiltonians. Due to the difficulty in analytical and numerical calculations for such models, these classes of systems are less well understood compared with their commuting counterparts. However, interesting results have emerged in the field of {\em subsystem codes}, see for instance Refs.~\cite{Bacon06,BravyiTerhal09, Bravyi11}. Subsystem codes of particular recent interest with respect to finite-temperature stability include the three-dimensional Bacon-Shor code~\cite{Bacon06}, the gauge color code~\cite{Bombin13a, Bombin14}, and the sparse-circuit codes due to Bacon~{\it et al.}~\cite{Bacon14}. We discuss subsystem codes in Sec.~\ref{Sctn:Subsystems}. 

Finally, we note that the presented general no-go theorems only rule out the possibility of energy barriers in certain classes of models. One might try to sidestep these no-go theorems by finding an alternative method to prevent a finite temperature environment from corrupting information in the ground space of a quantum memory. We discuss work towards finding such alternatives in Sec.~\ref{Sctn:EntropicProtection}.

In the next Section we give consideration to both classical and quantum systems that are known to be thermally stable. Unfortunately, known quantum systems that are proven to be thermally stable are local in dimensions larger than three. Nevertheless, such analysis gives constructive insights into the properties that give rise to a finite-temperature quantum memory.

\section{Thermal Stability in High Dimensions}
\label{Sctn:HighD}

Thermal stability in classical systems was first understood by the discovery of Peierls' argument~\cite{Peierls}. It shows us that in statistical mechanics stability increases with dimensionality. In this Section we follow this trend in the quantum realm. We consider high-dimensional generalizations of well-studied quantum memories to arrive at systems that support finite-temperature stability. Disappointingly, this approach has not yet uncovered a stable model with dimensionality smaller than four. However, to take a positive outlook on the results summarized in this Section, it is demonstrated that finite-temperature stability is not fundamentally inhibited by quantum mechanics. Furthermore, the consideration of high-dimensional quantum models that support finite-temperature stability may enable us to develop a new intuition of thermal stability, and, may inspire the discovery of a stable quantum memory in lower dimensions. 

In this Section we review the seminal case of finite-temperature stability by studying the famous two-dimensional classical Ising model. We go on to describe the four-dimensional toric code; the first quantum model rigoroulsy proved to be thermally stable at finite temperature. We conclude this Section with a further high-dimensional generalization, namely, the six-dimensional color code model. This model supports both finite-temperature stability, and a set of fault-tolerant operations that can implement universal quantum computation.

\subsection{Stability in Classical Models}
\label{Sctn:TheIsingModel}

In this Subsection, we review the Peierls argument of stability in the two-dimensional classical Ising model~\cite{Ising}. The Peierls' argument~\cite{Peierls}, later refined by Griffiths~\cite{Griffiths}, shows that the critical phenomena of the Ising model are dependent on its dimensionality. For a modern overview of the Peierls' argument, and other important topics relating to the Ising model, we refer the reader to Ref.~\cite{Huang, McCoy14}. A modern numerical study of this model is given in the context of a classical memory  by Day and Barrett in Ref.~\cite{Day12}. 

While we have more sophisticated methods of extracting the phase diagram of the two-dimensional Ising model due to its exact solution by Onsager~\cite{Onsager, Yeomans}, the intuition developed from Peierls' original argument is a very useful tool for understanding the stability of models where no exact solution is known, see, for instance, Refs.~\cite{LebowitzMazel, Bonati, Campari}. Indeed, Peierls' argument is used to demonstrate the stability of the high-dimensional quantum systems. It is therefore instructive to give a detailed discussion of Peierls' argument applied to the simplest case. 

\begin{figure}
\includegraphics{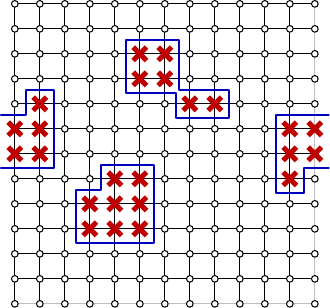}
\caption{(Color online) An example spin configuration of the two-dimensional Ising model. Spins lie on the vertices of a lattice with periodic boundary conditions, and interactions are shown as edges of the lattice. All the spins have the same orientation except for those that have been flipped, which we mark with red crosses. In general, we say that patches of flips occur in `droplets'. The energy penalty introduced by a droplet will scale like the length of its boundary. We mark boundaries that enclose droplets with thick solid blue lines which form closed loops. \label{PeierlsContours}}
\end{figure}

We consider the Ising model defined on an $L \times L $ periodic square lattice of $V = L^2$ spins on its vertices, as shown in Fig.~\ref{PeierlsContours}. The spin variables $\sigma_j$ take two values, $\pm 1$, and interact via nearest-neighbor interactions described by the classical Hamiltonian
\begin{equation}
E(\boldsymbol{\sigma}) = - {1 \over 2}\sum_{\langle j, k \rangle} \sigma_j \sigma_k,\label{IsingHam}
\end{equation}
where $\langle j, k \rangle$ denote pairs of vertices that are connected by edges of the square lattice, and $\boldsymbol{\sigma}$ is a configuration of all lattice spins. The model behaves as a classical memory that stores a single bit in its two-fold degenerate ground space, where the bit is encoded in the magnetization of the system, $\overline{\sigma}\equiv \sum_j\sigma_j / V $. The ground states of the model are $\overline{\sigma} = \pm 1$, such that the state $\overline{\sigma} = \pm1$ corresponds to the configuration where $\sigma_j= \pm 1$ for all $j$, respectively. 

In practice, the Ising model exists at finite temperature. At non-zero temperature the probability of finding the system in the ground state vanishes in the thermodynamic limit. However, for the purpose of storing a bit of information, the stored data will be maintained if the sign of the magnetization is constant. It is expected that the magnetization will maintain the correct sign in the ordered phases of the system. To this end, we must check that the thermal average of the absolute value of the magnetization, $\langle \left| \overline{ \sigma } \right| \rangle$, remains non-zero in the thermodynamic limit for some suitably low but finite temperature. 

The argument begins by considering a spin configuration with respect to Hamiltonian~(\ref{IsingHam}). In Fig.~\ref{PeierlsContours} we have a lattice of spins that are mostly in the $+1$ state, shown by white circles, with regions, or `droplets', of flipped spins in the $\sigma_j = -1$ state.  We show three such droplets in the example configuration by patches of red crosses. The Hamiltonian will impose a unit energy cost for each pair of nearest-neighbor spins that have opposite states. As such, the energy cost of a droplet is proportional to the length of its boundary. These boundaries are known as {\em Peierls contours}, indexed by $b$, and are marked in blue in Fig.~\ref{PeierlsContours}. We note that a contour is a {\em single} boundary of a droplet, and that in general a given configuration can contain many contours. The probability that a state is in a given configuration from a thermal Gibbs distribution is $p(\boldsymbol{\sigma}) = \exp(-\beta E(\boldsymbol{\sigma})) / \mathcal{Z}$, where $\mathcal{Z} = \sum_{\boldsymbol{\sigma}\in C} e^{-\beta E(\boldsymbol{\sigma})}$ is the partition function and $C$ is the set of all possible configurations. The expectation value of the absolute magnetization is then found by the expression
\begin{equation}
\langle | \overline{\sigma} | \rangle = \sum_{\boldsymbol{\sigma} \in C } | \overline{\sigma} | p(\boldsymbol{\sigma}),
\end{equation}
which we seek to bound using Peierls' argument.

To determine $\langle \left| \overline{ \sigma } \right| \rangle$, we begin by finding an approximation to the simpler value $\langle N_{-} \rangle$; the thermal average of the number of spins in the $-1$ state, for configurations where spins in the $+1$ state are dominant, i.e. states where $\overline{\sigma} > 0$. We refer to regions of $\sigma_j = -1$ as lying `inside' the boundary. In order to evaluate $\langle N_- \rangle$, we first find an upper bound for the number of spins found inside a given contour. Consider a contour of length $l$. The largest number of flipped spins within such a contour is $l^2/16$. This is because the largest number of flipped spins for a contour of fixed length $l$ occurs when the droplet is square with sides of length $l/4$. For the more general case on the lattice with periodic boundary conditions, one can find droplets that span the lattice with a disjoint boundary of two parts with $l \sim 2L$. In this case, we find an upper bound of $l^2/8$ flipped spins before the number of spins in state $+1$ become dominant. We thus obtain an upper bound for the number of spins in the $-1$ state for configurations with positive magnetization 
\begin{equation}
N_-(\boldsymbol{\sigma}) \le  \sum_{b}    {l_b^2\over 8} \delta_{b}(\boldsymbol{\sigma}), \label{NMinus}
\end{equation}
where $\delta_b(\boldsymbol{\sigma}) = 1$ if $\boldsymbol{\sigma}$ contains contour $b$, and 0 otherwise. The term $l_b$ denotes the linear size of contour $b$. 

For Eqn.~(\ref{NMinus}) to be meaningful, we must bound the number of contours that have length $l$. Of course, a boundary must be a closed loop. However, we can find an upper bound for the number of closed loops by calculating the number of random walks of length $l$ that can occur on the lattice, where a walk can begin from any initial position. A walk can begin from one of $V$ possible points. The first step moves in one of four possible directions, and subsequently, to avoid moving backwards, we choose from one of three possible directions. Under these conditions we find $4\cdot 3^{l-1}V$ possible paths. This method will count each closed loop $l$ times, as a given contour can begin from any of the $l$ faces that the contour crosses. We therefore arrive at an upper bound for the number of configurations $4\cdot 3^{l-1}V / l$. We thus have
\begin{equation}
\langle N_- \rangle \le  {V\over 6} \sum_{l = 4, 6, \dots}    l 3^l e^{-\beta l},
\end{equation}
where we have also used that the thermal average for configurations containing contour $b$ of length $l_b = l$ is suppressed by a Boltzmann factor $\langle \delta_b \rangle \le e^{- \beta l_b}$. We next take the infinite volume limit to obtain
\begin{equation}
\langle N_- \rangle \lesssim 27Ve^{-4\beta} {2 - 9e^{-2\beta} \over (1- 9e^{-2\beta} )^2}, \label{Vanish}
\end{equation}
for $e^{-2\beta} < 1$. By symmetry we find an equivalent value for $\langle N_+ \rangle = \langle N_- \rangle $ over configurations where $\overline{\sigma} < 0$. 

We return to the initial problem of obtaining $\langle |\overline{\sigma}| \rangle$. We divide the set of all configurations $C$ into two subsets; $C_+$ and $C_-$, where $C_\pm$ contains configurations with a greater number of $\pm 1$ spins. Configurations with $\overline{\sigma} = 0 $ will not contribute to magnetization, and we therefore neglect them. We then have that 
\begin{equation}
\langle |\overline{\sigma}| \rangle = \sum_{\boldsymbol{\sigma}\in C_+} \overline{\sigma} p(\boldsymbol{\sigma}) - \sum_{\boldsymbol{\sigma}\in C_-} \overline{\sigma} p(\boldsymbol{\sigma}) .
\end{equation} 
To complete the argument we use the fact that, by definition, configurations in $C_+$ have at least  $V/2$ spins in the $+1$ state. We can use that $\sum_{\boldsymbol{\sigma}\in C_+} \overline{\sigma} p(\boldsymbol{\sigma}) \ge 1/2 - \langle N_- \rangle / V $. Similarly, we use the relationship $ \sum_{\boldsymbol{\sigma}\in C_-} \overline{\sigma} p(\boldsymbol{\sigma}) = \langle N_+ \rangle/V =
 \langle N_- \rangle/V $ to arrive at 
\begin{equation}
\langle |\overline{\sigma}| \rangle \ge 1/2 - 2 {\langle N_- \rangle\over V}.
\label{Stability}
\end{equation}
We see from Eqn.~(\ref{Vanish}) that Eqn.~(\ref{Stability}) has solutions larger than zero for finite values of $\beta$, independent of system size, thus demonstrating an ordered phase where the magnetization of the Ising model remains stable in the infinite volume limit of the lattice.

\subsection{High-Dimensional Stable Quantum Models}
\label{Sctn:4Dtoric}

In the previous Subsection we studied suitable conditions for finite-temperature stability by considering the equilibrium state of the two-dimensional classical Ising model. This model is in stark contrast with its one-dimensional counterpart~\cite{Ising}, which does not have a finite-temperature phase transition. Instead, it has thermal dynamics akin to those of the two-dimensional toric code model. Indeed, it is a well understood principle of statistical mechanics that the stability of a model will increase with dimensionality. 

Following this reasoning, Dennis {\it et al.}~\cite{DennisKitaevLandahlPreskill} showed, using a Peierls' argument, that the generalized toric code in four dimensions has a finite critical temperature, below which the model is thermally stable. The four-dimensional toric code is defined on a hypercubic lattice. Qubits are placed on the faces of the lattice, $f$. The interactions of the model are six-body operators associated to the links $l$ and the cubes $c$ of the lattice. Link operators, $A_l$, are the tensor product of Pauli-X operators on the faces $f$ which include link $l$ in the boundary of each face $\partial f$. Similarly, cube operators, $B_c$, associated to cube $c$, are the tensor product of Pauli-Z operators on the face qubits that lie on the boundary of the respective cube $\partial c$. We define the 
four-dimensional toric code Hamiltonian as
\begin{equation}
H_{\text{4D toric}} = -\sum_l A_l - \sum _c B_c, \label{Ham:4DToric}
\end{equation}
where the link and cube operators are given by
\begin{equation}
A_l = \prod_{\partial f \ni l} X_f,\quad B_c = \prod_{f \in \partial c} Z_f, \label{Eqn:4DToricInteractions}
\end{equation}
respectively. The logical operators of the model are supported in two-dimensional planes which, in four dimensions, intersect at a single point. Each qubit in the model supports four $A_l$ operators, and four $B_c$ operators, such that excitations of the four-dimensional model are not point like particles, but instead are line-like particles created by two-dimensional membrane-shaped operators.  These geometric features reproduce the energetics of the two-dimensional Ising model that we described earlier in this Section. An environment must therefore overcome an $\mathcal{O}(L)$ energy barrier to decohere information encoded in the ground space of the model. The Authors of Ref.~\cite{DennisKitaevLandahlPreskill} used these features of the model and follow a Peierls' argument to show that there is a finite temperature, below which, the model lies in an ordered phase.

As an aside, we remark that the argument of~\cite{DennisKitaevLandahlPreskill} was constructed to show the discussed stable model could be decoded using a local algorithm, i.e. an algorithm that does not require the long-range propagation of classical information. Typically when we consider active error correction, such as the decoder described in App.~\ref{Sctn:Decoders}, we reasonably assume that the classical computation can propagate messages at an infinite velocity when compared with the frequency at which the underlying quantum hardware operates. We are therefore able to design effective decoding algorithms that use syndrome information obtained instantaneously from the entire quantum error-correcting code. Realistically, classical information is communicated at a finite rate bounded by the speed of light. The study of thermally stable quantum memories are therefore interesting from the point of view of decoding algorithms, where benefits might be gleaned from considering quantum error-correcting codes that are analogous to thermally stable memories. The local decoding scheme proposed by Dennis~{\it et~al.} has been studied numerically in Ref.~\cite{Pastawski11}. This direction of study has been extended in Ref.~\cite{Harrington, Herrold}, where a local decoder for the two-dimensional toric code is designed and numerically analyzed. Results in this direction may have important applications from the point of view of local thermally stable quantum memories in low dimensions.

Finally, we remark on extensions to the study of high-dimensional quantum systems. Consideration of Peierls' argument suggests that a phase transition occurs at the temperature where the Peierls contours percolate over the system with high probability. The recent work of Hastings {\it et al.}~\cite{HastingsWatsonMelko} shows, using mean-field arguments and supporting numerical evidence, that as dimensionality increases, the critical temperature of the transition diverges from the temperature at which Peierls contours percolate. This is well understood in the classical case of the $D$-dimensional Ising model~\cite{LebowitzMazel}. The study of high-dimensional quantum memories generalizes known classical results and offers new insights into the physics of phase transitions and critical phenomena.

\subsection{The Dynamics of the Four-Dimensional Toric Code}

Discovering ordered phases as we have discussed so far in the Section only provides a statement about the static equilibrium state of a system. To interrogate the memory time of a quantum memory, one must consider the dynamics of the memory under some realistic evolution.

In Ref.~\cite{AlickiHorodeckiHorodeckiHorodecki} Alicki~{\it et al.} rigorously proved that the memory time of the four-dimensional toric code grows exponentially with the size of the system when weakly coupled to a Markovian heat bath. Their results rely on {\em quantum dynamical semigroups}~\cite{AlickiLendi}. The tools that were developed to derive their results were built over a series of papers~\cite{AlickiFannes09} and~\cite{AlickiFannesHorodecki}.
We point out that the results we summarize in this Subsection are generalized and simplified in Refs.~\cite{ChesiLossBravyiTerhal} and~\cite{Bombin13}, where subsystem codes and high-dimensional color codes are respectively considered.

The thermal evolution of a many-body quantum state is very difficult to analyse. To simplify the problem, Alicki {\it et al.}~\cite{AlickiHorodeckiHorodeckiHorodecki} study the evolution of an anti-commuting pair of observables, $\tilde{X}$ and $\tilde{Z}$, that we specify shortly, which act on a two-dimensional subspace of the Hilbert space of the physical system. Specifically, the authors relate the fidelity of the qubit acted upon by the observables $\tilde{X}$ and $\tilde{Z}$ to their {\em decay rates}, $\lambda$, which is defined with respect to some Liouvillian, $\mathcal{L}$ as is given in Eqn.~(\ref{Eqn:GibbsState}). The decay rate of observable $O$ is defined
\begin{equation}
\lambda(O)  =  - \text{tr}(\rho_\beta O^\dagger \mathcal{L}(O)), 
\end{equation}
for observables satisfying $\text{tr}(\rho_\beta O) = 0$ and $\text{tr}(\rho_\beta O^\dagger O) = 1$, where $\rho_\beta$ is the Gibbs state of system, see Eqn.~(\ref{Eqn:GibbsState}). The interaction terms of the Liouvillian considered in Ref.~\cite{AlickiHorodeckiHorodeckiHorodecki} are single-qubit Pauli operators. 

The work of Alicki~{\it et al.} then shows that the fidelity, $F$, of an encoded state $\rho(t)$ decays over time like
\begin{equation}
F(\rho(t)) \equiv \langle \psi | \rho(t) | \psi \rangle \ge \frac{1}{2}\left(e^{-\lambda(\tilde{X}) t  } + e^{-\lambda(\tilde{Z}) t  } \right), \label{Eqn:Fidelity}
\end{equation} 
with respect to the initial pure state $\rho(0) = |\psi \rangle \langle \psi |$.

The above discussion reduces the problem of finding the coherence time of the four-dimensional toric code to finding an upper bound for $\lambda(\tilde{X})$ and $\lambda(\tilde{Z})$. In order to do so, we must first describe the {\em dressed logical operators}, $\tilde{X}$ and $\tilde{Z}$ for the four-dimensional toric code. A dressed logical operator takes the form 
\begin{equation}
\tilde{X} = \overline{X}  C^X, \quad \tilde{Z} = \overline{Z} C^Z,
\end{equation}
where $\overline{X}$ and $\overline{Z}$ are the two-dimensional logical operators of the four-dimensional toric code. The operators $C^X$ and $C^Z$, defined rigorously in Refs.~\cite{AlickiHorodeckiHorodeckiHorodecki, ChesiLossBravyiTerhal, Bombin13}, effectively perform the role of error correction, as we have discussed in Subsec.~\ref{Sctn:ErrorCorrection}. Specifically, operator $C^X$ first projects the encoded state onto an eigenstate of the $B_c$ operators, defined in Eqn.~(\ref{Eqn:4DToricInteractions}), and subsequently applies a {\em low-weight} Pauli correction operator that returns the system to the $+1$ eigenspace of the $B_c$ operators. Similarly, $C^Z$ projects the system onto an eigenstate of the $A_l$ operators, and subsequently applies a low-weight correction operator that returns the system onto the $+1$ eigenspace of the $A_l$ operators. The low-weight correction operators of the $C^X$ and $C^Z$ operators are obtained efficiently from a configuration of eigenvalues of $A_l$ or $B_c$ operators using the clustering decoder described in App.~\ref{Sctn:Decoders}.

Having defined the dressed observables of the four-dimensional toric code, it remains only to upper bound the decay rates, $\lambda(\tilde{X})$ and $\lambda(\tilde{Z})$ to show that the fidelity of an encoded qubit decays slowly if the system is large. Due to the symmetry between $\tilde{X} $ and $\tilde{Z}$, we restrict our attention to only the $\tilde{X}$ operator. An equivalent discussion holds for the $\tilde{Z}$ operator. 

To bound $\lambda(\tilde{X})$, an extension of Peierls' argument, discussed in Subsec.~\ref{Sctn:TheIsingModel}, is employed. First, it is shown that the equilibrium state of the four-dimensional toric code at a suitably low temperature is dominantly populated by configurations of small loop excitations that are created by low-weight configurations of errors. Such an equilibrium state can be successfully corrected by the $C^X$ operator with arbitrarily high probability. Indeed, it is easily checked using Peierls' argument that the probability of observing loop-excitations that are larger than a specified length that is a fraction of the linear size of the system is exponentially suppressed~\cite{DennisKitaevLandahlPreskill}. 

Finally, to understand the dynamics of the equilibrium state of the four-dimensional toric code one must consider the action of the Liouvillian on the dressed observable. Once again, an extension of Peierls' argument is used to show that the probability that local errors will introduce large loop-like excitations to the Gibbs state is exponentially suppressed in the size of the system. It follows that the decay rate of both $\tilde{X}$ and $\tilde{Z}$ operators are exponentially suppressed, thus providing the desired result by application of the decay rates to Eqn.~(\ref{Eqn:Fidelity}).

The above argument demonstrates that the four-dimensional toric-code Hamiltonian given in Eqn.~(\ref{Ham:4DToric}) is self correcting at a sufficiently low temperature. Moreover, the four-dimensional toric code satisfies the conditions required to demonstrate perturbative stability at zero temperature by the proof given in Ref.~\cite{BravyiHastingsMichalakis10}. It will be interesting to show that the four-dimensional toric code, or indeed a quantum system of any dimensionality, is self correcting at finite temperature, even in the presence of weak local perturbations. In Ref.~\cite{Hastings11}, the author proposes a definition for topological order at finite temperature, and shows that it is satisfied by the four-dimensional toric code. One approach to demonstrating that the self-correcting properties of a model are preserved under local perturbations might be to determine if the definition proposed by Hastings implies that a system is perturbatively stable.

\subsection{Thermally Stable Quantum Computation}
\label{Sctn:6D}

We conclude this Section with a discussion on the more general problem. What is the smallest dimensionality where we obtain both thermal stability and the desirable feature of a gate set that can be executed fault-tolerantly to realize universal quantum computation? This problem has been approached by Bombin~{\it et al.}~\cite{Bombin13}. The authors consider $D$-dimensional generalizations of the color code models~\cite{BombinMartin-Delagado07}. These models are of particular interest due to the extended set of gates they can achieve on their ground states {\em transversally}.

A logical gate on the ground space of the code is executed transversally when one can make a logical rotation on the code space of a code by applying {\em local} rotations to its physical degrees of freedom. This is a favorable approach to performing gates as local operations on individual degrees of freedom do not propagate errors during their application. In general, the available transversal gates of a given model are limited by its microscopic details. Notably, the two-dimensional color code~\cite{Bombin06} can perform the Clifford gate-set transversally. Together with the noisy preparation of magic states~\cite{BravyiKitaev05}, the Clifford gate set achieves universal quantum computation. It has also been discovered that a three-dimensional color code can achieve fault-tolerant universal quantum computation~\cite{Bombin07}. Transversally, this three-dimensional model can achieve a $\pi/8$-gate, and a controlled-not gate. Together with the ability to prepare and measure the ground space in the logical $X$- and the $Z$-basis, this model achieves universal quantum computation. Sadly however, the three-dimensional color code does not support finite temperature stability, as is shown by the no-go theorem due to Yoshida~\cite{Yoshida}.

To achieve a universal gate-set and have stable excitations akin to those of the two-dimensional Ising model within the color code family of models one needs $D=6$~\cite{Bombin13}. Together with preparation and measurement in both the Pauli-X, and Pauli-Z basis, the six-dimensional color code is compatible with the transversal application of the $\pi/8$-gate, and the controlled-not gate, which gives rise to universal fault-tolerant quantum computation.

Six dimensions are by no means a tight lower-bound on the system dimensionality where both of these features coincide. Instead, this result is to be understood as a first estimate on lowest spatial dimensionality that is to be improved upon. The result is obtained for the restricted case of models of commuting two-level physical systems. One may indeed be able to reduce the discovered critical dimension by considering many-body systems composed of higher-dimensional spins or fermionic degrees of freedom~\cite{Bombin13}. Moreover, this result is restricted to models that give rise fault-tolerant quantum computation by transversal gates. Indeed, quantum coding theory has shown that universal transversal operations are known to be incompatible with stabilizer error correction~\cite{Eastin, Zeng11, Anderson14}. The reader may question this remark as to how the six-dimensional color code achieves a universal set of operations given known restrictions on universal transversal gate sets. Indeed, its transversal gate set is not truly universal, but, as pointed out earlier, its universal set of operations are completed by the ability to prepare and measure in both the Pauli-Z and Pauli-X basis. We finally remark that we may find stable low-dimensional systems with universal fault-tolerant operations by considering different types of fault-tolerant operations other than transversal gates.

\section{Interacting Anyon Models}
\label{Sctn:InteractingAnyons}

As we have discussed in Subsec.~\ref{Sctn:AnyonicPicture}, the syndrome of two-dimensional topological stabilizer codes can be interpreted in terms of point-like anyonic quasiparticles. This leads to some favorable properties, such as the simple structure of the stabilizer space and the intuitive means by which the syndrome may be decoded. However, as explained in Sec.~\ref{Sctn:NoGoTheorems}, it is precisely this point-like nature that prevents the realization of self-correcting memories in two dimensions.

One path towards self correction is to consider models with all the advantages of two-dimensional topological codes, but which are nevertheless able to realize self-correcting behavior. Interacting anyon models have been proposed to this end. All consider coupling a toric or planar code~\cite{BravyiKitaev98, DennisKitaevLandahlPreskill}, the toric code Hamiltonian defined with open boundary conditions, to an external system, and then using this to mediate interactions between the anyons. These interactions change the energy landscape of the anyons, and can lead to models with diverging energy barriers. Here we review the different types of anyonic interactions. We go on to review proposals to generate them.

\subsection{Forms of Anyonic Interaction}

To understand interacting anyon models, it is useful to first make a distinction between stabilizers and projective anyon number operators. The toric code stabilizers, introduced in Subsec.~\ref{Sctn:2dTC}, are known as star and plaquette operators, denoted $A_v$ and $B_p$ respectively. Star and plaquette operators have eigenvalues $\pm 1 $. A state where a vertex $v$ or a plaquette $p$ of the toric code lattice holds an anyon lies in the $-1$ eigenspace of its corresponding stabilizer. Vacuum states, where there is no anyon on a plaquette or vertex, lie within the $+1$ eigenspace of all the stabilizers.

By convention the stabilizer space corresponds to the $+1$ eigenspace of all stabilizers. We can use stabilizer operators to define projectors onto the common $-1$ eigenspace of the code, and hence onto anyon states, such that
\be
n_v = {1\over 2} ( \openone - A_v ) , \quad n_p = {1\over 2}(\openone - B_p).
\ee
We call these projectors anyon number operators, as their eigenvalues are the anyonic occupation of vertex $v$ or plaquette $p$.

For local Hamiltonians, the replacements $A_v \rightarrow -n_v$ and $B_p \rightarrow -n_p$ defines an equivalent Hamiltonian, up to a constant shift in energy. However, interacting Hamiltonians that use $n_j$ projectors lead to different physics. To illustrate this point, we consider the example of two different Hamiltonians that describe interactions between a single pair of vertices, which we index 1 and 2. These Hamiltonians are
\be
H_s = -A_1 - A_{2} - {1\over 2}A_1 A_{2}, \quad H_n = -A_1 - A_{2} + n_1 n_2.
\ee
The first two terms ensure that the ground state is that of anyonic vacuum. The remaining term is an interaction.

The $A_1$ and $A_{2}$ terms contribute an energy penalty of $1$ for each anyon present in both cases. The $A_1 A_{2}/2$ term contributes an energy penalty of $1$ when there is a single anyon, but nothing when there are two. The $n_1 n_2$ term contributes nothing for a single anyon, but an energy penalty of $1$ for a pair of anyons.

The different behavior of $H_s$ and $H_n$ lead to different interpretations about what form the interactions take. For $H_n$ the $n_1 n_2$ only contributes when multiple anyons are present. We therefore call it an anyon-anyon interaction. The $A_1 A_{2}/2$ term of Hamiltonian $H_s$ contributes only when there is both an anyon on one vertex and vacuum on the other, so we call it an anyon-vacuum interaction. In what follows we consider both of these interaction types extended over the entire lattice.

\subsection{Interacting Anyon Hamiltonians}

The first proposals for interacting anyon models considered anyon-anyon interactions of the form
\be
H_{\text{AA}} = \Delta H_{\text{toric}} + V \sum_{k} \sum_{k' \neq k} n_k n_{k'} \, U(r_{kk'}) ,
\ee
with $H_{\text{toric}}$ defined in Eqn.~(\ref{Ham:Toric}), values $\Delta$ and $V$ are arbitrary coupling constants, indices $k$ denote all stabilizers, both vertices and plaquettes, and $r_{kk'}$ is the Euclidean distance between $k$ and $k'$. The first term here is the standard stabilizer Hamiltonian and the second is the anyon-anyon interaction.

\subsubsection{Logarithmic Potential}

In Ref.~\cite{HammaCastelnovoChamon} the following attractive potential was proposed
\be
U(r_{kk'}) = \ln r_{kk'}.
\label{eqn:loga}
\ee
Note that this diverges with distance, leading to a diverging energy barrier of $\mathcal{O}(\ln L)$. This potential has a confining effect for temperatures of $T<V/2$. In this regime, all anyons present in the code will typically be within an $\mathcal{O}(1)$ distance of each other. A logical error is caused when a single anyon breaks free and winds around the torus. This corresponds to a random walk in an $\mathcal{O}(\ln r_{kk'})$ potential formed by the other anyons, from which the typical coherence time $\tau = \mathcal{O}(L^{V\beta})$ can be found. This diverges polynomially with $L$, and has an exponent that increases as temperature decreases.

For higher temperatures the model continues to have a diverging coherence time. This is due to a different mechanism, described in more detail below. The scaling in this case is $\tau = \mathcal{O}(L^{2})$ \cite{Wootton13}. Note that, though this remains polynomial, the exponent no longer depends on temperature.

This analysis applies only to the toric code. For the planar code~\cite{BravyiKitaev98, DennisKitaevLandahlPreskill} a single anyon can be created at a boundary. Since this has no other anyons to confine it, it does not experience a $\tau = \mathcal{O}(L^{V\beta})$ coherence time for any temperature. Instead the coherence time scales as $\tau = \mathcal{O}(L^{2})$ for all temperatures.

An interaction of this form can be mediated by coupling the anyons to a two-dimensional lattice of hopping bosons. Only local couplings are needed to produce interaction (\ref{eqn:loga}). However, the coupling strength must diverge with system size to realize the potential. This violates a requirement for a self-correcting memory; that of bounded interactions. Nevertheless, note that these diverging couplings are not directly responsible for the increasing lifetime, since the physical energy scales remain finite. Moreover, the model needs to be fine tuned. Small perturbations in the hopping couplings of the bosons can cause the long-range interactions to become short-range and thus stop the effectiveness of the model.

\subsubsection{Power-Law Potential}

In Ref.~\cite{ChesiRothlisbergerLoss} the interactions have the following repulsive potential
\be
U(r_{kk'}) = r_{kk'}^{-\alpha}, \,\, \alpha \geq 0.
\ee
At first glance this would appear to be ineffective. The potential does not diverge with distance and in fact it decays in general such that the energy barrier for the creation and separation of a pair is finite. Furthermore, the potential is repulsive, making anyons less likely to annihilate once created than in the non-interacting case. Nevertheless, when the interactions are sufficiently long range, $\alpha<2$, they have a strong beneficial effect.

The simplest case to consider is that of $\alpha=0$, where the potential does not decay over distance but remains constant. The energy of the system does not depend on the positions of anyons in this case, only their number $N$. Every anyon repels every other one with an energy $V$. The energy above the ground state is then
\be
\varepsilon_N = N \Delta + \frac{V}{2}N(N-1).
\ee
Note that this grows quadratically with $N$, rather than simply linearly as in the non-interacting case. A similar superlinear scaling of energy with anyon number occurs for other $\alpha<2$ as well as the logarithmic potential above.

By using simple arguments, it is possible to find a lower bound on the coherence time for the $\alpha=0$ case. Since we can realistically only expect a finite energy density, i.e. states with $\varepsilon_N = \mathcal{O}(L^2)$, the system is limited to states with $N = \mathcal{O}(L)$ anyons at most. This would correspond to an infinitely sparse anyon configuration, with a distance of $\mathcal{O}(\sqrt{L})$ between each pair of anyons created by the thermal noise. The time required to cause a logical error is then no less than that required for random walks over this length scale, and so $\tau = \mathcal{O}(L)$.

Using a more careful treatment it can be shown that the number of anyons is suppressed further than the above argument implies. This leads to a longer coherence time which, for general $\alpha<2$, scales like $\tau = \mathcal{O}(L^{2-\alpha})$. This is a polynomial scaling that is quadratic in the best case. This effect is also responsible for the $\tau = \mathcal{O}(L^{2})$ scaling of coherence time for the logarithmic potential we met above.

The power law potential can be mediated by interaction with cavity modes. This is a non-local interaction, but is nevertheless reasonable up to a cut-off system size. The self-correcting behavior is therefore not truly scalable, as we require. However, it may still be possible to achieve system sizes that are useful in practice.

\subsection{Anyon-Vacuum Interactions}
\label{Sctn:AVinteractions}

More recently, interacting anyon model proposals have focussed on engineering couplings between anyons and the vacuum rather than interactions between anyons. It is found that this approach allows significantly more powerful self-correction. The first studies of such models are found in Refs.~\cite{PedrocchiChesiLoss,HutterWoottonRothlisbergerLoss}.


These proposals consider Hamiltonians of the form
\be
H_{\text{AV}} = \Delta H_{\text{toric}} - V \sum_{k} \sum_{k' \neq k} S_k S_{k'} \, r_{kk'}^{-\alpha}.
\ee
Here $S_k$ are the stabilizer operators, either $A_v$ or $B_p$ depending on whether each $k$ is a vertex or a plaquette. The only difference between this Hamiltonian and that of the power law potential above is the substitution of anyon number projectors $n_k$ with stabilizer operators $S_k$.

To compare $H_{\text{AV}}$ with that for the anyon-anyon case, $H_{\text{AA}}$, we can rewrite it in terms of anyon number operators. This yields
\be
H_{\text{AV}} = \mu(L) \sum_k n_k - 4V \sum_{k} \sum_{k' \neq k} n_k n_{k'} \, r_{kk'}^{-\alpha} \,\, + {\rm const.} \label{Ham:AnyonVacuumInt}
\ee
The first term here is the effective anyon gap, the energy that an anyon must overcome in order to be created
\be
\mu_k(L) = \Delta + 4 \sum_{k'} (1-\delta_{k,k'}) r_{kk'}^{-\alpha}.
\ee
The large energy penalty for creation is due to the anyons being repelled by the majority of plaquettes and vertices, which are in the vacuum state. The second term in Hamiltonian~(\ref{Ham:AnyonVacuumInt}) is an attractive and non-divergent anyon-anyon attraction.

The effects of thermal errors are suppressed first by the anyon gap, which significantly slows the creation of anyons. Once anyons have been created, errors are further suppressed by the attractive potential that favors reannihilation. Let us focus on the effects of the anyon gap.

To suffer a logical error, an error must first occur on a single qubit. This will create at least $n$ anyons, where $n=1$ for the planar code, for the case where a qubit is created on the boundary, and $n=2$ for the toric code. These $n$ anyons will feel a repulsion from all the $\mathcal{O}(L^2)$ plaquettes and vertices on which there is vacuum. This results in the energy gap $\mu(L) = \mathcal{O}(L^{2-\alpha})$ for each anyon when $\alpha < 2$. The typical time before the first error occurs on is then $e^{\beta n \mu(L)}$.

It is this timescale that dominates the lifetime of the code. Other factors must be taken into account to fully deduce the lifetime: the number of qubits on which it is possible for an error to occur; the probability that a pair, once created, will cause a logical error; and the time required for an anyon to diffuse across the code. However, these will contribute factors that are polynomial in $L$ at most. They are insignificant in comparison to the exponential timescale to overcome the anyon gap. As such, we can simply say that $\tau = e^{\beta n \mu(L)} = e^{\beta \mathcal{O}(L^{2-\alpha})}$. Note that the higher $n$ for the toric code will result in asymptotically longer lifetimes than the planar code, assuming all else is equal.

The main difference between proposals with this interaction is the physical system used to mediate it and the value of $\alpha$ that is achieved. Refs.~\cite{PedrocchiChesiLoss,HutterWoottonRothlisbergerLoss} achieve the optimal case of $\alpha=0$ by coupling to cavity modes. However, as above, this prevents the model from achieving the scalability that we require. A Hamiltonian simulation of this case was considered in \cite{BeckerTanamotoHutterPedrocchiLoss}.

\begin{figure}
\includegraphics[width = \columnwidth]{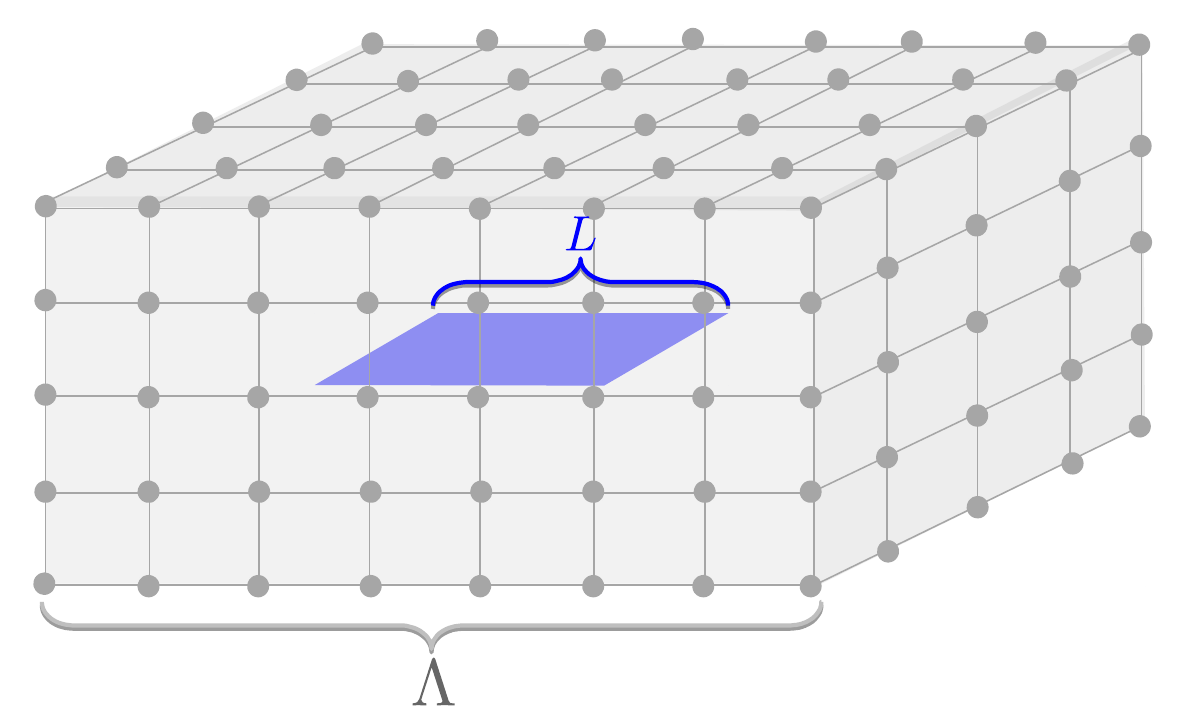}
\caption{(Color online) Schematic Figure of the planar code coupled to a bosonic bath. Anyon-Vacuum interactions in an $L \times L$ code can be induced by embedding it in a $\Lambda^3$ lattice of hopping bosons for $\Lambda> L$. The stabilizers of the code locally couple to the bosonic lattice. The interactions are mediated by the low-energy collective excitations of the bosons.
\label{BosonLattice}}
\end{figure}

In Ref.~\cite{PedrocchiHutterWoottonLoss} the interaction is mediated by coupling to a three-dimensional lattice of hopping bosons, as shown in Fig. \ref{BosonLattice}. The resulting interacting model corresponds to that of $\alpha=1$. This gives rise to an energy barrier of $\mu(L) = \mathcal{O}(L)$ and coherence time $\tau = e^{\beta \mathcal{O}(L)}$. This model uses only local, bounded strength and constant weight interactions, and yet is able to preserve the quantum information for an exponentially long time. 


A related model was also proposed in Ref.~\cite{PedrocchiHutterWoottonLoss12} by the same authors. Here the role of the bosons is played by magnons in a three-dimensional ferromagnet. The code is simply coupled to the spins of the ferromagnet, avoiding the need for unbounded operators. Perturbative gadgets~\cite{Bravyi08, Bravyi11b} are used in order to realize the entire Hamiltonian using only local two-body interactions on a three-dimensional system of spin-half particles.

The effective interacting anyon model in the case of the ferromagnet is the same as for the three-dimensional boson lattice. However, it is only expected to be valid in the regime for which the coupling of the code to a thermal bath is much weaker than the coupling to the ferromagnet. The Monte-Carlo simulation of thermal noise described in Subsec.~\ref{Sctn:NumericalMethods} therefore cannot be used. This approximates the true dynamics by allowing the thermal bath to instantaneously transfer large energies to the system, and so is unable to capture the subtle effects arising from the full time evolution of the system-bath interaction. It is in these effects that we would expect to observe self-correcting behavior. Since a study of this time evolution seems intractable, the full extent of the self-correcting behavior in this model is not known. This model can also be adapted to protect topological systems composed of superconducting qubits for which errors correspond to infinitely weak coupling to an infinite temperature bath \cite{Kapit14}.


\subsection{Open Questions in Interacting Anyon Models}

As we have discussed, interacting anyon models are capable of impressive memory times which grow polynomially or even exponentially with system size. However, due to the difficulties in solving the proposed interacting models, further study is required to better understand potential challenges we may need to overcome to realize effective long-range interactions. In this Subsection we outline some of the important directions of study that we must follow to understand the feasibility of realizing a self-correcting quantum memory by means of interacting anyons.

One such direction concerns the effects of local perturbations on interacting anyon models. The non-local terms of an interacting anyon Hamiltonian can cause local pertubations to affect the system non locally. Any perturbation that gaps the bosons, for example, will cause the anyonic interactions to become short range. This will induce a finite cutoff length scale, beyond which the lifetime no longer increases with system size. This was discussed for the ferromagnet based model in \cite{PedrocchiHutterWoottonLoss12}, and elaborated upon for the bosonic model in Ref.~\cite{Landon-Cardinal15}.

Such problems with perturbations are not unique to proposals for quantum memories. The same is true for the ferromagnetic systems currently used as classical memories. Perturbations in these systems limit the magnetic susceptibility, inducing an equivalent finite cutoff. However, this has not constrained the development of classical computation. It is therefore important to determine how large the cutoff will be for quantum memories when realistic perturbations are considered. Whether or not it is large enough to allow for large-scale quantum computation is a crucial test of these proposals, and remains an open area of study. Additionally, possible avenues towards the discovery of interacting anyon models that are stable against perturbations has also been discussed \cite{Landon-Cardinal15}. Concrete examples, however, remain to be found.

Another caveat of the interacting anyon proposals concerns the thermal bath used for their analysis. In all of the cases discussed in this Review, it has been assumed that the thermal bath will act locally on the anyonic system. However, it may be that thermalization will occur in a more complex manner, with the mediating system allowing thermal errors to become long-range. Analysis of the self-correcting properties with realistic thermalization dynamics therefore remains an open area of study.

\subsection{Finite-Temperature Topological Order}
\label{Subsec:FiniteTempTO}

It is important to determine if a system can maintain topological order at non-zero temperature. We might expect such a system to be stable against local perturbations which is a property that we require of a self-correcting quantum memory.  For many systems, topological order vanishes at non-zero temperature~\cite{NussinovOrtiz07, Nussinov, Nussinov09, ChesiLossBravyiTerhal}. Indeed, it has been shown by Hastings that local commuting Hamiltonians in two dimensions are incompatible with a condition for topological order at finite temperature~\cite{Hastings11}. Some models however demonstrate a finite-temperature phase transition between a phase with topological order and a disordered phase.

The relationship between topological order and self correction is not well understood in the general case~\cite{Yoshida}. So far,  we have only considered specific models to learn the correspondence between topological order and self correction, see for instance Ref.~\cite{Mazac12} where topological entanglement entropy and the coherence times of the toric code are studied on lattices of varying dimensionality. For the studied cases there is a strong coincidence between these two features. The interacting anyon models are also shown in Ref.~\cite{Wootton13} to provide an interesting perspective on this problem which we discuss in this Section.

The signature of topological correlations in systems such as the toric code are loop correlations. For finite temperature systems these can be found using the topological entropy \cite{Hamma05, KitaevPreskill06, LevinWen06}, topological mutual information~\cite{Iblisdir09} or anyonic topological entropy \cite{Wootton13}. In all cases one must consider a region of the system whose size is on the order of the system size which can be arbitrarily large. If topological correlations are detected for large regions, the state is said to be topologically ordered.

For power law anyon-anyon interactions, any finite temperature thermal state contains a diverging number of anyons \cite{ChesiRothlisbergerLoss}. These anyons will also be deconfined due to the repsulsive nature of the interactions. Such a diverging number of delocalized anyons ensures that topological order according to the above definition is not present for any finite temperature. However, the interactions are still known to support self-correction with a polynomial lifetime. This fact, along with the $\mathcal{O}(1)$ energy barrier, makes these models an interesting exception to widely held opinions about what is required for self-correction.

For the case of the logarithmic anyon-anyon interaction, it is found that the topological order persists up to a finite temperature of $T_c=V/2$. It therefore corresponds exactly to the confined anyon phase for which the lifetime is $\tau = \mathcal{O}(L^{V\beta})$. Beyond this temperature the topological order is no longer present, though the system is still self-correcting with a lifetime of $\tau = \mathcal{O}(L^{2})$. The phase transition does not have the effect of destroying the self correction, as we might expect, but it does alter the scaling of the lifetime.

To regain some semblance of our intuition that topological order is required at finite temperature for self correction, we can redefine what we mean by topological order. Instead of simply considering whether topological correlations can be detected for regions of an arbitrarily large size, we can determine the exact range of these correlations. This means considering regions of different sizes, calculating how the topological correlations decay as the size is increased, and then determining a correlation length to quantify this. This correlation length is denoted $\lambda$.

Even when a system is not topologically ordered, it is possible for topological correlations to be present on an $\mathcal{O}(1)$ length scale. The case of $\lambda = \mathcal{O}(1)$ therefore corresponds to topologically trivial states. Standard topological order requires that the correlations do not decay at all, even up to the linear system size, $L$. This corresponds to a super-extensive $\lambda > \mathcal{O}(L)$. In between these extremes there exists the possibility for the range $\lambda$ to increase with system size, but just not as quickly as is required for standard topological order. We refer to such states as {\em weakly topologically ordered}.

Studying the interacting anyon models from this perspective, it has been shown that all models at all temperatures are either in a standard or weakly topologically ordered phase~\cite{Wootton13}. All transitions between these two types of topological order correspond to a change in the way that the lifetime scales with system size. All known self-correcting memories correspond to phases that are either topologically ordered or weakly topologically ordered. As such, some relationship between finite-temperature topological order and self correction does persist. However, it would be interesting to determine whether counter examples exist even for the case of this weaker relationship.

\section{Commuting Three-Dimensional Models}
\label{Sctn:ThreeDimensions}
With limitations challenging the construction of finite-temperature quantum memories with commuting two-dimensional Hamiltonians, it is exciting to consider three-dimensional models. Indeed, as discussed in Sec.~\ref{Sctn:NoGoTheorems} there are still no-go results towards the feasibility of a finite-temperature quantum memory in three dimensions. However, these results are not as restrictive as their two-dimensional counterparts. Recently proposed models have shown positive progress, which, together with supporting numerical data offer promise for the discovery of good quantum memories at finite temperature. In this Section we provide an overview of the positive results found in three-dimensional models. We first review the concept of partial self correction; a new paradigm for macroscopic coherence time scaling that has emerged from the study of three-dimensional models. In Subsection~\ref{Sctn:CubicCode} we study the cubic code model, a quantum system demonstrating partial self correction. In Subsections~\ref{Sctn:WeldedToric} and~\ref{Sctn:Embed} we review other three-dimensional proposals that break translational invariance to achieve phenomena potentially important for self correction in quantum systems.

\subsection{Partial Self-Correction}
\label{Sctn:PartialSelfCorrection}

A new phenomenon to develop from the study of three-dimensional systems is that known as {\em partial self correction}. Partially self-correcting models are notable for polynomial coherence-time scaling with system size up to some cut-off size that depends on temperature. Moreover, they exhibit super-exponential inverse-temperature scaling. These features of known partially self-correcting models are attributed to its energy barrier, which grows logarithmically with the size of an error incident to a memory.

Partially self-correcting models have been discovered independently by both Haah~\cite{Haah}, and by Castelnovo and Chamon~\cite{CastelnovoChamon11} following remarkably different methods. In Ref.~\cite{Haah}, the author exhaustively searches over all translationally invariant stabilizer models on a cubic lattice with one or two qubits on each vertex of the lattice to find models that satisfy the `no-strings' condition, as is defined rigorously in~\cite{Haah}. Broadly speaking, the no-strings condition is a property of the excitation structure of a commuting Pauli Hamiltonian whereby a cluster of non-trivial excitations cannot be transported across a lattice over an arbitrary distance without introducing new excitations to the lattice. Models that satisfy the no-strings condition are thus expected to provide good protection against thermal interactions due to the large energy cost that is required to introduce logical errors to the lattice via local operations. Indeed, it was later proved that models satisfying the no-strings condition must have an energy barrier that scales at least logarithmically with the size of the system~\cite{BravyiHaah11}. Independent of the work of Haah, Castelnovo and Chamon~\cite{CastelnovoChamon11} looked to find a quantum generalization of known low-dimensional classical models with non-trivial energy barriers between ground states which are well known in the context of glassy systems~\cite{Newman99, GarrahanNewman}. The derived generalization is described locally in three dimensions. Similar models are studied in further generality in Ref.~\cite{Kim12}. 

We now give a heuristic analysis explaining partial self correction where we assume Arrhenius' law, Eqn.~(\ref{Eqn:Arrhenius}). Known partial self-correcting memories are characterized by an energy barrier that grows logarithmically with the size of an error, $\xi$. The energy $\varepsilon$ that is required for an error to increase in size to occupy a volume of lattice of diameter $\xi$ must thus be at least
\begin{equation}
\varepsilon \sim \kappa \Delta \log \xi,
\end{equation}
where $\kappa$ is a positive constant and $\Delta$ is the gap of the model. Given the typical finite-temperature noise analysis we described in Sec.~\ref{Sctn:FiniteTemperatures}, we assume that errors are created with an average separation that scales with $\beta$ like $\Lambda \sim e^{\beta \Delta / D}$, where $D$ is the dimensionality of the system. Given then that we require $\xi  \sim \Lambda$ for the memory to decohere, we arrive at the typical excitation energy of the model as a function of $\beta$. Namely, we obtain $\varepsilon \sim \kappa \Delta^2 \beta / D$ at the point of decoherence. Applying this to Arrhenius' law, Eqn.~(\ref{Eqn:Arrhenius}), we obtain
\begin{equation}
\tau \sim e^{\kappa \Delta^2 \beta^2/D}. \label{PSCBeta}
\end{equation} 

We can follow a similar analysis to study small system sizes such that $L \lesssim \Lambda$. In this case a diffusing error must attain energy  $\varepsilon \sim \kappa \Delta \log L$. Once again, applying this expression to Arrhenius' law it follows that partially self-correcting quantum memories have a coherence time that grows polynomially in system size
\begin{equation}
\tau \sim L^{\kappa \Delta \beta}, \label{PSCSystemSize}
\end{equation}
whose exponent is linear in $\beta$. This scaling is effective up to some cutoff $L_{\text{opt.}}\sim \Lambda$. We thus obtain a cutoff, $L_{\text{opt.}} \sim e^{c\beta}$, that grows exponentially in $\beta$ for positive constant, $c$~\cite{BravyiHaah13}.

An interesting feature of the known partially self-correcting models is that they do not have a constant ground-state degeneracy, which is outside the assumptions of the no-go theorem of Yoshida~\cite{Yoshida}. As an aside, it is interesting to consider that the known partially self-correcting models are not {\em purely} topological models. While the ground space of these systems are topologically ordered in the sense that the degenerate ground states of the model cannot be locally distinguished, their ground-state degeneracy still depends on the microscopic physics of the model. A similar model whose ground-state degeneracy depends on microscopic details~\cite{BravyiLeemhuisTerhal} is introduced by Chamon in Ref.~\cite{Chamon}. We remark however that unlike the cubic code, the model due to Chamon has a constant energy barrier, and is well understood not to give rise to self-correcting properties~\cite{Chamon, CastelnovoChamon11, Nussinov12, Temme14}. We finally remark that the study of exotic partially self-correcting systems has led to new classifications of systems under the context of {\em fractal topological quantum field theories}~\cite{Yoshida13, Haah14}.

In the following Section we will review and reproduce previously obtained numerics of the rigorously studied cubic code model. We remark that the fragile glassy model introduced by Castelnovo and Chamon is expected to behave in a phenomenologically equivalent way to the cubic code model~\cite{CastelnovoChamon11}.

\subsection{The Cubic Code}
\label{Sctn:CubicCode}
\begin{figure}
\includegraphics[scale=2]{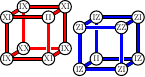}
\caption{(Color online) The stabilizers of the cubic code. The model is defined on a cubic lattice with two qubits on each vertex. Stabilizers, denoted $S^X_j$ and $S^Z_j$, are shown on the left and the right of the Figure, respectively. \label{CubicCodeStabilizers}}
\end{figure}
The cubic code model~\cite{Haah} is defined on a three-dimensional lattice of $L\times L \times L$ vertices, where two qubits lie on each vertex of the lattice. Associated to each of the unit cubes of the lattice, indexed $j$, we have two stabilizers $S^X_j$ and $S^Z_j$ shown in red and blue in Fig.~\ref{CubicCodeStabilizers}, respectively. We then write the Hamiltonian
\begin{equation}
H_{\text{cubic}} = -{1\over 2}\sum_j \left(S^X_j + S^Z_j\right).  \label{Ham:CubicCode}
\end{equation}
We take constant interaction strength $1/2$ such that excitations have unit energy cost. The model has a non-trivial ground state degeneracy that varies with $L$. The ground state degeneracy is studied in detail in Ref.~\cite{Haah13} using the language of commuting free modules. For simplicity, we consider only lattices of size $L < 200$ that do not have factors 2, 15 or 63. All of these system sizes have a four-fold ground state degeneracy~\cite{BravyiHaah11A}. We do not discuss the complex fractal structure of the logical operators of the model here, but we refer the interested reader to Ref.~\cite{Haah13}. We point out however that for lattices of odd $L$, we find two logical operators by taking $\overline{X}_1 = \prod_{k \in \mathcal{L}} X_k$, $\overline{Z}_1 = \prod_{k \in \mathcal{L}} Z_k$, $\overline{X}_2 = \prod_{k \in \mathcal{R}} X_k$ and $\overline{Z}_2 = \prod_{k \in \mathcal{R}} Z_k$, where $\mathcal{L}$ and $\mathcal{R}$ denote the subset of all the left and right qubits of each vertex, respectively. It is easily checked that these operators satisfy a suitable algebra for the logical qubits of the model. These of course are not the minimum-weight logical operators~\cite{BravyiTerhal09}. However, we find these logical operators particularly convenient for numerical simulations.

\subsubsection{Excitations of the Cubic Code}
The excitations of the cubic code have a more complicated structure to those of the two-dimensional models considered in earlier Sections. Indeed, the model was designed such that its excitations are created by operators that satisfy the no-strings rule~\cite{Haah}, and instead have a fractal-like structure. In Fig.~\ref{CubicCodeStabilizers} we can observe a symmetry over the $S^X_j$ and $S^Z_j$ stabilizers, such that both Pauli-X and Pauli-Z type errors act with equal effect on the $S^Z_j$ and $S^X_j$ stabilizers of the model. Analysis of only Pauli-X type errors therefore serves for a sufficient study of the excitations of the cubic code.

\begin{figure}
\includegraphics{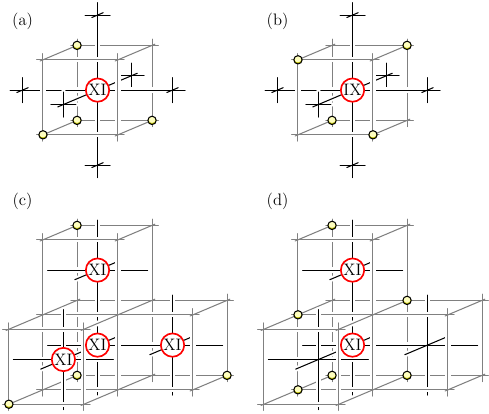}
\caption{(Color online) The excitations of the cubic code. The lattice supporting the cubic code is shown in black where two qubits lie on the vertices of the lattice. The dual lattice that supports excitations is shown in grey. (a) and (b) show the excitations generated by an $XI$ error and an $IX$ error acting on a single two-qubit vertex, respectively. Both errors generate four point-like excitations, marked by yellow points on the vertices of the dual lattice. (c) The error configuration that creates four excitations delocalized over two lattice spacings. (d) An example of a high-energy intermediate error configuration that must be achieved to delocalize four excitations over a long distance as in~(c). \label{CubicCodeExcitations}}
\end{figure}

Pauli operators acting on single qubits of the lattice in the ground state create four localized excitations to the dual lattice of the model, shown in Figs.~\ref{CubicCodeExcitations}(a) and~(b). We mark the stabilizers violated by the error by yellow points on the vertices of the dual lattice, where vertices of the dual lattice correspond to fundamental cubes of the primal lattice. Similar to the excitations of the string-like models we have already discussed, these particles are delocalized. Also, the excitations are their own anti-particles, which are transported by applying additional error operators that annihilate excitations, and create additional excitations at other locations on the lattice. In this way, it is possible to delocalize these excitations over arbitrary distances. We show an error configuration in Fig.~\ref{CubicCodeExcitations}(c) where four fundamental excitations have delocalized over two lattice spacings.

An important distinction between the excitations of the cubic code model and excitations in two-dimensional models is that the delocalization of these particles cannot be achieved using string-like operators. Instead, if we wish to delocalize the excitations of the cubic code model over arbitrary distances, we have to use transport operators that have a fractal-like support. As a finite-temperature noise model will only apply transport operators to the lattice via local single qubit operations, such operators are only achieved by temporarily increasing the energy of the error configuration, which reduces the propagation of excitations. In Fig.~\ref{CubicCodeExcitations}(d), we show an intermediate error configuration necessary to delocalize excitations over two lattice sites. 

The error configuration creates six excitations, which increases the energy of the system. Models where excitations are not propagated by string-like operators satisfy the `no-strings rule'. This is an important concept for partial self correction. An extensive program of analytical study from Bravyi and Haah has proved that models satisfying the no-strings rule necessarily has at least a logarithmic energy barrier~\cite{BravyiHaah11}. Further work in this program of research showed numerically that the cubic code model behaves as a partially self correcting memory with a logarithmic energy barrier~\cite{BravyiHaah13}. Beyond the study of self-correcting memories, the cubic code is also noteworthy from the point of view of localization. It is shown in Ref.~\cite{Kim15} that the glassy nature of the cubic code localizes the excitations of the model. This is particularly interesting since localization is typically attributed to disorder~\cite{Anderson}. Localization in the cubic code however is achieved with a frustration-free Hamiltonian with uniform interactions.

In the remainder of this Subsection we numerically simulate the cubic code at finite temperature to demonstrate its partial-self correcting behavior.

\subsubsection{Numerical Simulations}

\begin{figure}
\includegraphics[width = \columnwidth]{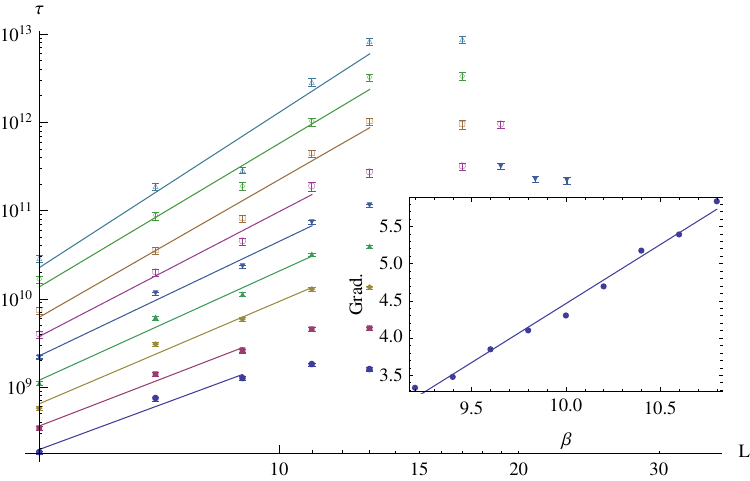}
\caption{(Color online) Numerically evaluated coherence times, $\tau$, for the cubic code shown as a function of system size, $L$. We show separate lines for $\beta = 9.2,\,9.4,\,\dots,\,10.8$, where $\beta = 9.2$ is shown by the lower dark blue line and $\beta = 10.8$ is shown by the light blue line at the top of the Figure. \label{CubicCodeCoherenceTimeWithSystemSize} The Inset shows the gradients of each of the lines in the main Figure which, in this temperature regime, grow like $\tau \sim L^{1.58\beta - 11.38}$.}
\end{figure}
In this Subsection we simulate the cubic code coupled to a finite-temperature environment using the numerical Monte Carlo methods described in Subsec.~\ref{Sctn:NumericalMethods}. We reproduce the phenomenological behavior demonstrated by Bravyi and Haah in Ref.~\cite{BravyiHaah13}. We remark that the results we present differ from those given in~\cite{BravyiHaah13} due to the choice of rate equation used in the simulation. Where we use the rate equation discussed in Subsec.~\ref{Sctn:RateEqn}, Bravyi and Haah use the rate equation of Ref.~\cite{Bortz}. Both rate equations satisfy the detailed balance given in Eqn.~(\ref{Eqn:DetailedBalance}), and ultimately will reproduce the same physics, up to some variation in the obtained constant factors.

We simulate Hamiltonian~(\ref{Ham:CubicCode}) under rate Eqn.~(\ref{secIIpartCeqnRate}). To obtain coherence times, we periodically attempt to decode the state of the evolving lattice using a variant of the clustering decoder described in App.~\ref{Sctn:Decoders}. The first time at which the decoder fails gives the coherence time of the sample. We find the coherence time by averaging over $N$ samples. Errors are determined by taking the standard deviation of the samples, divided by $\sqrt{N}$.

\begin{figure}[b]
\includegraphics[width = \columnwidth]{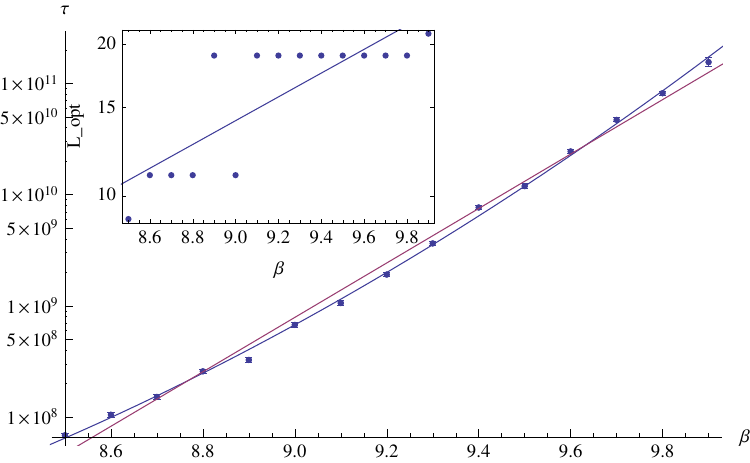}
\caption{(Color online) Coherence time data $\tau$ for the cubic code plotted as a function of inverse temperature, $\beta$. The plotted data is that found using system size $L = L_{\text{opt.}}$ for each value of $\beta$. We also show $L_{\text{opt.}}$ as a function of $\beta$ in the inset. We fit the data shown in the main Figure to the function $\tau \sim e^{1.05 \beta^2 - 13.7\beta + 58.5}$. We compare this fitting to a fitting based on Arhenius' law where we find $\tau \sim e^{5.6\beta - 30.3 }$, also shown in the Figure. The inset shows $L_{\text{opt.}}\sim e^{0.54\beta - 2.2}$.\label{CubicCodeInverseTempScalingPlot}}
\end{figure}

Identifying coherence times for the cubic code using numerical simulations is particularly challenging due to the glassy nature of the model~\cite{Chamon, CastelnovoChamon11}. Specifically, as the system evolves towards the equilibrium state, the simulation frequently finds states that are local minima of the considered Hamiltonian. At the computational level we must simulate many events to escape these metastable configurations, which we find to be numerically intensive. To overcome this, for the small system sizes we study, we find that decoding with very high frequency reduces the number of error events that we must simulate as the state decoheres. We attempt to decode the system at time intervals $ \sim 10^{-10} \cdot e^{4\beta} $. The time units $e^{4\beta}$ are natural as this the frequency at which four excitations are created from the vacuum which then mutually annihilate themselves shortly afterwards with high probability. This behavior is typical in the small system size and low-temperature regime.

To identify partial self correction in the cubic code we plot the coherence times as a function of $L$ in Fig.~\ref{CubicCodeCoherenceTimeWithSystemSize}. Here, we consider many different temperatures to identify polynomial coherence time scaling with system size whose exponent depends on $\beta$. We plot the gradients of the linear fittings, shown in the inset of Fig.~\ref{CubicCodeCoherenceTimeWithSystemSize}. The gradients we obtain show good agreement with polynomial coherence time scaling whose exponent grows linearly with $\beta$, as we expect for partial self correction, derived in Eqn.~(\ref{PSCSystemSize}). We also plot the optimal coherence time as a function of system size for a given $\beta$, as shown in Fig.~\ref{CubicCodeInverseTempScalingPlot}. We identify super-exponential inverse-temperature scaling, as we expect for a partially self-correcting model due to Eqn.~(\ref{PSCBeta}).

We use the fittings we obtain from the presented numerical data to obtain $\Delta$ and $\kappa$ of Eqns.~(\ref{PSCBeta}) and~(\ref{PSCSystemSize}) for the cubic code model
\begin{equation}
\Delta_{\text{CC}} = 2.0, \quad \kappa_{\text{CC}} = 0.79, \label{CubicCodeCharacteristics}
\end{equation} 
thus identifying the partial self-correcting behavior described in the previous Subsection. It is interesting that diverging coherence times at low temperatures are achieved here via a glassy mechanism~\cite{CastelnovoChamon11}, and not with some ordered phase of matter such as those we considered in Subsec.~\ref{Subsctn:CurieWeiss} and Sec.~\ref{Sctn:HighD}. The glassy nature of the model may introduce new difficulties in encoding information to the cubic code since cooling the system to its ground space will be very slow~\cite{Chamon}. Instead, we might consider some manual method of state preparation by measurement or otherwise. Work in this direction has been conducted in Ref.~\cite{Lodyga15}.

It is shown in Ref.~\cite{Haah13} that by imposing translational invariance on three-dimensional commuting Pauli Hamiltonians we cannot expect to find a system where the energy barrier scales better than logarithmically with system size. In the remainder of this Section we consider commuting Pauli-Hamiltonian models that surpass the result of Haah~\cite{Haah13} by breaking translational invariance. The models of interest are the welded three-dimensional toric code~\cite{Michnicki14}, and embeddable fractal product codes~\cite{Brell}.

\subsection{The Welded Toric Code}
\label{Sctn:WeldedToric}

The welded toric code~\cite{Michnicki, Michnicki14}, due to Michnicki, is the first explicit example of a three-dimensional commuting Pauli Hamiltonian with a power-law energy barrier. Remarkably, the model surpasses the no-go result of Haah~\cite{Haah13} by breaking the translational invariance assumption that is required to complete the theorem.

The model is found using an idea called {\em welding}, described in~\cite{Michnicki}. Welding gives a procedure to combine stabilizer codes. The advantage of welding codes is that logical operators are also combined non-trivially over a weld. We follow the exposition of Michnicki showing how he arrived at the welded toric code. 

\begin{figure}
\includegraphics{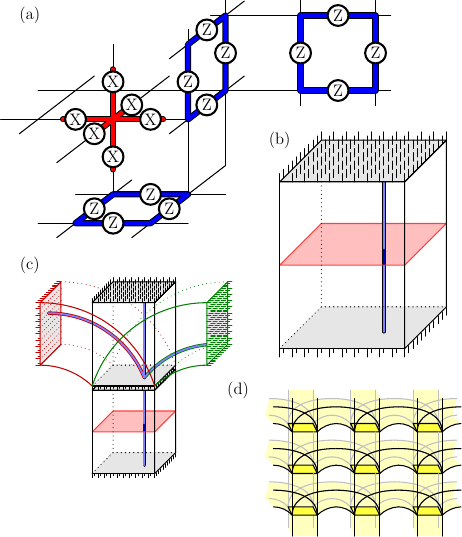}
\caption{\label{Fig:WeldedCode}(Color online) The welded toric code. (a) The stabilizers of the three-dimensional toric code where qubits lie on the edges of the lattice. A vertex operator is shown in red in the top-left corner of the figure. Three examples of face operators are also shown on the lattice, each with a different orientation relative to the cubic lattice. Vertex operators are the tensor product of Pauli-X operators supported on all the edges adjacent to a vertex. Face operators are the tensor product of Pauli-Z operators supported on the edges that bound a face of the lattice. (b) The three-dimensional toric code with two disjoint rough faces. The model supports one encoded qubit described by a membrane logical operator, shown as a red horizontal plane, and a string-like operator, shown by a vertical blue line that runs between the two disjoint rough edges. (c) A single weld between the rough faces of four blocks of three-dimensional toric code. We outline one block in red and one block in green where the blocks are overlapping. The string-like logical operators of the original code blocks must now overcome the a high energy barrier at the interface between the welds. (d) Many blocks of three-dimensional toric code are welded into a lattice. In this welded configuration, the string-like logical operators combine to give a two-dimensional logical operator that spans the lattice.}
\end{figure}

The welded toric code is achieved by welding a macroscopic number of copies of the three-dimensional toric code~\cite{HammaZanardiWen}. The three-dimensional toric code is defined on a cubic lattice with qubits on the lattice edges. The model has two types of stabilizer, vertex operators, and face operators, as shown in Fig.~\ref{Fig:WeldedCode}(a). Vertex operators are six-body operators of Pauli-X operators supported on the edges incident to a vertex. Face operators are four-body Pauli-Z operators supported on edges that bound faces of the cubic lattice. 

With a suitable choice of boundary conditions the three-dimensional toric code will support one logical qubit. The model has two types of boundaries, rough boundaries and smooth boundaries. We show a macroscopic picture of the code in Fig.~\ref{Fig:WeldedCode}(b), where we have a rough boundary on the top and bottom face of the lattice. Later, we will refer to one copy of the three-dimensional toric code that encodes a single logical qubit as a {\em block}.

There are two types of logical operators, membrane logical operators, and string logical operators, shown in red and blue respectively in Fig.~\ref{Fig:WeldedCode}(b). The membrane logical operators have dynamics akin to those of the two-dimensional Ising model, and as such are stable below a critical temperature,~\cite{CastelnovoChamon07, AlickiHorodeckiHorodeckiHorodecki}. However, string-like logical errors introduce point-like excitations whose creation and transport need only overcome a constant energy barrier. It is for this reason that a thermal noise model is able to introduce string-like logical errors in constant time. The three-dimensional toric code therefore does not behave as a self-correcting quantum memory.

Michnicki surpasses the problem of string-like logical operators using welding. He shows that it is possible to weld blocks of three-dimensional toric code along rough faces to generate a large energy barrier. We show an example of a weld in Fig.~\ref{Fig:WeldedCode}(c), where four copies of the three-dimensional toric code are welded along a common rough boundary. The diagram shows that the string-like logical operator divides at the weld. A single point-like excitation therefore cannot pass through a weld from one code block to another, but must instead split into $v-1$ excitations where $v$ is the valency of the weld, i.e. the number of code blocks that meet at the weld. Since these excitations must overcome an energy barrier to propagate across the weld, the diffusion of errors is suppressed.

Michnicki combines many three-dimensional toric code blocks in a lattice-like structure with dimensions greater than one to generate a macroscopic energy barrier over the string-like logical operators of the model. We show such a lattice of two dimensions in Fig.~\ref{Fig:WeldedCode}(d). Now, the string-like logical operators of the composite parts of the code are combined into a coarse lattice whose vertices are welded faces and whose edges are code blocks. Interestingly, the model has interactions resembling high-dimensional Ising-like interactions across the welded boundaries, separated by three-dimensional blocks of toric code lattice. With this lattice configuration, thermal fluctuations must overcome a polynomial energy barrier in the number of welds of the macroscopic welded lattice to introduce a logical error to the model.

The membrane logical operator of a code block is not extensive with the number of welded interfaces of the lattice. In fact, the membrane logical operator can be supported on a single block of the code. To scale the power-law energy barrier correctly, the volume of the blocks must grow with the size of the welded lattice. Michnicki~\cite{Michnicki14} suggests scaling the volume of the block size polynomially with the number of welds in the lattice. It is with this point that we see the model breaks translational invariance; we can vary the size of the model over two different length scales, the block size, and the volume of the lattice of welds.

\subsubsection{Excitations of the Welded Toric Code}
\label{Sctn:WeldedToricLimits}

The welded toric code model shares many similarities to the Ising model, with the code blocks in the former serving a similar function as the nearest neighbor ferromagnetic couplings, or `bonds', in the latter. We now discuss this analogy to find a better understanding of the finite temperature behavior of the welded toric code. 

We consider the simple case in which we do not know the full details of the excitation configuration within each code block. Instead we know only the total parity of the number of point-like excitations that each block contains. Each code block then has two possible states: even or odd parity. These correspond to the two possible states of a bond of the Ising model: aligned and anti-aligned. In general we are able to determine the locations of all the excitations in a code block. However, since the excitations of the code blocks become disordered after a constant time, information regarding their locations becomes irrelevant. For this reason the total parity of each code block will be the only useful syndrome information after thermalization occurs. The error correction procedure of the welded toric code is then equivalent to that of the corresponding Ising model.

The parity of a code block is changed only if a point-like excitation moves through one of its welds, which occurs only if an error occurs on one of the qubits involved in a weld. We refer to a weld that has suffered an odd number of errors as `broken'. Breaking a weld is equivalent to flipping the spin on a vertex of its corresponding Ising model. A region of broken welds (flipped spins) will be surrounded by a surface of odd parity codes (anti-aligned bonds). To perform a logical error, these surfaces must be removed by breaking all welds or flipping all spins, respectively.

Given this correspondence between the welded toric code and the Ising model, we may expect the former to inherit the exponential lifetime and finite temperature phase transition of the latter. However, this is not the case, since the welds of the welded toric code experience temperature differently from the spins of its corresponding Ising model. We observe that due to the $\mathcal{O}(L^2)$ area of each weld, for code blocks of size $L \times L \times L$, the probability that any given weld is broken after the constant thermalization time converges exponentially to $1/2$ as $L \rightarrow \infty$.  It follows from this that the probability of finding any code block in odd parity also quickly converges to $1/2$ as the model approaches the thermodynamic limit. For the Ising model at effective inverse temperature $\tilde{\beta}$, the probability of finding a bond in anti-alignment approaches $1/2$ only as $\tilde{\beta} \rightarrow 0$. Equating these probabilities it is clear that $\tilde{\beta}$ vanishes as $L \rightarrow \infty$. Hence, the welded toric code at finite temperature corresponds to an Ising model with an effective temperature that diverges with system size, and so it does not fare well as a memory in the thermodynamic limit.

We next consider the low-temperature behavior of the welded code. To decay encoded information, the thermal environment must overcome a macroscopic number of welds. The most energetically challenging process for a weld to break is for an error to occur on a qubit involved in a weld. This event happens at a constant rate $\sim e^{-\beta v}$ where $v$ is the valency of the welded lattice and excitations have unit mass. A weld breaks then at a rate $r_1 \sim L^2 e^{-\beta v}$ where we include an $L^2$ term to account for the size of the welded surface. High-energy processes such as creation on a weld are exponentially suppressed with inverse temperature compared with processes that reduce the energy of the system and we might thus expect the Hamiltonian to reverse the effects of thermal errors. Indeed, this is ultimately why the two-dimensional Ising model performs as an effective memory in its ordered phase.

\begin{figure}
\includegraphics{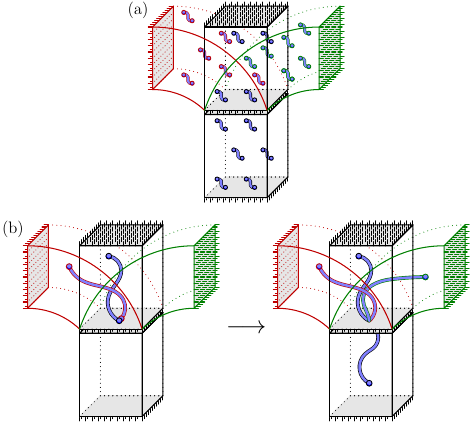}
\caption{(Color online) The excitations of the welded toric code. (a)~The blocks that make up the welded code contain an excitation gas of density $\rho \sim e^{-\beta}$ at thermalization. (b)~A zero-energy process; if $v/2$ excitations of different blocks meet at a common point on a $v$-valent weld, a weld can be broken at no energy cost.\label{BrokenWelds}}
\end{figure}

In spite of having favorable energetics that are analogous to those of the Ising model, at low temperatures we expect to observe processes that break the weld at a rate much quicker than $r_1$, which are due to the large volume of qubits within each code block involved in a weld. These processes are summarized in Fig.~\ref{BrokenWelds}. To analyze the low-energy processes we examine the surface of a weld. A single excitation created within a code block cannot pass the weld without incurring a high energy cost. However, if multiple excitations from different blocks meet at the same point along the weld, the excitations can pass through at a low energy cost. We make use of the ideal gas equation, $PV = nRT$, to show that we can expect these low energy processes to occur most commonly in the low-temperature limit.

We model the excitations that have occurred in a single block of volume $V = L^3$ as an ideal gas of point excitations of density $e^{- \beta}$, as shown in Fig.~\ref{BrokenWelds}(a), which is achieved quickly as the model approaches equilibrium. Using that the number of point excitations in a given block is $n = V e^{-\beta} $ and that $T = 1/\beta$, the excitation pressure on the boundaries of a given block follows from the ideal gas equation
\begin{equation}
P \sim e^{-\beta} / \beta,
\end{equation}
where we take the gas constant, $R$, as unity. We use the pressure to estimate the rate at which excitations fall on a particular point on the boundary to find the rate at which multiple excitations from different blocks of the weld meet at a common point, as shown in Fig.~\ref{BrokenWelds}(b). If $v/2$ point particles from $v/2$ different blocks that share a welded face meet at a common point, a weld can be overcome at no energy cost. This event occurs at a rate $ r_2 \sim L^2 P^{v/2}$ where the $L^2$ term comes from the size of the welded face. We compare this rate with $r_1$ to find 
\begin{equation}
{r_1 \over r_2} \sim {\beta^{v/2} e^{-\beta v/2}}  ,
\end{equation}
which vanishes in the limit that $\beta \rightarrow \infty $. This shows that low energy processes are the dominant processes in the low-temperature limit and we thus argue that the energy barrier will be ineffective in this regime. We obtain the same conclusion by modeling the free motion of excitations using other physically reasonable dynamics.

By consideration of low-energy processes and heuristic calculations we have argued that we cannot easily predict the thermal behavior of the welded model by simple understanding of its non-trivial energy barrier or by application of Arrhenius' law. As such, it is certainly interesting to understand the thermal behavior of the model at intermediate size and temperature regimes. Indeed, while the present manuscript was under peer review, results in Ref.~\cite{Siva16} emerged indicating that the welded code will demonstrate superexponential coherence-time scaling with inverse temperature through the study of finite temperature topological order.. More generally, a careful study of the dynamics of the model may allow us to derive new no-go theorems for finite-temperature stability that rely on a clearer understanding of entropic mechanisms that decohere encoded quantum information. Recent work following this direction has been conducted by Temme in Ref.~\cite{Temme14} by consideration of the case of commuting Pauli Hamiltonians.

\subsection{Fractal Product Codes}
\label{Sctn:Embed}

We finally remark on the new result of Brell~\cite{Brell}, where {\em fractal product codes} are introduced. In this Reference, the proposed family of models are mapped onto a classical model known to have a finite-temperature phase transition to argue that the model will be stable below a finite critical temperature.

The model is found using the formalism of homological product codes \cite{FreedmanHastings, BravyiHastings}. The homological product is an operation that combines pairs of codes to find new codes that in general are locally embeddable only in a larger number of spatial dimensions than their composite parts. The homological product of two two-dimensional toric codes for instance returns the four-dimensional toric code.

The model presented in Ref.~\cite{Brell}, is the homological product of two two-dimensional toric codes defined on the {\em Sierpi\'{n}ski carpet graph}. This gives a code that resembles the four-dimensional toric code defined on a fractal-like sublattice of the four-dimensional hyper-cubic lattice. It is conjectured that the choice of graph enables a local embedding of the product code in three dimensions. To demonstrate the stability of the model, it is shown that the model can be mapped onto the partition function of the product code onto an Ising model defined on a Sierpi\'{n}ski carpet; a model which has been rigorously proved to have a finite-temperature phase transition~\cite{Shinoda, Vezzani, Campari}.

The discovery of fractal product codes has opened a new avenue of research, and as such they have raised many new questions. For instance, it is yet to be shown that the model can be efficiently decoded. Moreover, the model has an extensive ground state degeneracy. This means we cannot easily apply known results to prove that it is perturbatively stable. Indeed, it is not even clear what the geometry of such a code might look like in three spatial dimensions if it is indeed embeddable. Further study of this model may lead to exciting insights towards stable quantum memories that can be realized in the laboratory.

\section{Other Protection Mechanisms}
\label{Sctn:EntropicProtection}

Following the restrictive no-go theorems described in Sec.~\ref{Sctn:NoGoTheorems} that forbid non-trivial energy barriers for large classes of two-dimensional commuting systems, it is interesting to consider other mechanisms that inhibit the long-range propagation of errors. Such a study is well motivated as protection mechanisms that do not require a macroscopic energy barrier may be compatible with experimentally amenable two-dimensional models. The purpose of this Section is to discuss other proposed mechanisms for the preservation of quantum information that do not rely on macroscopic energy barriers.

In Subsec.~\ref{SubSctn:EntropicSuppression} we discuss a model introduced in Ref.~\cite{Brown07} that is designed to exploit entropic effects to suppress thermal errors from developing. We go on to discuss known limitations of entropic protection in Subsec.~\ref{SubSctn:EntropicProtection}. Finally, in Subsec.~\ref{SubSctn:CoherentSuppression} we briefly discuss mechanisms to protect quantum information where the dominant noise source is coherent in nature.

\subsection{Entropically Suppressed Thermal Errors}
\label{SubSctn:EntropicSuppression}

It is interesting to ask if it is possible to protect quantum information from thermal errors by optimising the entropic term of the free energy in Eqn.~(\ref{Eqn:FreeEnergyCoherence}) to increase the coherence time of a system~\cite{LandonCardinalPoulin}, particularly in low-dimensional systems where the energy barrier between orthogonal ground states is necessarily constant. In this Subsection we describe a two-dimensional model~\cite{Brown07} that relies exclusively on entropic effects to protect encoded quantum information at finite temperature. We see by consideration of the dynamics of the system that the propagation of the commonly occurring excitations is suppressed. Here we give a qualitative picture of the mechanism that gives rise to entropic behavior by consideration of the anyonic excitation spectrum of the model. For a technical description of the underlying Hamiltonian, we refer the reader to the original Reference,~\cite{Brown07}.

The entropically protected model makes use of the fusion space of a generalized toric code model defined on a lattice of  $L\times L$ vertices with $N$-level spins on the edges of the lattice. For the reader familiar with quantum double models~\cite{Kitaev03}, we are considering the quantum double of the group $\mathbb{Z}_N$.

The relevant feature of the generalized toric code model that we discuss here is its anyonic excitation structure. Like the toric code, the generalized model has two types of anyons, electric charges and magnetic fluxes. However, in the generalized model they carry integer charges $1 \le k \le N-1$. We label excitations $e_k$ and $m_k$. The relevant fusion rules for the model are 
\begin{equation}
e_k \times e_l = e_{k\oplus l } \text{ and }m_k \times m_l = m_{k\oplus l }. \label{QuditFusionRule}
\end{equation}
where `$\oplus$' denotes addition modulo $N$. It follows from these fusion rules that the anti-particles of $e_k$ and $m_k$ excitations are $\overline{e_k} = e_{N-k}$ and $\overline{m_k} = m_{N-k}$. Henceforth, we will restrict the discussion to only $e_k$ particles. Due to the symmetries of the model an equivalent discussion holds for $m_k$ particles.

\subsubsection{Thermal Dynamics}

\begin{figure}
\includegraphics{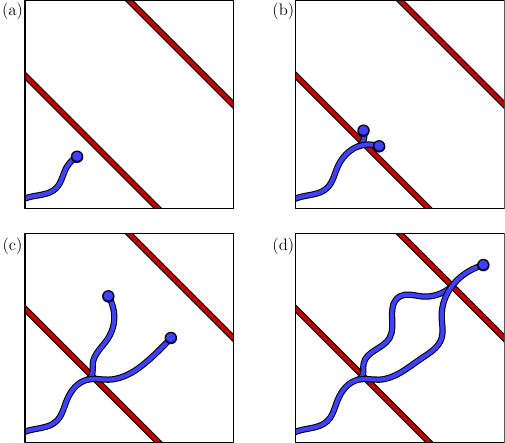}

\caption{\label{EntropicProcess}(Color online) Entropically suppressed excitations (a) A single $e_1$ excitation is marked by a blue circle. Red defect lines are drawn in red along diagonal lines of the lattice. (b) The $e_1$ particle propagates across the defect line to become an $e_2$ particle. This process is energetically suppressed by the choice of Hamiltonian. (c) A common process for the entropically protected model is for high-mass excitations to decay into pairs of low-mass excitations. The low-mass excitations are confined between the defect lines. (d) In the limit of very low temperatures, the lowest energy process for pairs of low mass excitations to pass a defect line is by recombination. This limits the entropic effects we describe at low temperatures.}

\end{figure}

As with the case of the toric code, excitations of the generalized model are free to propagate long distances across the lattice at no energy cost. This introduces large errors to the underlying physical lattice and thus rapidly decoheres information encoded in the ground state. Unlike the toric code, the excitations of the generalized model have a splitting structure where excitations $e_k$ can divide into two spatially separated excitations $e_j$ and $e_l$ provided the charge is conserved, i.e. $j \oplus l = k $, see Eqn.~(\ref{QuditFusionRule}). We write this splitting process $e_k \rightarrow e_j \times e_l$. This additional structure follows immediately from the fusion rules~(\ref{QuditFusionRule}). Errors incident to the lattice can achieve high energy configurations of many excitations due to the splitting structure of the model. We adapt the generalized toric code to exploit this splitting structure.

To encourage splitting processes to occur when coupled to the thermal environment, we write a Hamiltonian that assigns different masses to excitations of different charges. We choose the masses of the model such that it is energetically favorable for a subset of excitations to split. At this point we consider the explicit case for $N=5$. We set the Hamiltonian such that masses $M_k$ of particles $e_k$ are such that 
\begin{equation}
2M_1 = 2M_4 \le M_2 = M_3. \label{MassImbalance}
\end{equation}
With this setup, it is energetically favorable for the decay processes $e_2 \rightarrow e_1 \times e_1$ and $e_3 \rightarrow e_4 \times e_4$ to occur.

With the described setup at moderately low temperatures, and given that the model is initialized in the ground state, the most common process that we expect will decohere the information encoded in the ground space is the creation of an $e_1\times e_4$ pair, that will subsequently propagate rapidly across the lattice. The innovation of Ref.~\cite{Brown07} is the introduction of defect lines that entropically inhibit the long-range low-energy propagation of excitations by encouraging high-energy splitting processes. Defect lines are studied in generality in Ref.~\cite{KitaevKong, Barkeshli14}. Loosely speaking, in the general theory of defect lines, Hamiltonian terms are modified along a defect line such that when anyonic excitations cross the defect line, they modify their particle type according to some mapping.

The entropically protected model uses defect lines that modify the charge of crossing excitations. We show two defect lines lying on the lattice in red in Fig.~\ref{EntropicProcess}(a). Importantly, a defect line maps $e_k$ excitations crossing the line in the negative direction onto $e_{k\oplus k}$ excitations, where addition is carried out modulo $N$. Conversely $m_k$ excitations multiply their charge to become $m_{k \oplus k}$ excitations if they cross the defect lines in the opposite direction. The inverse operation occurs if excitations move over the defect line in the reversed direction. Importantly, the defect lines are designed such that if the commonly occurring low-mass $e_1$ and $e_4$ excitations cross defect lines in either direction their charge will always be modified to those of high-mass excitations.

We consider the long-range propagation of excitations in a moderately low temperature regime, where low-mass excitations are generated sparsely across the lattice. In Fig.~\ref{EntropicProcess}(a) an $e_1$ excitation is shown in the bottom-left corner of the lattice. Due to the choice of masses~(\ref{MassImbalance}), commonly occurring low-mass excitations are energetically suppressed from moving right across a defect line. The energy penalty reduces the rate of decoherence. The energetically suppressed process is shown in Fig.~\ref{EntropicProcess}(b).

Once the excitation crosses the defect line, it can continue to propagate across the lattice and over the next defect line as an $e_2$ particle with no energy penalty. However, we can configure the model such that this process is highly improbable. Due to the energetics of the model, it is favorable for the decay process $e_2 \rightarrow e_1\times e_1$ to occur. This process is shown in Fig.~\ref{EntropicProcess}(c). Following the decay process, the two $e_1$ excitations are confined between its enclosing defect lines, as they are subject to an energy penalty to propagate beyond a defect. Given the freedom to choose the defect line separation, we can optimize the system to make this process highly probable, thus commonly suppressing low-energy diffusion of errors. In what follows we discuss the numerics demonstrating the entropically favorable process that we have described thus far.

\subsubsection{Numerical Simulations}

In Ref.~\cite{Brown07} the entropically protected model is simulated in a thermal environment to show that the typical excitations increase in energy as they diffuse in some appropriate parameter regime. The model is set up with a square grid of defect lines with separations alternating between one and two lattice units where the excitation masses $M_1 = M_4 = 0.38$ and $M_2 = M_3 = 1$ are taken. These values best embellish the entropic effects with numerically tractable system sizes.

The thermalization data obtained in Ref.~\cite{Brown07} is compared to the partial self correction hypothesis introduced in Subsec.~\ref{Sctn:PartialSelfCorrection} in the temperature regime $ 6\le \beta \le 9$. By comparison to Eqns.~(\ref{PSCBeta}) and~(\ref{PSCSystemSize}), the values
\begin{equation}
\Delta_{\text{EP}} \sim 0.5 \text{ and } \kappa_{\text{EP}} \sim 0.2, \label{EntropicProtectionData}
\end{equation}
are obtained for the entropically protected model. The positive $\kappa_{\text{EP}}$ value is indicative of error dynamics with excitation masses that grow with the total size of the error. 
We observe that the entropic error suppression shown here is significantly weaker than those found in three-dimensional partially self correcting models. This is reflected by comparing the obtained data~(\ref{EntropicProtectionData}) with the data found for the cubic code~(\ref{CubicCodeCharacteristics}). This can be explained by the fact that entropic protection relies on probabilistic effects for energetic suppression, unlike the cubic code model where an energy barrier is inherent in the system.
We further remark that unlike the cubic code model these effects are limited to the regime $\beta \le 9$, and as such, do not satisfy the conditions we require of a quantum memory. This is because at very low temperatures, the thermal environment will find low energy paths to propagate excitations such as that shown in Fig.~\ref{EntropicProcess}(d). This low temperature behavior is reminiscent of that which occurs in the welded code, as discussed in Subsubsec.~\ref{Sctn:WeldedToricLimits}. In the following Section we discuss known fundamental limitations on entropic protection.

\subsection{On Entropic Protection}
\label{SubSctn:EntropicProtection}

It is interesting that the entropic protection of the model introduced in the previous Subsection is only effective above a certain moderately-high temperature. Following the introduction of the entropically protected model, significant work has been conducted to learn the limitations of error suppression by entropic effects.

In Refs.~\cite{Temme14, Temme15}, it is shown for commuting Pauli Hamiltonians that a macroscopic energy barrier is necessary to achieve a coherence time that diverges with system size. This work is extended in Ref.~\cite{Komar16} to show the necessity of an energy barrier for a more general class of models, namely, Abelian quantum double models~\cite{Kitaev03}, which include the entropically protected model discussed in Subsec.~\ref{SubSctn:EntropicSuppression}. It is thus clear that improvements in coherence time that are achieved by the introduction of defect lines are strictly finite in nature, and will never lead to diverging memory time, or super-Arrhenius-law coherence-time scaling in the $\beta \rightarrow \infty $ limit.

Despite the negative results towards self-correction via entropic effects, as two-dimensional quantum memories will almost certainly be more experimentally amenable than their higher-dimensional counterparts, it may still be worthwhile improving low-dimensional memories with entropic effects. As such, it may be interesting to optimize the parameters of entropically protected memories, such as that we reviewed in the previous Subsection, to boost their coherence times.

We finally remark that the known no-go theorems for entropic protection are applicable to Abelian quantum double models. It may be interesting to consider other mechanisms for entropic protection in non-commuting Hamiltonians, or even with commuting models that give rise non-Abelian anyon theories~\cite{Kitaev03, Levin05}. We finally point out another interesting proposal for a two-dimensional memory given in Ref.~\cite{Bardyn}. The authors consider coupling the toric code Hamiltonian to driven-dissipative ancilla systems~\cite{Pastawski11} to inhibit the propagation of excitations. It is the role of the dissipative systems to reduce the energy of the interaction terms of the Hamiltonian to introduce potential minima to the system to increase the likelihood of diffusing excitations retracing their steps which will thus reverse incident errors. It is shown in Ref.~\cite{Bardyn} that coupling the toric code to a dissipative ancilla system gives rise to super-exponential coherence-time scaling with inverse temperature.

\subsection{Coherent Noise Suppression}
\label{SubSctn:CoherentSuppression}

The majority of this Review has been concerned with finding systems that protect quantum information encoded in many-body systems from incoherent sources of noise, namely where the system of interest is coupled to a thermal bath. However, we might consider some implementation of a system where finite-temperature effects are negligible, and where coherent sources of noise are dominant. Specifically, we consider here weak local perturbations such as external magnetic fields. Protection against coherent sources of noise will become increasingly relevant as the temperature of the system becomes very low. In such a regime we might consider designing systems where we sacrifice self-correction and instead focus on experimental amenability. Here we discuss mechanisms that have been considered to defend low-dimensional Hamiltonian systems from various forms of coherent noise.

As we have discussed in Subsec.~\ref{subsec:stability}, perturbative stability at zero-temperature is a natural property of topologically ordered systems. It is this inherent stability that first motivated their use for quantum computation~\cite{Kitaev03}. In such systems, quantum information can be stored within locally indistinguishable degenerate ground states, which are separated from excited states by a finite gap.  Perturbations will cause some energy splitting between ground states, and can cause the gap to close to a degree. However, it has been shown that these effects will be suppressed with system size for perturbations whose strength is below some finite threshold~\cite{BravyiHastingsMichalakis10}. Such topological systems, including the simple toric code, would therefore provide a stable memory at zero-temperature.

Similar properties also arise in so-called symmetry protected topological phases~\cite{Chen11, Else12, Bonderson13}. These offer protection against perturbations that respect non-trivial symmetries of the model. As compensation for this, they are typically more experimentally accessible than topological models that do not rely on a symmetry. 

A well-known example is an open chain for spinless fermions with a superconducting pairing term~\cite{Kogut79,Kitaev01}. For a suitable choice of Hamiltonian parameters, this system can be brought to a phase supporting topological superconductivity~\cite{Kitaev01} with a perturbatively stable degenerate ground state and Majorana fermion zero-energy modes localized at opposite ends of the chain.

Quantum information encoded within this subspace spanned by these modes~\cite{Kitaev01,Bravyi06} will remain stable as long as the fermionic parity of the wire is conserved, which relies on only Cooper pairs being exchanged with the environment. Exchange of single unpaired fermions can cause fatal decoherence~\cite{Rainis12}. The system must therefore be engineered to ensure that such symmetry-violating processes are rare.

So far, low-dimensional symmetry protected phases have not been shown to offer self-correction against thermal noise. In fact, certain models have been found to decohere in a way that is not efficiently suppressed by reducing temperature~\cite{Campbell15}. In such cases, it has been argued that ctive error correction is the only hope of preserving the symmetry protected quantum memory at finite temperature~\cite{Pedrocchi15}.

The above discussion on the suppression of coherent noise assumes that information can be encoded directly into the ground state manifold. However, in practice we must expect that preparing the state will be noisy such that we may prepare an excited state with some mobile quasi-particle excitation on the lattice. It is thus problematic that the coherent noise can propagate these excitations and decohere stored information very quickly.

One mechanism studied to suppress this is disorder-assisted protection. In Refs.~\cite{Pastawski, WoottonPachos, Stark, Rothlisberger, BravyiKonig} it is shown that randomizing the Hamiltonian interaction strengths inhibit the coherent propagation of excitations across the lattice. This is attributed to the Anderson localization effect~\cite{Anderson}, where it is understood that randomized Hamiltonian interactions lead to `friction' in the motion of excitation dynamics due to quasi-particles becoming trapped in small potential minima of the random energy landscape. It is tempting then to believe that randomness might then be useful as a resource for error suppression. However, Anderson localization is not a well understood principle. Remarkably, in Ref.~\cite{BravyiKonig} it is shown that pseudo-random potentials outperform truly randomly chosen potentials.

Some of the randomness studied in Ref. ~\cite{Rothlisberger} was of the underlying lattice structure in the two-dimensional toric code model, rather than the coupling strengths. This demonstrates the important effect that the chosen lattice will have on coherence time, as expanded upon in Ref.~\cite{AlShimary}. Though this focusses primarily on thermal noise, it's main purpose is to optimize the lattice in the case that noise is biased towards certain kinds of errors. Specifically, either bit-flip or dephasing noise. It is identified that reducing the connectivity of the lattice geometry which embeds the toric code will reduce the rate of one type of noise from decohering the lattice. However, due to the dual structure of the toric code lattice, changing the lattice geometry to protect against one type of noise will detrimentally affect the performance of the lattice against the other type of noise. This approach therefore presents a tradeoff between protection against bit-flip and dephasing noise introduced by a general thermal noise model. Therefore these effects provide a constant improvement in coherence in the presence of an asymmetric noise model that may be present in a realistic experimental setting~\cite{Doucot}.

\section{Beyond Commuting Hamiltonians - Subsystem Codes}
\label{Sctn:Subsystems}

In this Section we discuss progress towards the study of a class of non-commuting models that has developed over the last decade, namely, the {\em subsystem codes}. This framework provides a natural extension to the stabilizer formalism that we have relied upon throughout this Review. Amongst the subsystem code literature are three-dimensional models that are conjectured to be self-correcting quantum memories.

Subsystem codes were introduced in Refs.~\cite{Kribs05, Kribs06}. The language of subsystems was introduced to find a unifying language for decoherence free subspaces~\cite{MassimoPalma, Duan, Lidar, Zanardi97, Beige00, PachosBeige2004} and noiseless subsystems~\cite{Knill00, Zanardi, Kempe01}. Initially coined operator quantum error correction, subsystem codes principally encode logical information in a quantum error-correcting code that is embedded in a {\em subsystem} of a larger Hilbert space. In a subsystem code the remainder of the Hilbert space of the total system is referred to as the {\em gauge subsystem}. 

A general language for subsystem codes is provided in Ref.~\cite{Poulin05}. A subsystem code is uniquely defined by its gauge group, $\mathcal{G}_n  $, a subgroup of the Pauli group of $n$ qubits. Given a gauge group, a stabilizer code is defined on the subsystem acted upon by the centralizer of the gauge group $\mathcal{N} (\mathcal{G}_n)$, i.e. operators that commute with elements of $\mathcal{G}_n$. The stabilizers of the code are also members of the gauge group, $\mathcal{S} = \mathcal{N}(\mathcal{G}_n) \cap \mathcal{G}_n$, whereas logical operators, $\mathcal{L} = \mathcal{N}(\mathcal{G}_n) \backslash \mathcal{G}_n$, commute with, but are not themselves members of the gauge group. It is easily checked that for the special case that $\mathcal{G}_n$ is Abelian, we recover the stabilizer formalism where $\mathcal{G}_n = \mathcal{S}$ up to phases. 

Throughout this Section we give examples which show that this innocuous abstraction of the stabilizer formalism behaves qualitatively differently from the stabilizer error-correcting model. First of all, by definition, the gauge subsystem can take any state. It can, for instance, become arbitrarily mixed due to incident noise or otherwise, and encoded quantum information will remain robust. One might also consider using the gauge subsystem for other practical purposes. The authors of Ref.~\cite{HerreraMarti14} make frequent measurements on the subsystem surface code presented in Ref.~\cite{Bravyi13} to suppress the coherent diffusion of excitations at zero temperature. It is also shown that the subsystem structure of certain codes can be used to realize universal quantum computation~\cite{Paetznick13}, or to design advantageous error-correcting protocols~\cite{Bombin14} which we discuss in Subsec.~\ref{Subsctn:GCC}.

We additionally remark that subsystem codes may have some differences in their practical realization compared with their stabilizer counterparts which must be taken into account if one is to build a quantum memory based on a subsystem code. In particular, certain local subsystem codes give rise to high-weight stabilizer measurements that become increasingly error prone in the large system-size limit. We urge the reader to find a comprehensive review of this topic in Ref.~\cite{Terhal13} and references therein to better understand the practical difficulties in realizing certain subsystem error-correcting codes. Conversely, some subsystem codes~\cite{Bombin10, Bravyi13} are specifically designed to reduce the weight of syndrome measurements such that they are less demanding to implement from an experimental perspective.

Little is known about the fundamental features or thermal characteristics of subsystem codes, though progress has been made in this area in Ref.~\cite{ChesiLossBravyiTerhal} where bounds on their relaxation times are obtained. Much of the work in this area has been a constructive search for models that we might expect to give rise to favorable properties for self correction. Of particular interest is the three-dimensional Bacon-Shor code~\cite{Bacon06}. This model has drawn significant attention to subsystem codes as it is conjectured to behave as a self-correcting memory. This argument is made as its Hamiltonian has key features in common with thermally stable classical models, namely, the Ising model, discussed in Subsec.~\ref{Sctn:TheIsingModel}. We introduce the three-dimensional Bacon-Shor code by first reviewing the two-dimensional Bacon-Shor code; a simple example of a subsystem code that has some qualitatively different features from local stabilizer codes. We conclude our discussion of the two-dimensional Bacon-Shor code by considering its generalizations and other subsystem codes that are shown to surpass constraints physically imposed on general commuting models.

Subsystem codes are of further interest when considered in the context of topological order. For the reader familiar with topologically ordered lattice models, we remark that Kitaev's honeycomb model~\cite{Kitaev06} provides an illustrative example of a subsystem code~\cite{Suchara11}; the model is studied by considering loop degrees of freedom that commute with its non-commuting parent Hamiltonian. The honeycomb model falls into a broader subclass of subsystem codes, known as {\em topological subsystem codes}~\cite{Bombin10, Suchara11}. Recently a three-dimensional generalization of topological subsystem codes, namely the gauge color code model, has recently been conjectured to be self-correcting. We conclude this Section with an overview of this model, and its features that have led to this conjecture. We discuss also other interesting results that have arisen by consideration of the gauge color code as a self-correcting quantum memory.

\subsection{The Bacon-Shor Code}
\label{BaconShor}

The Bacon-Shor code~\cite{Bacon06}, otherwise known as the quantum compass model~\cite{Kugel, Dorier05}, provides a non-trivial example of a subsystem codes that demonstrates physical features not accessible with local stabilizer codes. For an extensive review of compass models we direct the reader to Ref.~\cite{Nussinov14}. In the two-dimensional model, we observe that its local gauge generators that give rise to non-local stabilizer generators. Moreover, we see how the non-trivial gauge group affects the error-correction procedure. We conclude this Subsection by reviewing results that show the fundamental differences between subsystem codes and commuting models. 

\begin{figure}
\includegraphics{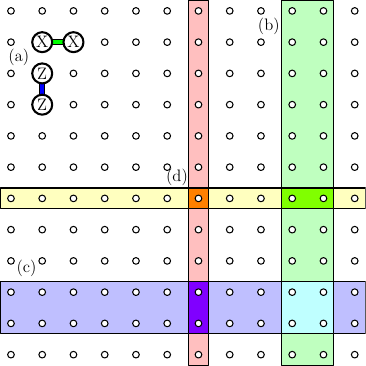}
\caption{\label{BSCode}(Color online) The Bacon-Shor code in two dimensions. Qubits are represented by white circles. (a) Example of gauge generators. The gauge group $\mathcal{G}_n$ is generated by nearest-neighbor two-body operators including the horizontal $A_x$ operators and the vertical nearest-neighbour $B_x$ operators which are shown, respectively, in green and blue. The shaded regions by (b) and (c) show the support of stabilizers of the Bacon-Shor code. (d) The support of $\overline{X}$ and $\overline{Z}$ are shown by the red shaded vertical strip and yellow shaded horizontal strip, respectively. The logical operators intersect at a single qubit marked by dark orange shading near (d).}
\end{figure}

The two-dimensional Bacon-Shor code is defined on an $L \times L$ square lattice with qubits on its vertices, as shown in Fig.~\ref{BSCode}. The gauge group $\mathcal{G}_n$ where $n = L^2$ is generated by two types of operator. It has two-body nearest-neighbor interactions $A_x$ and $B_x$ associated to lattice sites $x = (j,k)$, where $A_{j,k} = X_{j,k} X_{j+1,k} $ are aligned along the horizontal direction and $B_{j,k} = Z_{j,k} Z_{j,k+1}$ operators are aligned in the vertical direction, as shown in Fig.~\ref{BSCode}(a) in green and blue, respectively. We write the corresponding Hamiltonian of this model
\begin{equation}
H_{\text{2DBS}} = - \sum_x  \left(A_x + B_x \right). \label{BSHam}
\end{equation}

In general, $A_x$ and $B_x$ operators do not commute. The stabilizers of this code, $\mathcal{S} = \mathcal{N}(\mathcal{G}_n) \cap \mathcal{G}_n$, are non-local operators of the gauge group generated by products of $A_x$ and $B_x$ operators. They are $S^X_j = \prod_k A_{j,k}$ and $S^Z_k = \prod_j B_{j,k}$. Stabilizers $S_j^X$ and $S_k^Z$ are supported on bands, two vertices wide, such as those shown in green and blue in Fig.~\ref{BSCode}(b) and~(c). The horizontal bands that support $S_k^Z$ stabilizers have four common qubits with the support of the vertical bands supporting $S^X_j$ stabilizers. As such, it is easily checked that all stabilizers commute. The model encodes a single logical qubit. Its logical operators, $\mathcal{L} = \mathcal{N}(\mathcal{G}_n) \backslash \mathcal{G}_n$, are $\overline{X} = \prod_k X_{j,k}$ for fixed $j$ and $\overline{Z} = \prod_j Z_{j,k}$ for fixed $k$. The support of $\overline{X}$ and $\overline{Z}$ are shown in red and yellow on Fig.~\ref{BSCode}. They cross at a single point by Fig.~\ref{BSCode}(d).

The two-dimensional Bacon-Shor code shows how subsystem codes respond differently to noise. We consider a noise model that introduces Pauli-X errors. An equivalent discussion holds for Pauli-Z errors where the lattice is rotated by $\pi / 2$. Remarkably, for any given row of the lattice, we only need to correct the {\em parity} of errors. A pair of Pauli-X errors along a horizontal strip are elements of $\mathcal{G}_n$, and thus commute with all $\mathcal{S}$ and $\mathcal{L}$. Errors of this type therefore do not affect information encoded in the subsystem code. As in the case for stabilizer codes, we make stabilizer measurements to identify the parity of errors between pairs of strips, and we correct the parity of a given row using a single-qubit Pauli-X operation along that row.

Importantly, as is pointed out in the original paper~\cite{Bacon06}, it has been argued persuasively that the gap of Hamiltonian~(\ref{BSHam}) vanishes in the thermodynamic limit. This is shown in Ref.~\cite{Dorier05} using extensive numerical methods in an extended region of phase space. The model is therefore only likely to perform well as a quantum memory in some limited regime where the system size is relatively small. Nevertheless, the example shows that a clever choice of gauge enables us to ignore large classes of errors that affect only the gauge subsystem. Further to this, the Bacon-Shor code provides an example of a local subsystem code model that gives rise to a non-local stabilizer code, thus indicating a fundamental difference between local stabilizer and subsystem codes.

Indeed, the study of local subsystem codes that give rise to non-local stabilizer codes has been significantly extended due to the recent work by Bacon {\it et al.}~\cite{Bacon14}. In this work they show a very general scheme where one can find a local subsystem code that gives rise to a stabilizer code, that is not necessarily local, given the quantum circuit that measures the stabilizer generators of the stabilizer code. The Authors use their formalism to find codes with favorable code distance scaling that saturate known bounds for commuting projector codes. As remarked in Ref.~\cite{Bacon14}, perhaps one could consider using their formalism to construct local subsystem codes that correspond to non-local stabilizer models with favorable properties at finite temperature to find stable quantum memories. Certainly, it is shown that the presented construction can give rise to models with no string-like logical operators, even if one considers the more general class of {\em dressed} logical operators, i.e. logical operators with non-trivial action on the gauge subsystem of the model.

We finally remark on an extension of the Bacon-Shor code due to Bravyi~\cite{Bravyi13}, where he shows that the fundamental storage capacity of a two-dimensional subsystem code can surpass the storage capacity of commuting models~\cite{Bravyi10}. In the commuting case, it is known that $kd^2 \le \mathcal{O}(n)$, where $k$ is the number of qubits of the code, and $d$ is the minimum weight of the lowest weight logical operator of the code. Bravyi shows that in the case of subsystem codes we can obtain scaling like $kd \le \mathcal{O}(n)$ with a local gauge group, and that randomly chosen codes saturate this bound asymptotically in the limit of large $n$. Unfortunately, this bound cannot be saturated for codes with constant $k$. Indeed it is also known from Ref.~\cite{BravyiTerhal09, Haah12} that distance of two-dimensional subsystem codes must satisfy $d^2 \le \mathcal{O}(n)$. Results such as these provide further intrigue and motivation towards the study of subsystem codes, as we clearly see that these non-commuting codes are capable of supporting encoding properties provably unattainable by their commuting counterparts.

\subsection{The Three-Dimensional Bacon-Shor Code} 

Subsystem codes have attracted a lot of interest from the point of view of thermal stability due to the three-dimensional Bacon-Shor code. In Ref.~\cite{Bacon06}, Bacon conjectures that the model gives rise to self-correcting behavior in a thermal setting. 

The model is defined on an $L \times L \times L$ lattice for odd $L$ with sites labeled by $x = (j,k,l)$. The corresponding Hamiltonian is given by
\begin{equation}
H_{\text{3DBS}} = -\sum_x \left(A_x + B_x +C_x + D_x\right), \label{Eqn:3DBS}
\end{equation}
where the interaction terms are two-body nearest neighbor Pauli operators $A_{j,k,l} = X_{j,k,l}X_{j+1,k,l}$, $B_{j,k,l} = X_{j,k,l}X_{j,k+1,l}$, $C_{j,k,l} = Z_{j,k,l}Z_{j,k+1,l}$ and $D_{j,k,l} = Z_{j,k,l}Z_{j,k,l+1} $ which generate the gauge group. The logical operators of the code are two-dimensional plane like operators $\overline{X} = \prod_{k,l} X_{1,k,l}$  and $\overline{Z} = \prod_{j,k} Z_{j,k,1}$ that anti-commute due to the condition that $L$ is odd. The stabilizer generators are two-dimensional plane-like operators that are two vertices wide, such that $S^X_j =  \prod_{k,l} X_{j,k,l}X_{j+1,k,l}$ and $S^Z_l =  \prod_{j,k} Z_{j,k,l}Z_{j,k,l+1}$. 

\begin{figure}
\includegraphics{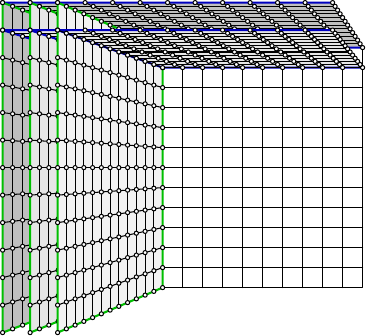}
\caption{\label{3DBS}(Color online) The gauge group of the three-dimensional Bacon-Shor Hamiltonian. The three-dimensional Bacon-Shor code is supported on qubits arranged on cubic lattice. The gauge group consists of two-body nearest-neighbor terms, where different planes of the lattice support different types of gauge operators. Horizontal planes such as those we outline in blue in the Figure support two-body Pauli-X gauge operators between their qubits. Similarly, the vertical planes outlined in green that protrude out of the page support two-body Pauli-Z gauge terms along its edges. The full model is described by an array of $L$ of these planes arranged in parallel along their respective directions. The locations of the full set of planes are marked by black lines at the back of the Figure. Indeed, the gauge group of the three-dimensional Bacon-Shor code can be regarded as intersecting copies of the two-dimensional Ising model supported on each of these planes.}
\end{figure}

The Hamiltonian of the three-dimensional Bacon-Shor code has the gauge terms of many intersecting two-dimensional Ising models. It has two-body Pauli-X type interactions along one two-dimensional plane, and Pauli-Z type interactions along an orthogonal plane. We show the construction in Fig.~\ref{3DBS}. It is shown in Ref.~\cite{BaconCasaccino} that in general one can design generalized Bacon-Shor codes where one takes an arbitrary classical code input to find its gauge group. Its overlying stabilizer code then inherits the properties of the input classical code. The intuition of the three-dimensional Bacon-Shor code is that it inherits the stability properties of its input code, the two-dimensional Ising model, as discussed in Subsec.~\ref{Sctn:TheIsingModel}. In Ref.~\cite{Bacon06}, the author considers a simple Pauli noise model and mean-field arguments to show that the model may demonstrate a macroscopic power-law energy barrier. The question of stability of the three-dimensional Bacon-Shor code remains an open problem.

The challenge of interrogating this conjecture, as pointed out in Ref.~\cite{Bacon06}, is that the noise model the author considers is not representative of all possible channels under which encoded information can decohere. It may be possible that a finite-temperature environment may find a low energy path to introduce errors to the encoded logical information. Indeed, no exact diagonalization of Hamiltonian~(\ref{Eqn:3DBS}) has been found and without one it is difficult to make strong statements about the model. 

Certainly, flaws have been identified with the model that must be overcome to satisfy the required conditions of a stable memory. First of all, the model provably has no error-correction procedure in the thermodynamic limit~\cite{Pastawski}. 
Further, given that we are relying on the model to inherit the stability properties of the two-dimensional Ising model, we may also expect it to inherit its perturbative instability, described in Refs.~\cite{Richards, Cirillo, Grinstein, Pastawski}. One might support this assertion by extrapolating results regarding the gap from the well studied two-dimensional case~\cite{Dorier05}. If the model is indeed gapless, like the two-dimensional code, we may encounter issues with the perturbative stability of the three-dimensional Bacon-Shor code as we increase the size of the system towards the thermodynamic limit.

\subsection{The Gauge Color Code}
\label{Subsctn:GCC}
We conclude this Section with a discussion of the {\em gauge color code}~\cite{Bombin13a}, the smallest realization of which was first discovered by Paetznick and Reichardt in Ref.~\cite{Paetznick13} by consideration of Reed-Muller codes. The model represents a three-dimensional topological subsystem code. Topological subsystem codes are discussed in two dimensions in Refs.~\cite{Bombin10, Suchara11, Brell11, Sarvepalli, Bravyi13, Kubica14}. The model is of particular interest as it is conjectured to give rise to finite-temperature stability~\cite{Bombin13a}. For a comprehensible overview of the gauge color code see Refs.~\cite{Kubica14, Watson15}.

We briefly elaborate on the fault-tolerant computational properties of the color codes. Notably, the color code models are favorable for their implementation of transversal gates. The two-dimensional color code has a transversal implementation of the Clifford gate set~\cite{Bombin06} which, importantly here, includes the Hadamard gate. In the two-dimensional model, the feature that enables its implementation is the {\em self-duality} of its stabilizers. Explicitly, a self-dual stabilizer group is such that for every stabilizer $S^X = \prod_{j\in \mathcal{T}}X_j$ supported on subset of qubits $\mathcal{T}$, there exists also the stabilizer $S^Z = \prod_{j\in \mathcal{T}}Z_j$.

The three-dimensional stabilizer color code~\cite{Bombin07} gives rise to a transversal implementation of the controlled-not gate and the $\pi / 8$-gate. Supplemented by the Hadamard gate, the gate set of the three-dimensional color code is capable of universal fault-tolerant quantum computation. However, the three-dimensional stabilizer color code is not self dual, and as such, does not support a universal transversal gate set.

\begin{figure}
\includegraphics{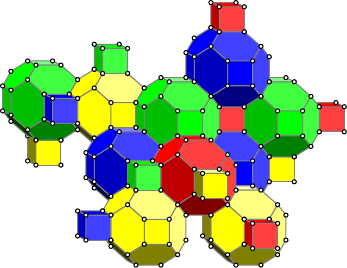}
\caption{\label{GaugeColorCodeLattice}(Color online) The lattice of the gauge color code. Qubits reside on the vertices of a four-valent lattice. The lattice is also {\em four colorable}. This means we can assign each of the three-dimensional cells of the lattice one of four colors such that no two cells of the same color touch. The cells in the Figure are colored dark blue, red, light green and yellow to show the four colorability of the cells.}
\end{figure}

In Ref.~\cite{Bombin13a} Bombin shows that a universal gate set can be realized with a subsystem generalization of the three-dimensional color code. Importantly, the stabilizer group of the three-dimensional color code contains a self-dual {\em subset} of stabilizers that are capable of successfully identifying an arbitrary set of errors below a certain weight. Given a suitable choice of gauge generators, we can restrict the gauge color code stabilizers to its self-dual subset, such that we have a code with a transversal implementation of the Hadamard gate.

The proposal of Bombin represents an explicit lattice realization of {\em gauge fixing}, introduced in Ref.~\cite{Paetznick13}. Gauge fixing avoids resource costly methods of achieving a universal gate set using, for instance, magic state distillation~\cite{BravyiKitaev05}.  Instead, gauge fixing enables us to fault-tolerantly move information between different codes by changing the gauge operators of a given  subsystem code. Specifically, one can `promote' certain elements of the gauge group of a code to elements of the stabilizer group by imposing that the code states of the code take particular eigenstates of gauge operators, thus changing, or fixing, the code. The development of gauge fixing has shown that we can move between different error-correcting codes that collectively support a universal gate set. Gauge fixing is shown explicitly for Reed-Muller codes in Ref.~\cite{AndersonDuclosCianciPoulin14}. The idea of gauge fixing extends ideas presented in Ref.~\cite{Knill96} where fault-tolerant gates that do not preserve the code space are suggested. 

Here we consider the gauge color code for a candidate self-correcting quantum memory. A lattice suitable to describe the code is shown in Fig.~\ref{GaugeColorCodeLattice}. Another appropriate lattice geometry is given in Ref.~\cite{Kim11}, see also Ref.~\cite{Brown15}. Importantly, the lattice is four valent and four colorable, i.e. the lattice is such that we can assign to each cell one of four colors in such a way that no two neighbouring cells are of the same color. It follows that three-dimensional lattices that satisfy these properties have faces that contain an even number of vertices~\cite{Kubica14}. The generators of the gauge group are associated to the faces, $f$, of cells of the lattice. Specifically, for every face of a cell, we have gauge generator $A_f = \prod_{v\in f} X_v$ and $B_f = \prod_{v\in f} Z_v$, where we use shorthand $ v \in f $ to mean the vertices adjacent to face $f$. On the lattice shown, gauge generators are either four- or six-body terms. We therefore have the Hamiltonian 
\begin{equation}
H_{\text{GCC}} = -\sum_f \left(A_f + B_f \right).\label{Ham:GCC}
\end{equation}
The stabilizers are then associated to the cells, $c$, of the lattice, such that $A_c  = \prod_{v\in c} X_v$ and $B_c  = \prod_{v\in c} Z_v$, where $v \in c$ are qubits associated to vertices adjacent to cell $c$. 

The conjecture due to Bombin~\cite{Bombin13a} is based on the structure of the $A_f$ and $B_f$ Hamiltonian terms. Specifically, if one considered the simpler Hamiltonian $H^X = -\sum_f B_f$ subject only to Pauli-X error channel, then the excitations of the model are exclusively akin to Peierls contours, as discussed in Sec.~\ref{Sctn:HighD}. An equivalent argument holds for Pauli-Z noise for Hamiltonian $H^Z = -\sum_f A_f$, and as such one can argue that there may be a macroscopic energy barrier for arbitrary local quantum noise channels. Of course, given the difficulty in diagonalizing Hamiltonian~(\ref{Ham:GCC}) this argument is not rigorous.

\begin{figure}
\includegraphics{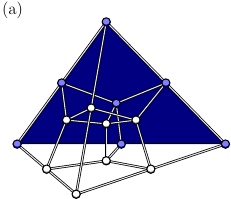}
\includegraphics{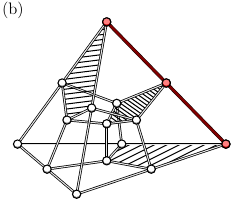}
\caption{\label{BareAndDressedLogicalOperators}(Color online) A fifteen-qubit tetrahedral lattice that supports the gauge color code. (a)~The two-dimensional logical operator of the gauge color code, marked in blue with shaded vertices at the back of the lattice. (b) The support of the dressed logical operator is shaded red. The faces which support Hamiltonian terms that anti-commute with the dressed logical operator are shaded.}
\end{figure}

When defined on a tetrahedron, the logical operators that commute with the gauge generators are two dimensional. We show the smallest tetrahedron that embeds a gauge color code in Fig.~\ref{BareAndDressedLogicalOperators}. Both $\overline{X}$ and $\overline{Z}$ are supported on one face of the tetrahedral lattice. Their support is shown in blue on the example lattice in Fig.~\ref{BareAndDressedLogicalOperators}(a). The geometry of the color code ensures that the support of a face of the tetrahedron remains odd for any size of lattice. 

While it is a promising feature with respect to self correction that the logical operators are two dimensional, it is not clear that such logical operators are sufficient for self correction in subsystem codes. Indeed, the support of logical operators of subsystem codes can be reduced by multiplication by gauge generators. Such a logical operator will not necessarily commute with the gauge group, but still commutes with the stabilizer group, as is required. A logical operator of this type is known as a {\em dressed logical operator}. 

In contrast to the models proposed by Bacon {\it et al}.~\cite{Bacon14} mentioned in Subsec.~\ref{BaconShor}, the gauge color code has one-dimensional dressed logical operators. We show the support of the dressed logical operators of the gauge color code in Fig.~\ref{BareAndDressedLogicalOperators}(b). We can conclude little from discovering this operator, but we make this point only to illustrate some of the additional complexity involved in studying subsystem codes. Indeed, due to the non-trivial commutation relations of the dressed operator with respect to the interaction terms of Hamiltonian~(\ref{Ham:GCC}), it is not clear that such an operator can be achieved at constant energy cost under a local noise channel. As such, it is not known if the discovery of a low-dimensional dressed logical operator does rules out the possibility of self correction in the gauge color code model. We additionally remark that, like the three-dimensional Bacon-Shor code, it is not even obvious that the gauge color code is gapped. To this end, the thermal stability of the gauge color code remains an open problem.

We finally mention further progress in the study of the gauge color code with respect to its error-correcting capabilities. In~\cite{Bombin14}, Bombin shows that the gauge color code model has favorable properties for decoding when measured using unreliable laboratory equipment. He argues that we need to measure each face operator of the gauge color code code only once to obtain reliable fault-tolerant syndrome information. This phenomenon, coined {\em single-shot error correction}, has been demonstrated numerically in Ref.~\cite{Brown15}. This differs from the well-studied case of two-dimensional stabilizer models where syndrome information is read out using unreliable measurement apparatus. Known schemes for fault-tolerant error correction require that each stabilizer must be measured a macroscopic number of times to read out logical information reliably~\cite{DennisKitaevLandahlPreskill, WangHarringtonPreskill}. Bombin remarks that the special error-correcting properties he demonstrates are generic to known self-correcting stabilizer models~\cite{Bombin14}. Certainly, this is an exciting advance in active error correction that has been derived from the study of a candidate self-correcting model. 

\section{Discussions and Outlook}
\label{Sctn:Outlook}

Among the major discoveries that led to scalable classical information processing was the discovery of a transistor. The physics of these little solid-state devices ensure the reliable storage {\em and} robust processing of classical information that is easily scaled. The discovery of an experimentally amenable stable quantum memory presents a monumental hurdle, which, if overcome, will be invaluable for the discovery of fault-tolerant quantum information processing~\cite{Bacon06}. In the present Review we have given an overview of the analytical and numerical tools we use to approach the study of many-body quantum systems at finite temperature. We have examined the no-go theorems that have been discovered in this field, and we have presented many new physical models with certain properties suitable for the passive protection of quantum information. We conclude by highlighting some of the forthcoming challenges we face towards discovering and developing a quantum memory.

This Review has highlighted many open problems in this field. We have discussed models that arguably present favorable properties to make them resilient in a thermal environment. Such models include the ferromagnet-coupled toric code, the welded toric code, embeddable-fractal product codes, the three-dimensional Bacon-Shor code and the gauge color code. These conjectures need to be interrogated by numerical experiments or by rigorous proofs. To study these models we need to develop both analytical and numerical techniques in both condensed-matter physics and fundamental statistical mechanics. In addition to this, it needs to be checked that these models have other features that are required of a passive quantum memory, such as efficient decoding algorithms and perturbative stability. 

A noteworthy theme that has occurred frequently throughout this Review is that many of the candidate models for finite temperature stability are composed of simpler systems. For instance, the work in Sec.~\ref{Sctn:InteractingAnyons} shows that we can introduce anyonic interactions with a local model by coupling the toric code to either a system of bosons or a Heisenberg ferromagnet. Similarly, the non-translationally invariant models in Subsecs. \ref{Sctn:WeldedToric} and \ref{Sctn:Embed} are constructed by combining simpler topologically ordered models using either welding or by taking the homological product of many codes. Indeed, the entropically protected model presented in Subsec.~\ref{SubSctn:EntropicSuppression} can be regarded as a patchwork lattice of many different, albeit equivalent, topological phases. Moreover, the Bacon-Shor codes enable the combination of favorable classical models using the underlying gauge structure of subsystem codes. To develop new models one might consider adapting these composition tools further to construct new hybrid Hamiltonians with features we require of a quantum memory.

Finally, we conclude by emphasizing that all the promising theoretical models that are proposed must be developed until they are sympathetic to the engineering constraints of the laboratory. Certain proposals for a finite-temperature quantum memory are more amenable to experimentalists than others, for instance, the ferromagnetic coupled toric code was designed with physical media in mind. Moreover, subsystem codes offer an avenue to simplify the architectural challenges of building high-weight stabilizer models. However, in spite of a few exceptions, experimentally amenable quantum memories that can be realized using existing technology is an avenue of research which thus far remains untrodden. We consider the example of the celebrated cubic code model; a well studied model whose favorable partial self-correcting properties have been analyzed and numerically verified. We must work now to develop models such as this one into a form that an experimentalist can prepare in the laboratory. Undoubtedly, such an achievement will be directly incorporated in the design of the fault-tolerant quantum technologies of the future.

\begin{acknowledgements}
We thank R. Alicki, H. Bomb\'in, C. Brell, D. Browne, A. Caldeira, S. Flammia, A. Hamma, A. Hutter,  A. K\'{o}m\'ar, A. Kubica, O. Landon-Cardinal, Z. Nussinov, G. Ortiz, F. Pastawski, J. Preskill, S. Simon, K. Temme and B. Yoshida together with our three anonymous referees for insightful comments, interesting discussions and for directing us to important parts of the literature. BJB would also like to thank Sean Barrett for introducing him to this exciting area of research. We acknowledge the Imperial College HPC Service for computational resources.  BJB is supported by the EPSRC and the Villum Foundation, JKP and CS are supported by the EPSRC, JRW and DL are supported by Swiss NF, NCCR Nano and NCCR QSIT.
\end{acknowledgements}

\appendix
\section{Decoders}
\label{Sctn:Decoders}

Throughout this Review we make use of decoding algorithms for numerical analysis. Given an encoded quantum state that has been subject to noise, a decoder takes classical syndrome information, i.e. all the outcomes of stabilizer measurements, and returns a correction operation that returns the state to the code space~\cite{DennisKitaevLandahlPreskill}. Provided the errors occur at a suitably low density, the decoder will successfully find a correction operator that will recover the initially encoded state with probability that grows with the size of the system. Here we briefly discuss decoding, and a particular decoding routine, namely the clustering decoder. This present discussion make use of the stabilizer formalism described in Sec.~\ref{Sctn:HamiltoniansBasics}.

Even in the case that we find a self-correcting quantum memory, a decoding step will still be required to correct for errors caused by small thermal fluctuations when information is read from the system. To illustrate this, we briefly consider the two-dimensional Ising model, as discussed in Subsec.~\ref{Sctn:TheIsingModel}. We encode classical bits in the two-fold degenerate ground space, and we read out by measuring the magnetization; the average spin value of all the spins of the system. The ground states for the model have magnetization $\pm 1$. At finite temperature in the limit of large system sizes we cannot expect all the spins to be aligned. Instead it is suitable to take the sign of the magnetization measurement to readout the memory. This measurement corresponds to taking a majority vote over all the encoded spins of the lattice. Measuring the encoded ferromagnet in this way accounts for small thermal fluctuations that take the spin configuration out of the ground space. For robust quantum information storage we require a decoding algorithm to deal with small errors incident to a code during readout.

\begin{figure}
\includegraphics{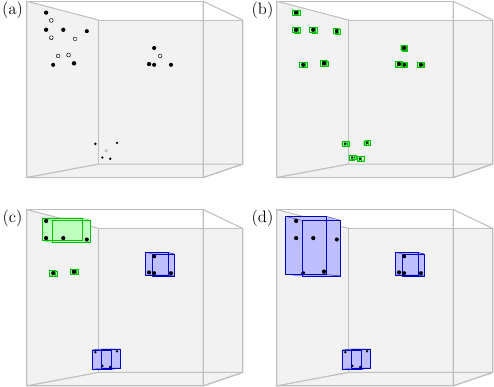}
\caption{\label{DecodingSketch}(Color online) The clustering decoding algorithm. (a) Unknown errors, marked in white on a three-dimensional lattice, are identified by syndrome measurements, marked in black. (b) The measured syndromes are initially contained in unit boxes. The unknown errors are not shown in this Figure. (c) Box sizes increase to contain other nearby syndromes within a small fixed radius of the existing boxes. Boxes that contain a correction operator are colored dark blue, they are otherwise colored light green. (d) The search increases the box size to find boxes large enough to contain correction operators for all the syndromes on the lattice.}
\end{figure}

As we cannot measure the state of individual physical qubits of a code, accounting for the errors during the readout of a quantum code is not as straight forward as the classical case we have described. Instead, for the quantum case, we perform stabilizer measurements to learn the errors that are incident to a code. The stabilizer measurements perform the task of collapsing the incident noise onto an error $E$ that is an element of the Pauli group, and provides syndrome information we can use to attempt to determine $E$. It is the task of the decoder to predict the error $E$ of the Pauli group, and return a correction operator $C$ such that $CE$ acts trivially on the encoded state.

Many approaches to decoding have been studied with tradeoffs between speed, performance, and versatility. Decoders have been designed that make use minimum-weight perfect matching~\cite{DennisKitaevLandahlPreskill}, renormalization group techniques~\cite{Duclos-CianciPoulin10, Duclos-CianciPoulin13}, and Monte-Carlo methods~\cite{WoottonLoss12, HutterWoottonLoss13}. Moreover, the study of decoding has foundations in the study of glassy statistical mechanical models~\cite{DennisKitaevLandahlPreskill, WangHarringtonPreskill, Bombin12a, Andrist}.

Here, we describe the clustering decoder which is commonly used throughout this Review. The clustering decoder is introduced in~\cite{Harrington} and developed by Bravyi and Haah specifically for the study of the cubic code at finite temperature in Refs.~\cite{BravyiHaah11A, BravyiHaah13}, as we have discussed in Subsec.~\ref{Sctn:CubicCode}. The clustering decoder is further refined in Refs.~\cite{Anwar14, Hutter14}. This simple algorithm is particularly suitable for the present work as it can be adapted for any translationally invariant local stabilizer code~\cite{BravyiHaah11A}.

To find a correction operator for the most-likely error configuration, the clustering decoder will implicitly make use of the locality constraint of commuting Pauli Hamiltonian models. In addition to this, we assume that a low-weight correction operator will approximate the inverse of the most probable error that has occurred in the limit of a low error rate.

Here we sketch the clustering algorithm routine. A rigorous explanation of the implementation of the decoder can be found in Refs.~\cite{BravyiHaah11A, Anwar14, Hutter14}. In Fig.~\ref{DecodingSketch}(a) we show a series errors, marked in white, and the syndrome that corresponds to the error. Violated stabilizers, i.e., stabilizers that return -1 outcomes, are marked by black points for some local code. We consider this example to demonstrate the clustering decoder.

To find a low-weight correction we find a set of small boxes that enclose all of the error syndromes. We search for a set of boxes that contains a correction operator that is consistent with the violated stabilizers in each box. The algorithm begins by putting all the violated stabilizers in individual boxes of unit size, as shown in Fig.~\ref{DecodingSketch}(b). The algorithm then proceeds by incrementally increasing the size of the boxes. This is achieved by combining pairs of boxes that lie within a some small radius of one another. The routine continues until the boxes are large enough to contain a correction operator that corrects for all the violated stabilizers contained within the box. 

\begin{figure}
\includegraphics[width=\columnwidth]{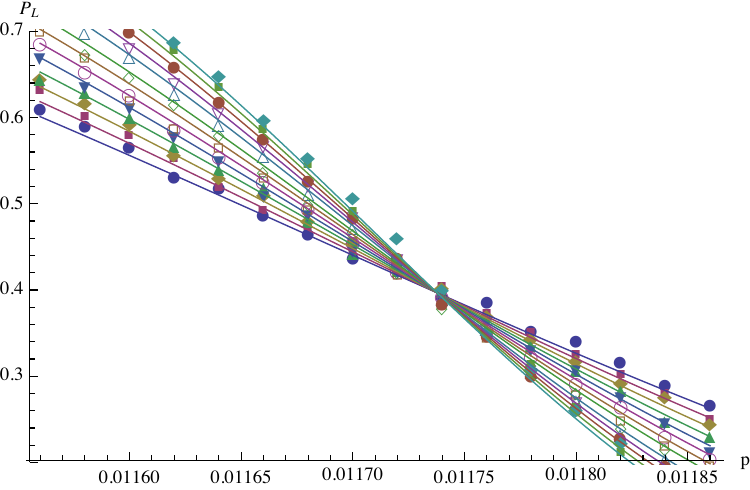}
\caption{(Color online) Threshold calculation using the clustering decoder for the cubic code model. We study an independent and indentically distributed bit-flip noise model using system sizes between $L = 101$ and $L = 201$ with Monte Carlo samples. We find the crossing at $p_{\text{th}} = 1.17\% $. \label{CubicCodeThreshold}}
\end{figure}

For the initially chosen boxes, where all of the boxes are of unit size, we cannot find a local correction operator that is consistent with the single violated stabilizer that is contained within each box. We increase the size of the boxes by checking within a fixed radius $r$ of each of the violated stabilizers of each box. In the event that another violated stabilizer contained in a different box is found within a distance $r$ of the violated stabilizer from which we are searching, their two respective boxes are combined giving a single larger box. Once it is confirmed that no pairs of disjoint boxes contain any violated stabilizers that lie within distance $r$ of one another, we check to find a correction operator that is consistent with all of the violated stabilizers in each box. 

If a box contains a correction operator that is consistent with all of its violated stabilizers, the box is considered neutral, and the violated stabilizers within the box are no longer considered in later searches of the routine. In the case that all the boxes are neutral, the algorithm terminates and the correction operator of each of the boxes is returned. If some boxes remain unneutralized, $r$ is increased and the algorithm is repeated for the violated stabilizers that are contained in boxes that are not yet neutral. 

We continue to follow the example syndrome in Fig.~\ref{DecodingSketch}. In Fig.~\ref{DecodingSketch}(c) we show the new boxes obtained after searching within a radius of $r =1$ of each of the violated stabilizers. After this search, not all the boxes contain a local correction operator. Two boxes, colored blue in Fig.~\ref{DecodingSketch}(c), are neutralized. Violated stabilizers in the blue boxes are no longer considered in the algorithm. Three boxes found at radius $r=1$, marked in green, do not contain a correction operator consistent with their respective violated stabilizers. To find a correction operator consistent with the violated stabilizers in the green boxes, the algorithm increases its box search radius once again to $r = 2$ for all the remaining violated stabilizers. At $r = 2$, all boxes contain a correction operator that is consistent with all of the violated stabilizers. Fig.~\ref{DecodingSketch}(d) shows the boxes that together contain a correction operator consistent with the syndromes of the error configuration. The decoder will return the correction operator contained within the blue boxes to return the encoded information to the ground state, enabling the readout of the encoded quantum information. 

We remark that this algorithm is suitable for any translationally invariant local stabilizer code. The description of the algorithm we give makes no reference to the underlying code. We need only assume that the code is local, such that violated stabilizers can be interpreted as lying within a fixed radius of an incident error. Moreover, it is shown in Ref.~\cite{BravyiHaah11A} that for transitionally invariant codes we can determine efficiently if a box contains a correction operator consistent with its enclosed violated stabilizers.

We can evaluate decoder performance by obtaining a {\em threshold} with respect to an identical and independently distributed noise model. This noise model is where each qubit suffers an error with probability $p$. The threshold value $ p_{\text{th}}$ is the value below which the logical error rate of a quantum error-correcting code decreases as we increase the size of the system to the thermodynamic limit. In this Review we use cluster decoding for the toric code model in Subsec.~\ref{SEC:TCfinitetemp}, and the cubic code in Subsec.~\ref{Sctn:CubicCode}. The toric code threshold is already found to be $\sim 8.3 \%$, shown in Ref.~\cite{Anwar14}. The threshold for the for the cubic code using the clustering algorithm has not been published to the best of the Authors' knowledge, though good estimates are given in Ref.~\cite{BravyiHaah11A}. We estimate a threshold $p_{\text{th}} \sim 1.17\%$ for the cubic code using a bit-flip noise model. We show the threshold data in Fig.~\ref{CubicCodeThreshold}.

\section{Simulating the Toric Code at Finite Temperature}
\label{Sctn:NumericsForTC}

Here we present additional numerical results for the finite-temperature behavior of the toric code, supporting the arguments we make in Sec.~\ref{SEC:TCfinitetemp}.

\begin{figure}[b]
\includegraphics[width=0.92\columnwidth]{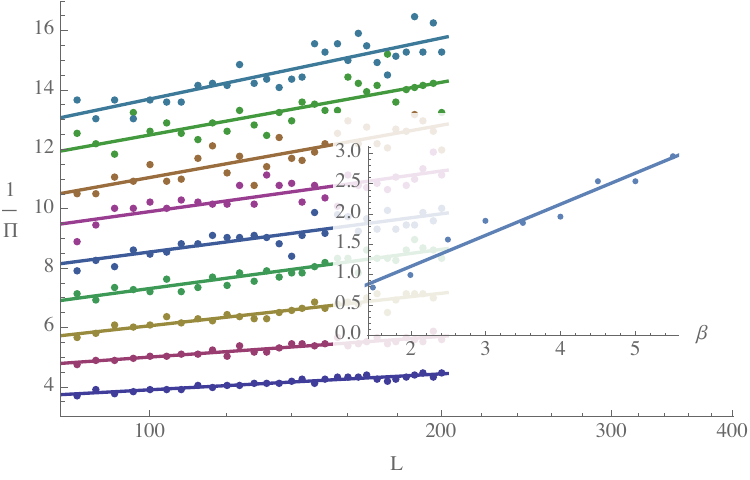}
\caption{ \label{Fig:TCLowTempPi}(Color online) Plot showing the reciprocal of the probability that a pair of toric code excitations reach critical separation $L/2$. 
Values of $\Pi$ are obtained by averaging over 10000 simulations. They are plotted against $L$ over a range of temperatures from $\beta = 1$ (bottom line) to $\beta = 6$ (top line). 
The linear fittings show that $1/\Pi$ grows linearly with $\ln(L/2)$, with a gradient that increases with $\beta$. The Inset shows the gradients found with the fittings shown in the main plot displayed as a function of $\beta$.}
\end{figure}

For system sizes $L$ that are small compared to the natural scale imposed by the finite-temperature dynamics, decoherence is typically the result of a single pair of anyons. In Sec.~\ref{SEC:TCfinitetemp} we predicted scaling with $L$ and inverse temperature $\beta = 1/T$ of three independent elements of the coherence time, which we called $\Pi(L,\beta)$, $\tau_c$ and $\tau_m$. We isolate each of these terms and estimate them numerically.

We first investigate $\Pi(L,\beta)$. This term quantifies the probability that after a pair of anyons is created they do not annihilate by mutually fusing together before reaching a significant enough distance to cause logical errors, $L/2$. 
To evaluate this function we alter the standard simulation scheme; rather than beginning in a ground state we initialize the system with a single pair of anyons present on the lattice, where the initialized anyons are separated by a single lattice spacing. We additionally set the rate of creation equal to zero such that no additional pairs of anyons are created. Indeed, we model only the random motion of a single pair of anyons walking across the lattice. We evolve the system until either the separation of the anyons reaches the Euclidean distance $L/2$, or the pair meet at a common point on the lattice and annihilate. The quantity $\Pi$ is the fraction of samples that reach separation $L/2$ rather than annihilating. We estimate $\Pi$ by sampling over $10^4$ trials. The results we obtain are shown in Fig.~\ref{Fig:TCLowTempPi}. We find the fitting 
\be
1/\Pi = (0.108 + 0.513 \beta) \ln(L/2) .
\label{EQN:NumericalPi}
\ee
This is in good agreement with the scaling we hypothesised in Eqn.~\eqref{EQN:TCPi}, where constant $A \approx 5$.

\begin{figure}
\includegraphics[width=0.92\columnwidth]{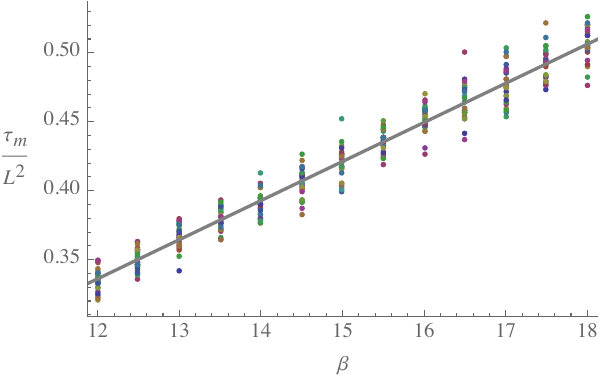}
\caption{ \label{Fig:TCLowTempTauM}(Color online) Contribution to the coherence time from anyon motion in the small system size limit.
Times $\tau_m$ are obtained by averaging over 1000 simulations. They are plotted against $\beta$ for a range of system sizes $L=$50, 52, 54, $\dots$, 100. 
The values of $\tau_m$ are divided by $L^2$ to show data points that are independent of the size of the system, thus validating the $L^2$ factor we derive on the right-hand side of Eqn.~(\ref{EQN:taum}). 
The linear fit displayed uses the average values of the fit parameters obtained from the different system sizes $\tau_m \sim 0.028 \beta L^2$.}
\end{figure}

We next numerically study how $\tau_c$ and $\tau_m$ scale with $L$ and $\beta$. Time $\tau_c$ is the typical time it takes for a pair of anyons to be created that will cause a logical error, and $\tau_m$ is the average time it takes for a pair of anyons to achieve separation $L/2$ after they have been created. To find these values we simulate the thermal evolution of the system prepared in a ground state. We attempt to decode the lattice after each simulated event of the thermal evolution, and the simulation is stopped once the decoder fails. Every time a pair of anyons is created its creation time is recorded, this information is discarded if the pair annihilates. 
Estimates of the creation timescale $\tau_c$ and diffusion time $\tau_m$ are obtained by averaging over $10^3$ simulation runs. This data is plotted in Fig.~\ref{Fig:TCLowTempTauC} and Fig.~\ref{Fig:TCLowTempTauM}, respectively. The data we obtain scales as we expect with fittings given in Eqns.~\eqref{EQN:taum} and~\eqref{EQN:tauc}, where we find numerical fits $\tau_m \simeq 0.028 \beta L^2$ and
\be
\tau_c = 0.150 \frac{e^{1.99\beta}}{L^{2.01}} \frac{1}{\Pi} \, ,
\ee
where we take the function $\Pi$ that we evaluated numerically, Eqn.~\ref{EQN:NumericalPi}.

For large systems, decoherence results due to the interaction of many thermally created anyons. In Sec.~\ref{SEC:TCfinitetemp} we describe a model of these dynamics that gives an estimate, Eqn.~\eqref{EQN:tauh}, of the coherence time $\tau_{\textrm{large}}$. We test some of the assumptions of this model and compare the predicted coherence time to numerical values.

\begin{figure}
\includegraphics[width=0.92\columnwidth]{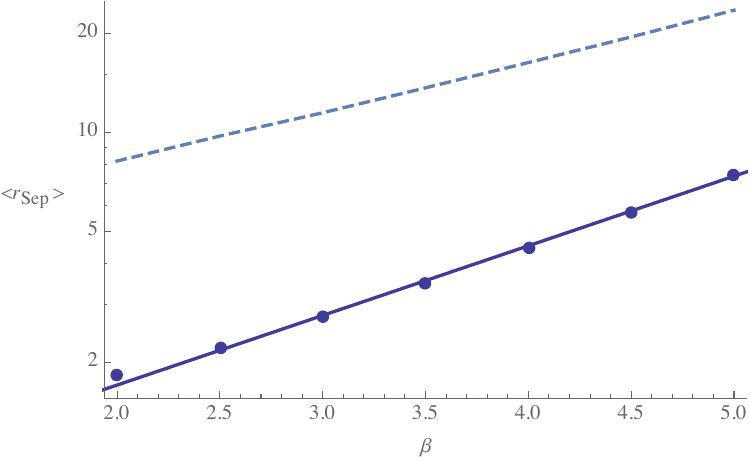}
\caption{ \label{Fig:TCHighTempAvgSep}(Color online) Average separation between paired anyons at the time the decoder fails in the large system-size limit.
The separation is plotted as a function of $\beta$ for fixed system size $L = 200$ where each data point is obtained by averaging over 1000 simulations.
The solid line is a linear fit to the data with gradient $\sim 0.49$ on logarithmic axes, consistent with a scaling of $\langle r_{\textrm{sep}} \rangle \sim e^{\beta \Delta/2}$. 
The dashed line is the maximum pair separation, averaged over the simulations, which is seen to be much smaller than $L/2$.}
\end{figure}

\begin{figure}[b]
\includegraphics[width=0.92\columnwidth]{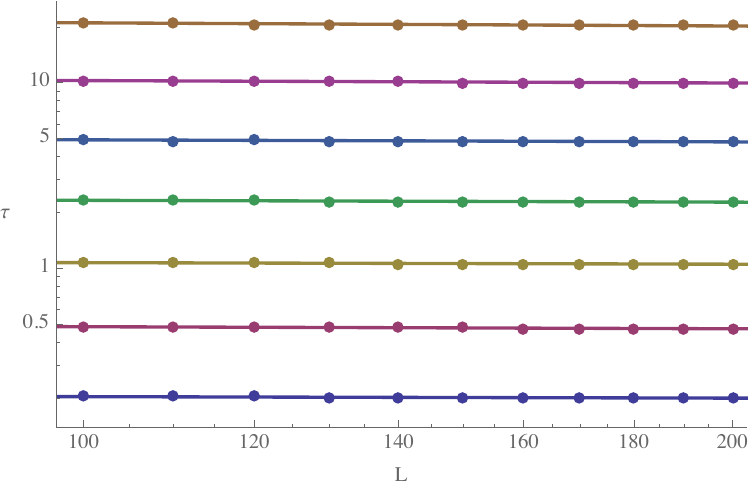}
\caption{ \label{Fig:TCHighTempTauVsL}(Color online) Plot showing that the coherence time of the toric code is independent of system size at high temperature. 
The quantity $\tau$ is shown on a log scale against $L$ for different temperatures $\beta = 2,\, 2.5,\ldots,\, 5$ with $\beta = 2 $ the bottom line and $\beta = 5$ the top line. Data points are found by taking an average over 1000 simulations. Linear fits show very small negative gradients ranging from $-0.03$ for the lowest value of $\beta$, down to $-0.06$ for the highest $\beta$. 
Indeed, the zero gradient is well within the $95\%$ confidence interval of each fit.}
\end{figure}

We simulate the system evolving in the large size regime. The decoder runs after every event is introduced to the system by the thermal evolution. At the earliest time the decoder fails we stop the simulation and record the time elapsed during the simulation. Results are obtained by averaging over $10^3$ simulations runs. We find average anyon numbers $\langle N \rangle$ which confirm that the anyon density at the time the decoder fails scales like $\rho \sim e^{-\beta \Delta}$. These numbers are also seen to satisfy $\langle N \rangle \gg 1$, indicating that our data is taken for sufficiently large systems.

In Sec.~\ref{SEC:TCfinitetemp} we argued that for large systems, in contrast to the smaller case, the important length scale is $\Lambda \sim e^{\beta \Delta/2}$ as opposed to system size. This value is the typical separation between creation events and thus corresponds to the average distance each anyon must diffuse to cause the decoder to fail.

To test that the described motion is the main decoherence mechanism for large systems
we study the distance between anyon pairs at the time the decoder fails. During the simulation each pair of anyons that are given a unique mark to indicate their pair created partner. At the end of the simulation we measure the Euclidean distance between each marked pair. If two anyons from two separate pairs fuse, the remaining two unpaired anyons on the lattice are marked as members of the same extended pair.

As one might expect, we find that extended pairs created from a fusion will typically achieve a greater separation than pairs that are initialized by creation from vacuum. However, its effect is small with respect to the average data. Our numerical results show that the average separation between all anyon pairs grows like $e^{\beta \Delta/2} $, as predicted. In addition we find that the maximum separation between any pair is always much less than $L/2$. The scaling of the average and maximum separations are shown in Fig.~\ref{Fig:TCHighTempAvgSep}. These observations confirm that the decoherence results from the average motion of anyons in local regions.

The scaling of coherence time with $\beta$ is shown in Fig.~\ref{Fig:TCHighTempTau}.
We see that the data reproduces the exponential dependence on $\beta \Delta$ predicted in Eqn.~\eqref{EQN:tauh}. Another prediction of the model is that $\tau_{\textrm{large}}$ is independent of system size. Fig.~\ref{Fig:TCHighTempTauVsL} plots the numerical values of coherence times against $L$ showing that the data is consistent with this hypothesis.

\bibstyle{plain}

\end{document}